\pdfoutput=1
\documentclass[a4paper,11pt]{article}
\usepackage[T1]{fontenc}
\usepackage{a4wide}
\usepackage{amsmath,amssymb,amsthm,mathtools,array,booktabs}
\usepackage{graphicx,hyperref}
\usepackage{xcolor}
\usepackage{cancel}
\usepackage{tikz}
\usetikzlibrary{shapes.geometric}

\hypersetup{
  unicode,
  bookmarksnumbered,
  linktoc = all,
  hidelinks
}

\def\mC{\mathbb{C}}
\def\mZ{\mathbb{Z}}
\def\mR{\mathbb{R}}

\def\cC{\mathcal{C}}

\def\cF{\mathcal{F}}
\def\cB{\mathcal{B}}
\def\cK{\mathcal{K}}

\def\cT{\mathcal{T}}
\def\cO{\mathcal{O}}
\def\cW{\mathcal{W}}

\def\CP{{{\mathbb{P}}}}

\let\Re\relax
\let\Im\relax
\DeclareMathOperator{\Re}{Re}
\DeclareMathOperator{\Im}{Im}
\DeclareMathOperator{\Star}{Star}
\DeclareMathOperator{\Ann}{Ann}
\DeclareMathOperator*{\res}{res}
\DeclareMathOperator{\Sym}{Sym}
\DeclareMathOperator{\Span}{Span}
\DeclareMathOperator{\Lie}{Lie}
\DeclareMathOperator{\Hom}{Hom}
\DeclareMathOperator{\Cone}{Cone}
\DeclareMathOperator{\rank}{rank}
\DeclareMathOperator{\ch}{ch}

\DeclareMathOperator{\sign}{sign}
\DeclareMathOperator{\Vol}{Vol}
\DeclareMathOperator{\End}{End}
\DeclareMathOperator{\Pexp}{Pexp}
\DeclareMathOperator{\id}{id}
\DeclareMathOperator{\Tr}{Tr}

\DeclarePairedDelimiter{\abs}{\lvert}{\rvert}
\newcommand{\CC}{{\mathbb{C}}}
\newcommand{\RR}{{\mathbb{R}}}
\newcommand{\QQ}{{\mathbb{Q}}}
\newcommand{\ZZ}{{\mathbb{Z}}}
\newcommand{\lie}{\mathfrak}

\newcommand{\tW}{{\widetilde{W}}}

\numberwithin{equation}{section}

\hypersetup
  {
    pdftitle={B-brane transport in anomalous (2,2) models and localization},
    pdfauthor={Joel Clingempeel, Bruno Le Floch, Mauricio Romo}
  }
\begin{document}
\thispagestyle{empty}
\centerline{\LARGE\bf\boldmath B-brane transport in anomalous $(2,2)$ models and localization}
\vspace{8mm}
\begin{center}
{\large Joel Clingempeel\footnote{{\tt jac634@math.rutgers.edu }}${}^{\dagger}$ Bruno Le Floch\footnote{{\tt lefloch@lpt.ens.fr}}${}^{\ddagger}$  and Mauricio Romo\footnote{{\tt mromoj@ias.edu}}${}^{\ast}$}
\end{center}
\vspace{6mm}
\begin{center}\footnotesize
${}^{\dagger}$ Department of Mathematics,
Rutgers University,
Hill Center - Busch Campus,\\
110 Frelinghuysen Road,
Piscataway, NJ 08854-8019, USA\\[2mm]
${}^{\ddagger}$ CNRS and LPTHE, Sorbonne Universit\'e,\\ 4 place Jussieu, 75005 Paris, France\\[2mm]
${}^{\ast}$ School of Natural Sciences, Institute for Advanced Study,\\ Princeton, NJ 08540, USA\\[2mm]
${}^{\ast}$ Yau Mathematical Sciences Center, Tsinghua University, Beijing, 100084, China
\end{center}
\vspace{15mm}

\begin{abstract}
\noindent\normalsize
We study how B-branes in two-dimensional $\mathcal{N}=(2,2)$ anomalous models behave as we vary the energy scale and bulk parameters in the quantum K\"ahler moduli space.
We focus on $(2,2)$ theories defined by abelian gauged linear sigma models (GLSM).
Guided by the hemisphere partition function we find how B-branes split in arbirary phases into components on the Higgs branch and other branches: this generalizes the band restriction rules of Herbst-Hori-Page to (abelian) anomalous models.

Secondly, we address divergences in non-compact models, through the central example of GLSMs for Hirzebruch-Jung resolutions of cyclic surface singularities.
For a brane with compact support we explain how to regularize and compute the hemisphere partition function and extract its Higgs branch component, which we match in the zero-instanton sector to the geometric central charge of the brane.
To this aim, we clarify the definition of zero-instanton geometric central charge for objects in the derived category of a non-compact toric orbifold.
\end{abstract}

\newpage
\setcounter{tocdepth}{2}
\tableofcontents
\setcounter{footnote}{0}

\section{Introduction}

The study of the dynamics of gauged linear sigma models (GLSMs) \cite{Witten:1993yc} has been a continuous source of new results in physics and mathematics.
GLSMs are two-dimensional ${\cal N}=(2,2)$ supersymmetric gauge theories that can describe the world-sheet theories of strings propagating on certain space-times.
In this context, boundary conditions of GLSMs can describe D-branes.

\subsection{B-branes: from GLSM to SCFT}

\paragraph{GLSM symmetries and B-branes.}
Classically, GLSMs have left and right $U(1)$ R-symmetries, or equivalently vector $U(1)_{V}$ and axial $U(1)_{A}$ R-symmetries,
and $U(1)_{V}$ can only be broken by the superpotential term.
For appropriate superpotentials~$W$ that are quasi-homogeneous under~$U(1)_{V}$, a non-renormalization theorem ensures that the quantum theory has this symmetry too.
Coefficients of F-terms (superpotentials) are protected by supersymmetry, which makes them invariant under the RG flow.
On the other hand, $U(1)_{A}$ may fail to be a symmetry of the quantum theory due to an anomaly, in which case we refer to the theory as an anomalous GLSM\@.
Explicitly, if the matter content of the GLSM transforms in a representation $\rho\colon G\to GL(V)$ of the gauge group $G$, the anomaly is proportional to the weight $b$ of the character $\det(\rho)\colon G\to \mathbb{C}^{*}$.
The theory is thus non-anomalous when $\rho$ factors through $SL(V)$.
The same weight $b$ controls the renormalization of twisted F-terms in the action: the FI-theta coefficient $t=\zeta_{(\text{FI})}-i\theta$ receives a 1-loop correction proportional to $b\log\mu$, where $\mu$ is the energy scale.
This renormalization is important to understand the infrared (IR) dynamics of anomalous GLSMs.

In a GLSM, we can consider boundary conditions that preserve a particular subset of the supersymmetries.
We will consider those that preserve the B-type supersymmetry algebra $\mathbf{2}_{B}$ generated by supercharges that have charge $+1$ under $U(1)_{A}$.
Boundary conditions invariant under this subalgebra of the $(2,2)$ supersymmetry algebra are termed B-branes.
While B-branes have been studied in detail in superconformal field theories, the same definition applies to any $\mathcal{N}=(2,2)$ theory.
In the context of GLSMs~\cite{Herbst:2008jq}, B-branes are known to form a category that can be defined and studied mathematically \cite{Segal:2010cz,halpern2015derived,ballard2012variation}.

One of the main insights of~\cite{Herbst:2008jq} was to define transport of B-branes as one varies the twisted chiral parameters~$t$ belonging to the so-called quantum K\"ahler moduli space $\mathcal{M}_{K}$:
twisted chiral terms in the bulk are D-terms of the boundary supersymmetry, so that B-branes can be transported modulo D-terms of $\mathbf{2}_{B}$.
In the abelian non-anomalous cases studied in~\cite{Herbst:2008jq} this notion is exploited to find equivalences between categories of B-branes, which may have different descriptions on the various K\"ahler cones in~$\mathcal{M}_{K}$.
B-brane transport functors found in this way have been extended to nonabelian non-anomalous GLSMs \cite{grr,procnonabelian} by the third-named author, and an equivalent approach to these functors has been considered recently in the mathematical literature \cite{Addington:2014sla,MR3223878,2016arXiv160904045V}, for the same classes of models.
B-brane transport has been explored much less in anomalous models using this approach, as one must account for the renormalization of the FI-theta parameters.

\paragraph{Phase structure.}

In the non-anomalous case, at fixed~$W$, each point in~$\mathcal{M}_{K}$ corresponds to an $\mathcal{N}=(2,2)$ SCFT defined by RG flow of the GLSM\@.
The SCFTs and their B-brane categories typically have concrete geometrical descriptions in different phases (K\"ahler cones) inside~$\mathcal{M}_K$, as depicted below.
Two natural questions are (1)~for each phase, determine the functor that maps a B-brane of the GLSM to its image under the RG flow, which is a B-brane of the SCFT, and (2)~determine the functor that describes how a B-brane is transported from one phase to another.
This second functor is an equivalence of categories.
\[
  \begin{tikzpicture}
    \node[draw,ellipse](GLSM) at (2,0) {\mbox{}\qquad GLSM\qquad\mbox{}};
    \node at (-.4,0) {\scriptsize $\mathcal{M}_K$};
    \node(phaseI) at (.55,-2) {Phase I};
    \node at (.55,-2.34) {\scriptsize SCFT};
    \node(phaseII) at (3.45,-2) {Phase II};
    \node at (3.45,-2.34) {\scriptsize SCFT};
    \draw[thick,->] (GLSM.south west) -- (phaseI) node [pos=.4,left]{\scriptsize\begin{tabular}{@{}r@{}}RG flow\\$t=\text{const \ }$\end{tabular}};
    \draw[thick,->] (GLSM.south east) -- (phaseII) node [pos=.4,right]{\scriptsize\begin{tabular}{@{}l@{}}RG flow\\\ \ $t=\text{const}$\end{tabular}};
  \end{tikzpicture}
\]

In anomalous GLSMs the situation is more elaborate.
There are two interesting limits:
\begin{itemize}
\item the RG flow decouples gauge degrees of freedom and moves the FI-theta parameter deep in specific K\"ahler cones;
\item the gauge-decoupling limit, as we call it, consists of decoupling the gauge degrees of freedom by taking the gauge coupling $g_{\text{ren}}(\mu)\to\infty$ while keeping the FI-theta parameter $t_{\text{ren}}(\mu)$ fixed at some fixed finite energy scale~$\mu$.
\end{itemize}
The second limit is more flexible than the first one, as it allows the FI-theta parameter to explore arbitrary K\"ahler cones.
Deep in a K\"ahler cone, an anomalous GLSM can have disjoint branches, for example a Higgs branch and some massive vacua, as we explain in \autoref{sec:section2}.
Its gauge-decoupling limit is described by a direct sum of one (possibly trivial) SCFT for each of these branches.
Schematically, the situation can be depicted as follows.
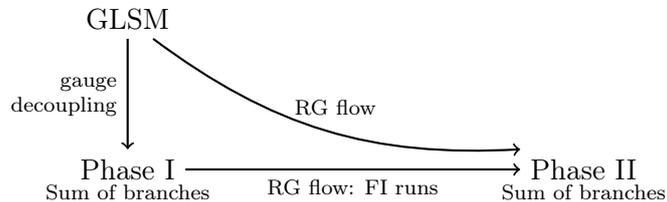
\[
  \begin{tikzpicture}
    \node[draw,ellipse](GLSM) at (3,0) {\mbox{}\qquad\qquad\qquad GLSM\qquad\qquad\qquad\mbox{}};
    \node at (-1.7,0) {\scriptsize $\mathcal{M}^{\text{bare}}_{K}$};
    \node(phaseI) at (0.02,-2) {Phase I};
    \node at (0.03,-2.3){\scriptsize Sum of branches};
    \node(phaseII) at (5.98,-2) {Phase II};
    \node at (5.98,-2.3){\scriptsize Sum of branches};
    \draw[thick,->] (GLSM.south west) -- (phaseI) node [midway,left]{\scriptsize\begin{tabular}{@{}r@{}}gauge decoupling\\$t_{\text{bare}}$ fixed, $g_{\text{bare}}\to+\infty$\end{tabular}};
    \draw[thick,->] (GLSM.south east) -- (phaseII) node [midway,right]{\scriptsize\begin{tabular}{@{}l@{}}gauge decoupling\\or RG flow\end{tabular}};
    \draw[thick,->] (GLSM.186) to [bend right=18] node [pos=.45,above right,inner sep=1pt]{\scriptsize\begin{tabular}{@{}l@{}}\hspace{-2em}RG flow with\\$g_{\text{bare}},t_{\text{bare}}$ fixed\end{tabular}} ([yshift=1ex]phaseII.west);
    \draw[thick,->] (phaseI) -- (phaseII) node [midway,below]{\scriptsize\begin{tabular}{@{}l@{}}RG flow with\\$t_{\text{bare}}$ fixed\end{tabular}};
  \end{tikzpicture}
\]
Each point in the classical space $\mathcal{M}^{\text{bare}}_{K}$ of bare FI-theta parameters ``flows'' in general to one of these direct-sum phases in the gauge-decoupling limit, and a B-brane can split in this limit into components on the Higgs branch and other branches.
There are now three natural questions, (1)~find the gauge-decoupling image of GLSM B-branes as B-branes of the SCFTs describing each branch, (2)~transport B-branes between phases, possibly restricted to particular branches,\footnote{Note that B-brane transport is only an equivalences of categories if all branches are taken into account.  If one projects out mixed and Coulomb branches to obtain categories of B-branes on Higgs branches, which have convenient geometric descriptions in each K\"ahler cone, one should at best expect brane transport to be an embedding of one category into the other instead of an equivalence as in non-anomalous models, due to some branes ``leaking out'' onto the other branches.} (3)~find the image of GLSM B-branes under the RG flow by combining the gauge-decoupling limit and transport (caused by renormalization of~$t$).

\subsection{Band restriction rule for abelian GLSMs}

\paragraph{An intertwined approach.}

We explore the gauge-decoupling limit and B-brane transport in the realm of abelian GLSMs with vanishing superpotential
\begin{equation}
W = 0 ,
\end{equation}
with an emphasis on anomalous models.
Their definition and bulk dynamics are reviewed in \autoref{sec:section2}, and we develop there a detailed description of all  Higgs and mixed and Coulomb branches.
Determining directly how B-branes of the GLSM decompose into these branches in the $g\to\infty$ limit is unpractical in abelian GLSMs with gauge groups of rank $r>1$, so we use a multi-step strategy.
\begin{itemize}
\item We work out in \autoref{ssec:Bbraneabelian} the gauge-decoupling limit $F_{g\to\infty}(\cB)$ of a GLSM B-brane~$\cB$ in a ``pure-Higgs phase''
  that solely features a Higgs branch.  It admits a geometric quotient construction denoted schematically as $F_{g\to\infty}(\cB)=F_{\text{geom}}(\cB)$.  In particular, we identify some B-branes of the GLSM (which we dub ``empty branes'') whose gauge-decoupling limit is empty in these phases, and thus in all phases by a holomorphy argument.

\item We then turn (in \autoref{sec:Band restriction rule}) to transporting B-branes across a given wall (phase boundary).  We find it useful to introduce a ``local model'' valid deep inside the wall, which is a GLSM with a very simple gauge group $U(1)\times\Gamma$, and which is non-anomalous exactly when the wall is parallel to the renormalization direction~$b$.
\begin{itemize}
\item By tracking some integration contours defining the brane, we find that B-branes are easily transported through such a phase boundary provided they obey a band restriction rule \eqref{intro-band-1} or~\eqref{intro-band-2} arising from the grade restriction rule~\cite{Herbst:2008jq,Hori:2013ika} of the non-anomalous or anomalous local model.
\item To transport B-branes that do not obey the band restriction rule, one first binds the B-brane with empty branes to make it obey the (big) band restriction rule.
\end{itemize}

\item Starting from pure-Higgs phases, where the gauge-decoupling limit of GLSM B-branes is known, one can reach all phases by crossing walls transverse to~$b$.
  The local model of each such wall is an anomalous $U(1)$ GLSM (up to discrete factors).
  Based on the $U(1)$ case analyzed in~\cite{Hori:2013ika}, we argue in \autoref{ssec:wallcrossingHiggsing} that the same (big/small) band restriction rule also controls which B-branes acquire contributions along the mixed and Coulomb branches.  In particular, for a GLSM B-brane~$\cB$ that obeys the band restriction rules associated to every wall between the pure-Higgs phase and the phase of interest, its gauge-decoupling limit has a Higgs branch part equal to $F_{\text{geom}}(\cB)$.\footnote{\label{foot:favorable-cases}In favorable cases, the set of GLSM B-branes that obey all band restriction rules is large enough that any other brane can be bound with empty branes so as to fit in that set.
    If the set is too small (for instance in the $U(1)^2$ GLSM of \autoref{ssec:C2Zn2 brane transport} with some choices of theta angles), it is not clear with our techniques how to determine the gauge-decoupling limit of arbitrary GLSM B-branes.
    We have also not developped our technique enough to determine images of GLSM B-branes on mixed and Coulomb branches.  Nevertheless, we expect that B-brane transport will eventually allow to determine the full gauge-decoupling limit of GLSM B-branes, and not only their Higgs branch contribution.}
\end{itemize}
Our analysis inevitably goes back and forth between a discussion of the gauge-decoupling limit and of B-brane transport.
To express the band restriction rules~\eqref{intro-band-1} and~\eqref{intro-band-2} below, we now set up some notation.

\paragraph{Branches of abelian GLSMs.}

Abelian GLSMs have gauge group $G=U(1)^r\times\Gamma$ for a discrete~$\Gamma$, hence have twisted chiral parameters $t_\alpha=\zeta_\alpha-i\theta_\alpha\in\CC/(2\pi i\ZZ)$, and a collection of chiral fields $X\in V$ transforming with charges $Q^i_\alpha$, with $\alpha=1,\dots,r$, $i=1,\dots,\dim V$.
As we explain in \autoref{sec:section2}, the space of FI parameters $\zeta\in\RR^r$ is partitioned into phases by classical walls, which disappear or acquire complex codimension~$1$ due to quantum effects.

When decoupling gauge degrees of freedom (taking $g\to\infty$), deep in a given phase, the GLSM reduces to a direct sum of SCFTs describing different branches of vacua, ranging from Higgs to mixed to Coulomb branches.
The Higgs branch is a quotient of the form $(V\setminus\Delta)/G_{\CC}$ where $G_{\CC}=(\CC^*)^r\times\Gamma$ and $\Delta$ is a union of vector subspaces of~$V$ that depends on the phase.
The other branches involve non-vanishing vector multiplet scalars, and we determine in \autoref{ssec:mixedbranches} a detailed procedure to find all branches in all phases.

\paragraph{Admissible contours.}

We use a construction of B-branes on the GLSM in terms of (equivalence classes of) algebraic data~$\mathcal{B}$ and a contour~$L$ (see \autoref{ssec:BbraneGLSM}).
The contour must be admissible in the sense that it obeys asymptotic conditions on the boundary effective twisted superpotential~$\tW_{\mathcal{B}}$, which can be easily read off from the hemisphere localization results of~\cite{Hori:2013ika}.
Deforming $L$~among admissible contours does not change the B-brane.
As we vary the FI parameters in~$\mathcal{M}^{\text{bare}}_{K}$ (at a fixed energy scale) and make them cross a wall (deep in the wall) between K\"ahler cones, $\tW_{\mathcal{B}}$~changes, so $L$~must be deformed to remain admissible.
Details of brane transport then depend on whether the renormalization direction~$Q^{\text{tot}}$ of the FI parameter $\zeta=\Re t$ is parallel to the wall or not.

\paragraph{Band restriction rule: walls parallel to~$Q^{\text{tot}}$.}

For a wall parallel to the renormalization direction~$Q^{\text{tot}}$, keeping the contour~$L$ admissible is only possible for a class of branes $(\mathcal{B},L)$ called \textbf{band restricted.}  The data $\mathcal{B}$ includes a complex of Wilson lines $\cW(q)$ with charge vector~$q$ under the GLSM's gauge group, and band restriction is the inequality
\begin{equation}\label{intro-band-1}
  \abs[\big]{(\theta\cdot u) + 2\pi (q\cdot u)} < \frac{\pi}{2} \sum_j |Q^j\cdot u| ,
\end{equation}
where $u$~is a vector normal to the wall, $\theta$~is the theta angle at which the wall is crossed, and the sum ranges over charge vectors~$Q^j$ of all chiral multiplets.
While \eqref{intro-band-1}~is formally identical to the band restriction rule for nonanomalous GLSMs~\cite{Herbst:2008jq}, our discussion also applies to anomalous models.
If a brane $(\cB,L)$ does not fit in the band~\eqref{intro-band-1}, one can bind it (non-uniquely) with some branes that are empty in the gauge-decoupling limit, in such a way that the resulting brane fits in the band.

\paragraph{Band restriction rule: walls transverse to~$Q^{\text{tot}}$.}

For a wall not parallel to~$Q^{\text{tot}}$ there exists a contour~$L$ that is admissible regardless of~$\mathcal{B}$ and of the FI parameter.
The decomposition of this contour in terms of contributions from the Higgs branch and other branches is however delicate.
In $U(1)$ models it relies on a saddle-point analysis in~\cite{Hori:2013ika}.
What we find in \autoref{ssec:wallcrossingHiggsing} for general abelian GLSMs is as follows.
Consider a B-brane described as a (complex of) Wilson lines $\cW(q)$ with charge vector~$q$ under the GLSM gauge group.
Let the FI-theta parameter cross a wall between K\"ahler cones, deep in the wall.
Denoting by $u$ a vector normal to the wall, we define two (nested) conditions on charge vectors~$q$:
\begin{equation}\label{intro-band-2}
  \begin{alignedat}{4}
    & \textbf{small band} & \qquad &
    \abs[\big]{(\theta\cdot u) + 2\pi (q\cdot u)} < \pi \min(N_{u,+},N_{u,-}) , \\
    & \textbf{big band} & &
    \abs[\big]{(\theta\cdot u) + 2\pi (q\cdot u)} < \pi \max(N_{u,+},N_{u,-}) ,
  \end{alignedat}
\end{equation}
with $N_{u,\pm} = \frac{1}{2}\sum_j (|Q^j\cdot u|\pm Q^j\cdot u)$ the total positive/negative charges in the direction~$u$,
where $Q^j$ are charge vectors of every chiral multiplet under the gauge group of the GLSM\@.
We completely characterize how the Wilson line brane $\cW(q)$ gets transported as the FI-theta parameter crosses the wall.
\begin{itemize}
\item For $q$ in the small band, the Higgs branch components of $\cW(q)$ are the same on both sides of the wall.
\item For $q$ in the big band but not the small band, the Higgs branch component of $\cW(q)$ on one side of the wall (the side defined by $(N_{u,+}-N_{u,-})(\zeta\cdot u)>0$) is transported on the other side to a combination of the Higgs branch and one mixed or Coulomb branch contribution.
\item For $q$ outside the big band, the brane can be bound with an ``empty brane'', and can be replaced in this way by a complex of Wilson lines with charges in the big band, to which the above rules apply.
\end{itemize}

\subsection{Comparisons of B-brane central charges}

Throughout the main text, we buttress our heuristic arguments by using the supersymmetric localization formula in~\cite{Hori:2013ika} for the hemisphere partition function $Z_{D^2}(\cB)$ of a GLSM with the given B-brane~$\cB$ at the boundary.
This involves decomposing the hemisphere partition function into terms and identifying them with geometrical quantities pertaining to the Higgs branch and other branches.

\paragraph{Hirzebruch-Jung model.}

We use as our central example Hirzebruch-Jung resolutions of singularities of the form $\mathbb{C}^{2}/\mathbb{Z}_{n}$ where $\mathbb{Z}_{n}$ acts diagonally with weights $(1,p)$. The singularity is Gorenstein only if $p=-1$, and otherwise the minimal resolution is always non-crepant. This is reflected in the fact that the GLSMs we use to study these singularities and their (partial) resolutions are anomalous for $p\neq -1$.
In these models, one should ask how B-branes in the $\mathbb{Z}_{n}$ orbifold phase are transported to B-branes in the various partial resolutions that we review in \autoref{sec:HJmodel}, and what is the map between them.
For resolution of quotient singularities this has been studied previously in the math literature.
For instance for quotients of $\CC^{2}$ by finite subgroups of $GL(2,\mathbb{C})$ \cite{ishii2015special}\footnote{Derived categories on quotients of $\CC^{3}$ by finite subgroups of $GL(3,\mathbb{C})$ and the relation with their resolutions have been studied in \cite{kawamata2018derived} and for certain cyclic quotients of projective varieties in \cite{krug2017derived}.}. In physics, the K-theory, i.e., the lattice of charges of B-branes, and the chiral rings have been studied using a GLSM approach for the different resolutions of $\mathbb{C}^{2}/\mathbb{Z}_{n}$ singularities~\cite{Martinec:2002wg,Moore:2004yt,Moore:2005wp} and for other nonsupersymmetric orbifolds~\cite{Morrison:2004fr,Sarkar:2004ry,Morrison:2004ja,Narayan:2005dz,Narayan:2009uy}.

The mathematical references~\cite{ishii2015special,kawamata2018derived,krug2017derived} deal with the map between categories while references~\cite{Martinec:2002wg,Moore:2004yt,Moore:2005wp,Morrison:2004fr,Sarkar:2004ry,Morrison:2004ja,Narayan:2005dz,Narayan:2009uy} mainly focus on the projection to K-theory.
In the former case the picture of transport of B-branes along the moduli space is lost and in the latter case, the analysis is limited only to K-theory.
We would like to unify these two approaches using the GLSM and recent results on supersymmetric localization as a guide.
B-branes on anomalous models have been studied in the context of mirror symmetry of massive theories~\cite{Hori:2000ck} and the interpretation in terms of flows in the moduli space in~\cite{Hori:2013ika} for Fano and general type hypersurfaces in~$\CP^{N}$.\footnote{In mathematics similar cases have been analyzed in \cite{Acosta:2014cma,Galkin:2014laa,galkin2015gamma}, but mostly limited to Fano varieties or general type hypersurfaces in $\CP^{N}$.}
We want to unify \cite{Martinec:2002wg,Moore:2004yt,Moore:2005wp} with the categorical approach of \cite{Segal:2010cz,halpern2015derived,ballard2012variation,Herbst:2008jq}.
The GLSM approach and the localization formula of \cite{Hori:2013ika} provide us with the perfect setup for this purpose.
One can say that we are making modest steps into extending the work of~\cite{Herbst:2008jq} to anomalous models.

\paragraph{Central charges in non-compact toric varieties.}

The hemisphere partition function $Z_{D^2}(\cB)$ computed in~\cite{Hori:2013ika} has been conjectured to reproduce the geometric central charge~\cite{Hosono:2004jp,Iritani:2009ab} with all its instanton corrections, for geometric phases corresponding to compact Calabi-Yau (CY) varieties.
The conjecture can also be extended to nongeometric phases and there are some checks and evidence for it in~\cite{Hori:2013ika,fjrwhpf,hemispherenotes}.
For anomalous models, when the Higgs branches corresponds to a compact non-CY variety, some generalization of such conjecture have been proposed~\cite{hemispherenotes}.
We study here the generalization to \emph{non-compact toric varieties} (not necessarily CY), which are Higgs branches of (possibly anomalous) abelian GLSMs without superpotential.
Besides the aforementioned difficulty of working with anomalous models, we are then faced with a second problem: both the hemisphere partition function and the geometric central charge must be properly regularized in the non-compact case.

We make sense in \autoref{sec:Ktheoreticaspects} of the zero-instanton geometric central charge~\eqref{cc-ours} of a compactly supported brane described by an object~$\mathcal{F}$ in the derived category of the Higgs branch~$X$ in an arbitary phase.
The (zero-instanton) central charge is then a well-defined map from (compactly supported) K-theory to~$\CC$, computed using differential topology on toric orbifolds:
\begin{equation}\label{intro-ZF}
  Z(\mathcal{F}) = \int e^{\tau}\hbar^{c_{1}(X)/(2\pi i)}\widehat{\Gamma}(TX)\ch^{\text{c}}(\mathcal{F}) ,
\end{equation}
with notation described in \autoref{sec:Ktheoreticaspects}.
In particular, $\widehat{\Gamma}(TX)$ is the gamma class of the tangent bundle, and $\hbar$~encodes the energy scale and accounts for renormalization of the FI-theta parameter~$\tau$.
The compactly-supported Chern character $\ch^{\text{c}}$ enables the integral~$\int$ to be well-defined, whereas a similar formula for non-compact branes is meaningless.

In models with non-compact Higgs branches, the hemisphere partition function $Z_{D^2}(\cB)$ of a GLSM B-brane naively blows up.  We argue in \autoref{sec:hpf} that it can be easily regularized if and only if the brane~$\cB$ has compact support on the Higgs branch.
We apply our regularization scheme to the Hirzebruch-Jung resolutions, and use it to compute the hemisphere partition function of B-branes corresponding to sheaves with compact support on the Higgs branch, namely (fractional) D0 and D2 branes.
In all phases, we find perfect agreement between the zero-instanton sector of $Z_{D^2}(\cB)$ and the geometric central charge $Z(\cF)$ of the corresponding brane~$\cF$ on the Higgs branch.
The match requires us to include twisted sectors of the orbifold's cohomology.

\subsection{Organization and open questions}

\paragraph{Plan of the paper.}

In \autoref{sec:section2} we review the basics of abelian anomalous GLSM and we perform a very careful analysis on how the mixed phases arise in the different K\"ahler cones. In \autoref{sec:HJmodel} we review the necessary background on Hirzebruch-Jung resolutions and their corresponding GLSMs that will accompany us throughout the paper. We apply results of the previous section to these models in order to have a complete picture of their phase structure.

In \autoref{sec:hpf} we define GLSM B-branes $(\cB,L)$ and their geometric projection $F_{\text{geom}}(\cB)$ into the Higgs branch of each phase.  This turns out to correctly describe the gauge-decoupling limit for a class of branes called band-restricted, determined later on.
We review there our main tool, the GLSM hemisphere partition $Z_{D^2}(\cB)$, which is also the central charge of $(\mathcal{B},L)$, and we compute it for several classes of branes in Hirzebruch-Jung models.  A careful regularization is needed to ensure that compactly-supported branes have finite central charge.
In \autoref{sec:Ktheoreticaspects} we review the necessary machinery of K-theory and cohomology of toric varieties to define the geometrical central charge of B-branes on the different Higgs branches and compare it to the hemisphere partition function computed in \autoref{sec:hpf}.

This brings us finally in \autoref{sec:Band restriction rule} to the intertwined questions of B-brane transport and the gauge-decoupling limit. These are addressed by studying the contours of $(\mathcal{B},L)$ as we vary the K\"ahler parameters, and we derive this way a version of the band restriction rule \cite{Herbst:2008jq} for anomalous abelian GLSMs.  Hirzebruch-Jung models provide a good illustration of the techniques involved.

\paragraph{Future directions.}

Our work is a starting point in the exploration of B-branes in anomalous and non-compact models.
Even within our focus on the Higgs branch, a limitation is that we only considered binding a brane~$\cB$ with empty branes, namely GLSM B-branes whose gauge-decoupling limit is empty.
As pointed out in \autoref{foot:favorable-cases}, there may be very few branes that are band-restricted with respect to all walls separating a pure-Higgs phase to a phase of interest.
Depending on the GLSM, it may not be possible to bind empty branes to~$\cB$ so as to get a band-restricted brane.
We expect that this is resolved by binding with ``Higgs-empty branes'' as we cross successive walls when moving from a pure-Higgs phase to the phase of interest.
These Higgs-empty branes would be defined in intermediate phases as GLSM branes that reduce purely to the mixed or Coulomb branches in the gauge-decoupling limit.
Specifically, for a given wall that is not parallel to~$Q^{\text{tot}}$, consider the set $J$ of chiral fields whose charges~$Q^j$ lie on the same side of the wall as~$Q^{\text{tot}}$.
It appears that the Koszul complex~$\cK_J$ defined in~\eqref{KoszulResolutionOfSheaf} is such a Higgs-empty brane (in the phase before the wall), and that binding with $\cK_J$~allows us to make any brane band-restricted with respect to the wall of interest.
Repeating the process should yield the Higgs branch part of every GLSM B-brane.
This deserves further exploration.

Beyond the Higgs branch, it would be interesting to describe mixed and Coulomb branch contributions to the gauge-decoupling limit of GLSM B-branes, to B-brane transport, and to the matching of $Z_{D^2}(\cB)$ with geometric integrals analogous to~\eqref{intro-ZF} for the other branches.
This last point will be particularly relevant when mixed and Coulomb branches have positive dimension and a non-trivial geometry.

Regarding our comparison of the GLSM hemisphere partition function $Z_{D^2}(\cB)$ with an integral of characteristic classes on the Higgs branch, it should be possible to perform this comparison in all phases of all abelian GLSMs.
For compactly-supported branes this requires a suitable regularization of the integral expressions involved in~$Z_{D^2}$.
For branes with non-compact support, it appears that one can introduce equivariant parameters for the symmetries at infinity and match the regularized~$Z_{D^2}$ to geometric integrals involving equivariant cohomology (or K-theory) of the Higgs branch.
Crucially, our comparison concerns the zero-instanton sector because a definition of worldsheet instanton corrections is lacking on the geometric side.

While abelian GLSMs without superpotential are already a very rich arena of study, adding a superpotential~$W$ allows to achieve compact Higgs branches.  The hemisphere partition function and B-brane transport only depend on~$W$ through the constraints that the superpotential places on the superconformal R-charge.  We expect most of our discussion to go through.  As the superpotential reduces the Higgs branch to a subvariety, some further complexes of Wilson lines flow to empty branes in the gauge-decoupling limit.  It would be interesting to understand their behaviour under B-brane transport.

Finally, one should develop tools for the wilder case of non-abelian GLSMs, and, beyond GLSMs, to explore the case of hybrid models, where the contour integrals of the localization calculation coexist with geometric integrals on the underlying manifold of the hybrid model.

\section{\label{sec:section2}Branches of abelian gauged linear sigma models}

In this section we review the general properties of gauged linear sigma models (GLSM).
We define a GLSM by specifying the following data.
\begin{itemize}
  \item \textbf{Gauge group}: a compact Lie group~$G$.
  \item \textbf{Chiral matter fields}: a faithful unitary representation $\rho_{m}\colon G\rightarrow U(V)$ of $G$ on some complex vector space~$V$.
  \item \textbf{Superpotential}: a holomorphic, $G$-invariant polynomial $W\colon V\rightarrow \mathbb{C}$, namely $W\in \Sym(V^{*})^G$.
  \item \textbf{FI-theta parameters}, or \textbf{stringy K\"ahler moduli}: a set of complex parameters $t$ such that $\exp(t)\in \Hom(\pi_{1}(G),\mathbb{C}^{*})^{\pi_{0}(G)}$ i.e.\@, $\exp(t)$ is a group homomorphism from $\pi_{1}(G)$ to $\mathbb{C}^{*}$ that is invariant under the adjoint action of~$G$.
  \item \textbf{R-symmetry}: a vector $U(1)_{V}$ and axial $U(1)_{A}$ R-symmetries that commute with the action of $G$ on~$V$.
    To preserve the $U(1)_{V}$ symmetry the superpotential must have weight~$2$ under it.
    As we explain below, $U(1)_{A}$ is anomalous in general.
  \item \textbf{Twisted masses}: an element of the Cartan algebra of the flavour symmetry group~$F$.  This group is the quotient by~$G$ of the normalizer of $G\times U(1)_V\times U(1)_A$ in $U(V)$.
\end{itemize}
In this paper we only consider \textbf{abelian GLSMs}, namely an abelian gauge group $G\simeq U(1)^r\times\Gamma$ with $\Gamma$ a finite abelian group.
Since we want non-compact Higgs branches, we also restrict to cases with \textbf{zero superpotential} and \textbf{zero twisted masses}.
There are no discrete $\theta$ angles because $\pi_{1}(G)\simeq\mathbb{Z}^r$ has no torsion.
Choosing a basis of $\lie{g}=\Lie(G)$ we can write coordinates of~$t$ as
\begin{equation}
  t_\alpha=\zeta_\alpha-i\theta_\alpha\in\mathbb{C}/2\pi i\mathbb{Z} , \qquad 1\leq\alpha\leq r .
\end{equation}
The action of $U(1)^r$ on~$V$ is characterized by a \textbf{charge matrix} with integer entries~$Q_\alpha^j$, where $1\leq j\leq\dim V$ is a flavour index and $1\leq\alpha\leq r$ is a gauge index.

We often take $\Gamma$ trivial.  Otherwise the charges of each chiral multiplet under~$\Gamma$ must also be specified, and the theory is an orbifold by~$\Gamma$ of the theory with gauge group~$U(1)^r$.

\subsection{Classical phases [review]}

Classical vacua are common solutions of the \textbf{mass}, \textbf{D-term}, and \textbf{F-term} equations
\begin{equation}\label{classicalvacuumequations}
  \sum_{\alpha=1}^r \sigma^\alpha Q_\alpha^i X_i = 0\ \ \forall i, \qquad \sum_{i=1}^{\dim V} Q_\alpha^i \abs{X_i}^2 = \zeta_\alpha\ \ \forall \alpha , \qquad \frac{\partial W}{\partial X_i} = 0\ \ \forall i ,
\end{equation}
modulo gauge transformations, namely $G$ acting on the~$X_i$.
Here, $\sigma^\alpha$ are vector multiplet scalars and $X_i$ are chiral multiplet scalars and we sometimes denote the chiral multiplet itself in the same way.

\paragraph{Cones and phases.}

Let us introduce some notation.
We denote the set of non-negative linear combinations of the $Q^i$ for $i$ in some subset $I\subset\{1,\dots,\dim V\}$ by
\begin{equation}
  \Cone_I = \biggl\{ \sum_{i\in I} \lambda_i Q^i \biggm| \lambda_i\in\mathbb{R}_{\geq 0}\ \ \forall i\in I \biggr\} .
\end{equation}
Each set $I$ of $r-1$ linearly independent charge vectors~$Q^i$ defines a codimension~$1$ \textbf{wall} (phase boundary) $\Cone_I$ in FI parameter space.
The complement of the union of all walls is typically disconnected, and each connected component is called a (classical) \textbf{phase} of the GLSM\@.

The D-term equation expresses~$\zeta$ as a non-negative linear combination of charge vectors~$Q^i$.
When $\zeta$~does not belong to a wall, the charge vectors with non-zero coefficient necessarily span $\mZ^r$,
hence the mass equations for the corresponding $X_i\neq 0$ impose linear constraints on~$\sigma$ that set all $\sigma^\alpha=0$.
In addition, the non-zero $X_i$ Higgs the gauge group down to a (possibly trivial) discrete subgroup because they are not fixed by any infinitesimal gauge transformation.
This set of vacua is called the \textbf{Higgs branch} ($\sigma=0$, $X\neq 0$).
Within a phase, the possible sets of non-zero $X_i$ do not change, and only the magnitudes of various $\abs{X_i}^2$ are affected by the precise values of~$\zeta_\alpha$.

\paragraph{Branches for vanishing superpotential.}
We focus on the case of a zero superpotential,
\begin{equation}
  W=0.
\end{equation}
Then the Higgs branch is a GIT (geometric invariant theory) quotient.
As a complex manifold or orbifold it is a complex quotient $(V\setminus\Delta)/G_{\CC}$ where $G_{\CC} = (\CC^*)^r\times\Gamma$.
The \textbf{deleted set}~$\Delta$ is a union of complex subspaces of~$V$ that are intersections of hyperplanes $\{X_i=0\}$.  This set and its complement are:
\begin{align}\label{DeletedSetAbelianGLSM}
  \Delta & = \bigcap_{I\mid\zeta\in\Cone_I} \bigcup_{i\in I} \{X\mid X_i = 0\}
  = \bigcup_{I\mid\zeta\not\in\Cone_I} \bigl\{X\bigm| \forall i\in {^{\complement}I},X_i=0\bigr\}, \\
  V\setminus\Delta & = \bigcup_{I\mid\zeta\in\Cone_I} \bigl\{X\bigm|\forall i\in I,X_i\neq 0\bigr\} .
\end{align}
Since every $\Cone_I$ that contains~$\zeta$ is a union of phases and phase boundaries, the deleted set, hence the complex manifold or orbifold, only depends on the phase in which~$\zeta$ is.  (The K\"ahler structure of the Higgs branch depends on~$\zeta$ even within a phase.)
When $\zeta$ crosses a phase boundary, the Higgs branch typically undergoes a change of toplogy called \textbf{flop}.
The Higgs branch may even be empty in some phases.

For $\zeta$ on a wall there are solutions of~\eqref{classicalvacuumequations} where only $r-1$ chiral multiplet scalars~$X_i$ are non-zero.
The mass equation then allows $\sigma$ to take a non-zero value transverse to the wall.
Such a branch of vacua with $\sigma\neq 0$ and $X\neq 0$ is called a \textbf{mixed branch}.
It opens up at walls in FI parameter space, and at intersections of walls there are further mixed branches in which $\sigma$ can vary in a higher-dimensional subspace of $\mR^r$, culminating in a \textbf{Coulomb branch} ($X=0$, $\sigma$ arbitrary) at $\zeta=0$.
Therefore classically one expects the theory to be singular whenever $\zeta$ belongs to any wall.

\subsection{Quantum effects [review]}

The classical phases get corrected in several ways by quantum effects.

\paragraph{Gauge-decoupling limit.}

As a warm-up to understand one of the energy scales involved, we consider Higgs branches.  Classically, for $\zeta$ not in a wall, solutions of the D-term equations (Higgs branch vacua) are such that the charge vectors~$Q^i$ of non-zero chirals~$X_i$ span~$\mC^r$, and the mass equation sets $\sigma=0$.
The quantum version is that these chirals get vevs (vacuum expectation values)~$\langle X_i\rangle$ which make all~$\sigma$ massive, and fluctuations of~$X$ transverse to the Higgs branch are also massive.  We now show that both masses are of order~$g\abs{\zeta}^{1/2}$ for $\zeta$~deep in a phase, where $g$ is the gauge coupling, of mass dimension~$1$.
It is useful to display the classical potential from which~\eqref{classicalvacuumequations} derive:
\begin{equation}\label{classicalpotential}
  U = \sum_{i=1}^{\dim V} \abs{Q^i\cdot\sigma}^2 \abs{X_i}^2 + \frac{g^2}{2} \sum_{\alpha=1}^r \biggl(\zeta_\alpha-\sum_{i=1}^{\dim V}Q^i_\alpha \abs{X_i}^2\biggr)^2 + \sum_{i=1}^{\dim V} \abs*{\frac{\partial W}{\partial X_i}}^2 ,
\end{equation}
in which $Q^i\cdot\sigma=\sum_\alpha Q^i_\alpha\sigma^\alpha$ and we have already integrated out the vector multiplet's auxiliary field~$D$.
The vector multiplet scalar with a canonical kinetic term is $\sigma^\alpha/g$, to which the first term in~$U$ schematically gives a mass $2gQ\langle X\rangle\sim g\abs{\zeta}^{1/2}$.
More precisely, the mass-squared of the scalars $\sigma^\alpha/g$ is the positive-definite matrix
\begin{equation}
  (m_{\sigma,\text{eff}}^2)_{\alpha\beta} = 4g^2 \sum_{i=1}^{\dim V} Q^i_\alpha Q^i_\beta \abs{\langle X_i\rangle}^2 .
\end{equation}
For $\zeta$~deep in a phase, Higgs branch vacua are such that $\abs{\langle X_i\rangle}\gtrsim\abs{\zeta}^{1/2}$ for a set of indices~$i$ such that the corresponding~$Q^i$ span~$\mC^r$.
Eigenvalues of the mass-squared matrix of the canonically normalized~$\sigma/g$ are thus all of order $g^2\abs{\zeta}$.  A similar calculation
shows that fluctuations of chiral multiplets transverse to the Higgs branch have mass-squared of order $g^2\abs{\zeta}$ too.  We conclude that at energies well below $g\abs{\zeta}^{1/2}$ the theory is well-described by a non-linear sigma model with target the Higgs branch.\footnote{The gauge field $A_\mu$ brings no additional local degree of freedom, nor non-local ones because we are eventually interested in the theory on a disk, on which all gauge bundles are topologically trivial.}
A convenient way to ensure this regime is to take the gauge-decoupling limit $g^2\to\infty$ with all other parameters fixed, at some fixed energy scale.

To keep the notation simple we took all gauge couplings to be equal to some~$g$.
Upon a $GL(r,\mZ)$ change of basis on~$\lie{g}$, which is useful in concrete models, $g^2$~is replaced by a quadratic form on $\lie{g}^*\simeq\mR^r$, dual to the quadratic form $1/g^2$ on $\lie{g}$ that appears in gauge kinetic terms.
The second term in~$U$ becomes schematically $\frac{1}{2} \sum_{\alpha,\beta} (g^2)^{\alpha\beta} (\zeta_\alpha-\cdots) (\zeta_\beta-\cdots)$.
As explained just above we eventually only care about the gauge-decoupling limit $g^2\to\infty$, unaffected by such changes of basis.

\paragraph{Axial R-symmetry anomaly.}

The first quantum effect is that \textbf{the FI parameter is renormalized}:
\begin{equation}\label{FIrenorm}
  \zeta_{\text{ren}}(\mu) = \zeta_{\text{bare}} + Q^{\text{tot}} \log\Bigl(\frac{\mu}{\Lambda}\Bigr)
\end{equation}
where $\zeta_{\text{bare}}$ is the bare FI parameter, measured at the UV energy scale~$\Lambda$, and $Q^{\text{tot}}=\sum_i Q^i$ is the $U(1)_A$ (axial R-symmetry) anomaly.
Models with $Q^{\text{tot}}=0$ are called Calabi-Yau models because their Higgs branch is a Calabi-Yau orbifold.
At very low energies, such that $\mu\ll g\abs{\zeta_{\text{ren}}(\mu)}^{1/2}$, the gauge theory flows to a nonlinear sigma model (NLSM) with target space the Higgs branch.
Further RG flow is expected to change the K\"ahler metric given by the GIT construction to one that gives a conformal NLSM\@.

In non-Calabi-Yau models, flowing to the IR shifts the FI parameter in the direction $-Q^{\text{tot}}\neq 0$.
The deep IR limit can thus only explore some of the phases, specifically the phases whose closure contains the vector $-Q^{\text{tot}}$, interpreted as a point in FI parameter space.
For example, if $-Q^{\text{tot}}$ is not parallel to any wall, then it belongs to one specific phase, and the deep IR limit is described by that phase of the GLSM regardless of~$\zeta_{\text{bare}}$.
Nevertheless, for every phase we can arrange parameters so that the phase gives a good description of the physics at some intermediate energy scale~$\mu$:
tune $\zeta_{\text{bare}}$ so that the renormalized $\zeta_{\text{ren}}(\mu)$ lies deep in the given phase, then take $g$ sufficiently large to ensure large masses $g\abs{\zeta_{\text{ren}}(\mu)}^{1/2}\gg\mu$.

A counterpart to the fact that FI parameter varies under scale transformations as~\eqref{FIrenorm} is that the theta angle varies by $\alpha Q^{\text{tot}}$ under $e^{i\alpha}\in U(1)_A$ due to that symmetry's anomaly.
Scaling and $U(1)_A$ act on twisted chiral parameters such as twisted masses (vector multiplet scalars) by scaling and phase rotations.
As a result, the anomalous transformation of $t=\zeta-i\theta$ can be repackaged into a dependence of all observables on a \textbf{complexified energy scale}~$\mu$.
Both $\zeta$ and $\theta$ have one unphysical component (along the renormalization direction~$Q^{\text{tot}}$) that can be traded for this complexified energy scale~$\mu$, but we find it more convenient to take the point of view of fixing~$\mu$ and keeping all components of~$t$.

\paragraph{Complexified walls.}

A second quantum effect makes some classical walls (\textbf{Calabi-Yau walls}) into \textbf{complex codimension~$1$} singular loci in the FI-theta parameter space while others correspond to \textbf{no wall} quantum mechanically.
A classical wall is a real codimension~$1$ cone in the FI parameter space, where a mixed (or Coulomb) branch opens up.  Let $u_0$ be some nonzero vector orthogonal to that wall.  In this branch, $\sigma=\sigma_0u_0$ can be an arbitrary multiple of~$u_0$.
There may then be vacua whose wave-function explores large values of~$\sigma_0$.
Such large values give mass to all chirals for which $Q^i\cdot u_0\neq 0$,
and integrating these out gives an effective action for~$\sigma_0$ that is given by the twisted superpotential
\begin{equation}\label{Weffwall}
  \tW_{\text{eff}} = - (t_{\text{ren}}(\mu)\cdot u_0)\sigma_0 - \sum_{j\mid Q^j\cdot u_0\neq 0} (Q^j\cdot u_0)\sigma_0 \biggl( \log\biggl((Q^j\cdot u_0)\frac{i\sigma_0}{\mu}\biggr) - 1 \biggr) ,
\end{equation}
which is actually $\mu$-independent due to~\eqref{FIrenorm}.
Vacua are critical points of this twisted superpotential, namely solutions of
\begin{equation}\label{tmusol}
  t_{\text{ren}}(\mu)\cdot u_0
  = - (Q^{\text{tot}}\cdot u_0) \log\biggl(\frac{i\sigma_0}{\mu}\biggr)
  - \sum_{j\mid Q^j\cdot u_0\neq 0} (Q^j\cdot u_0) \log(Q^j\cdot u_0) .
\end{equation}
The $2\pi i$ ambiguity of $\log$ has no effect since $t$ is defined modulo $2\pi i\mathbb{Z}^r$ and all $Q^j\in\mZ^r$.
Then we have two very different cases: the classical walls parallel to $Q^{\text{tot}}$ typically correspond to walls in the quantum theory (up to some shift), while others are not quantum walls.
\begin{itemize}
\item Calabi-Yau walls are those for which $Q^{\text{tot}}\cdot u_0=0$, that is, $Q^{\text{tot}}$ is parallel to the wall.  Then there is a whole mixed (or Coulomb) branch of vacua at a specific locus in the FI-theta parameter space:
  $t_{\text{ren}}(\mu)\cdot u_0=-\sum_j (Q^j\cdot u_0)\log(Q^j\cdot u_0)$.\footnote{\label{foot:electric}A more pedestrian approach (to see that the singularity is at specific values of $\theta\cdot u_0$) is that the Lagrangian includes terms $\frac{1}{2g^2}E^2-\frac{i}{2\pi}(\theta\cdot u_0)E$ where $E$ is the electric field $\partial_1 A_2-\partial_2 A_1$ in the direction~$u_0$.  Completing the square suggests a naive energy contribution proportional to $(\theta\cdot u_0)^2$, then one should take into account the freedom to shift the theta angle along a lattice.}
  Quantum effects thus shift the wall away from $\zeta\cdot u_0=0$ and give it complex (rather than real) codimension~$1$.

  More precisely, our analysis is valid infinitely deep in the wall.
  In Calabi-Yau models, we typically have a collection of loci that asymptote to this wall and at which the theory is singular due to a non-compact Coulomb branch opening up.
  In some non-Calabi-Yau models, the Coulomb branch that opens up has a finite size controlled by the FI parameters along the wall,
  and the theory actually has no singular locus near the classical wall.\footnote{For example in theories that have no Calabi-Yau walls in the UV (namely no wall containing the direction $Q^{\text{tot}}$) it seems that none of the classical Calabi-Yau walls survive in the quantum theory.  We hope to explore later this subtle issue that seems to have been missed previously.}

\item
  If $Q^{\text{tot}}\cdot u_0\neq 0$, there are $\abs{Q^{\text{tot}}\cdot u_0}/\gcd\{u_0^\alpha\mid 1\leq \alpha\leq r\}$
  solutions\footnote{By $\gcd$ we mean here the greatest real number $x$ such that all $u_0^\alpha/x$ are integers.  This is well-defined because $u_0$ is proportional to the wedge product of $r-1$ charge vectors defining the wall, which has integer components.} at
  \begin{equation}\label{sigma0sol}
    \sigma_0 \simeq -i \mu \, \exp\frac{-\bigl(t_{\text{ren}}(\mu)+2\pi i k\bigr)\cdot u_0-\sum_{j\mid Q^j\cdot u_0\neq 0} (Q^j\cdot u_0) \log(Q^j\cdot u_0)}{Q^{\text{tot}}\cdot u_0}
  \end{equation}
  where $k\in\mZ^r$ parametrizes solutions with redundancy.
  The approximation used to derive these is good at energies well below $\abs{\sigma}$, namely provided $(t_{\text{ren}}(\mu)\cdot u_0)/(Q^{\text{tot}}\cdot u_0)\ll 0$.
  (For any UV FI parameter, the model flows to a phase that obeys this.)
  In other words these solutions should be ignored in the other phase $(t_{\text{ren}}(\mu)\cdot u_0)/(Q^{\text{tot}}\cdot u_0)\gg 0$ as they merge smoothly into the $\sigma_0=0$ Higgs branch.
  In fact there is no wall between the two phases, even though the low-energy descriptions are quite different.

  In $U(1)$ GLSMs the solutions~\eqref{sigma0sol} are isolated \textbf{quantum Coulomb vacua}.
  Excitations around these vacua are all massive: indeed, chiral multiplets have mass $\abs{Q^i\sigma}\gg\mu$, while fluctuations of~$\sigma/g$ have a mass $g^2/\abs{\sigma}$, which can be taken much larger than~$\mu$ by choosing a sufficiently large~$g$.  This is a further condition on~$g$ besides the condition $g\abs{\zeta_{\text{ren}}(\mu)}^{1/2}\gg\mu$ needed for the Higgs branch NLSM to give a good approximation.
  In GLSMs with $U(1)^r$ gauge groups, \eqref{sigma0sol}~may also describe \textbf{mixed branches}, which we analyze further below.
\end{itemize}

\paragraph{Pure-Higgs phases.}
\label{para:pureHiggs}

Depending on the GLSM, the vector $Q^{\text{tot}}=\sum_i Q^i$ may lie in a phase or a wall.
Consider phases whose closure contains~$Q^{\text{tot}}$, namely phases where $t_{\text{ren}}$ may lie in the deep~UV\@.
These \textbf{pure-Higgs phases}, as we dub them, do not feature any mixed or Coulomb branch, which will make them useful starting points for our analysis of brane transport.
The intuition is that shifting $t_{\text{ren}}(\mu)\to t_{\text{ren}}(\mu)+\alpha Q^{\text{tot}}$ in~\eqref{sigma0sol} scales the solution $\sigma_0\to e^{-\alpha} \sigma_0$.
For $\alpha\to+\infty$, which makes $t_{\text{ren}}(\mu)$ lie deep inside such a phase, any Coulomb or mixed branch simply merges with the Higgs branch at $\sigma_0=0$.
To be precise, this limit should be applied to the mixed branch equation~\eqref{subtheoryCoulomb} instead of~\eqref{sigma0sol} but the same logic applies.

All other phases contain mixed or Coulomb branches.
Indeed, one can track what happens upon wall-crossing, as we do in \autoref{ssec:wallcrossingHiggsing}.
Starting from the pure-Higgs phases and moving in the $-Q^{\text{tot}}$ direction (by following the RG flow or by varying parameters of the theory at fixed~$\mu$) one can reach all other phases.
Along the way, $t_{\text{ren}}(\mu)$ does not cross Calabi--Yau walls since they are parallel to~$Q^{\text{tot}}$, and all other walls are crossed from the phase $(t_{\text{ren}}(\mu)\cdot u_0)/(Q^{\text{tot}}\cdot u_0)\gg 0$ to the phase $(t_{\text{ren}}(\mu)\cdot u_0)/(Q^{\text{tot}}\cdot u_0)\ll 0$, so that new Coulomb or mixed branches appear.

\subsection{Interlude: nonlinear twisted superpotential}

Before studying in detail the possibility of mixed branches, let us consider a slight generalization of usual GLSMs.
Usually, the (bare) twisted superpotential of a GLSM is taken to be linear: $\tW=-t\cdot\sigma$ with $t$ the FI and $\sigma$ the twisted chiral field strength of the vector multiplet.
We now consider a gauge theory with a more complicated unspecified~$\tW(\sigma)$.
We allow the charges $Q^j$ not to span~$\RR^r$, namely the gauge group action not to be faithful.
In other words, the charge lattice $\ZZ^r=\Hom(U(1)^r,U(1))$ of the gauge group contains all integer linear combinations of the~$Q^j$, but may contain more elements.

As we explain shortly, vacua in which $\sigma$ gives a mass to none of the chirals (analogous to Higgs branches) are solutions of
\begin{align}
  \label{strangeGLSMeq1}
  (Q^j\cdot\sigma) X_j & = 0, \\
  \label{strangeGLSMeq2}
  \zeta_{\text{eff}}\coloneqq - \Re\biggl(\frac{\partial \tW}{\partial\sigma}\biggr) & = \sum_i\bigl(Q^j|X_j|^2\bigr), \\
  \label{strangeGLSMeq3}
  \theta_{\text{eff}}\coloneqq \Im\biggl(\frac{\partial \tW}{\partial\sigma}\biggr) & \in \Span_{\RR}\bigl(\{Q^j\}\bigr) + 2\pi \ZZ^r ,
\end{align}
modulo gauge transformations.  While the equations are real, this is a complex orbifold because the $d=\dim(\Span\{Q^j\})$ ``missing'' constraints on $\Im(\partial \tW/\partial\sigma)$ are accounted for by the $U(1)^d$ (times discrete factor) gauge transformations that act non-trivially on the chirals.

Equations \eqref{strangeGLSMeq1}--\eqref{strangeGLSMeq2} come from the same classical potential as~\eqref{classicalpotential}, with $\zeta\to\zeta_{\text{eff}}$.
Equation \eqref{strangeGLSMeq3} is found by considering the action for the gauge field components along directions $u\in\lie{g}$ such that all $Q^j\cdot u=0$.
These components only appear in the gauge kinetic term and in the twisted superpotential term:
$L = \frac{1}{2g^2} E^2 - \frac{i}{2\pi} \bigl(\theta_{\text{eff}}\cdot u\bigr)\, E$,
where $E$ is the electric field $\partial_1 A_2-\partial_2 A_1$ in the direction~$u$, and one can then follow the logic of \autoref{foot:electric}.

These equations reduce to well-known ones when $d=0$ or~$r$.
When there are no charged chiral multiplet ($d=0$) they state that $\partial \tW/\partial\sigma$ vanishes modulo $2\pi i$.
Instead, when charges span $\RR^r$ ($d=r$), the equations reduce to the mass equation $(Q^i\cdot\sigma)X_i=0$ and to D-term equations
$\sum_j\bigl(Q^j\abs{X_j}^2\bigr) = \zeta_{\text{eff}}$, modulo gauge transformations.

\subsection{\label{ssec:mixedbranches}Mixed branches}

We now go back to a standard GLSM and generalize our earlier discussion from $U(1)$ to $U(1)^r$ models to find mixed branches.
We learn that mixed branches are essentially products of Coulomb and Higgs branches.
Furthermore some phases get subdivided beyond the classical analysis.

We discussed previously how in one phase of a $U(1)$ GLSM in which $Q^{\text{tot}}\neq 0$ there are quantum Coulomb branch vacua at~\eqref{sigma0sol}.
In $U(1)^r$ models, Coulomb branch vacua ($\sigma\neq 0$, $X=0$) are found as follows.
Assume $\sigma$ has a generic large vev so that all chirals are massive.
Integrate out all the massive chirals to get an effective twisted superpotential $\tW_{\text{eff}}(\sigma)$.
Find classical solutions for~$\sigma$ (critical points of $\tW_{\text{eff}}$).
Check whether chiral multiplets in these solutions are indeed all massive, or not: if yes we found a Coulomb branch vacuum.

We follow a similar procedure to find all branches.
In each branch we expect some set of chiral multiplets to be made massive by~$\sigma$, and some set not to be.
Let us search for vacua in which a set $I\subset\{1,\dots,\dim V\}$ of flavours are given no mass by~$\sigma$, so $Q^j\cdot\sigma=0$ for all $j\in I$.
Equivalently, $\sigma\in\lie{q}_\CC^\perp\subset\lie{g}_\CC$ where
\begin{equation}
  \lie{q}\coloneqq\Span_{\RR}(\{Q^j\mid j\in I\}) \subset \lie{g}^* .
\end{equation}
There could be further chiral multiplets with charge $Q^j\in\lie{q}$.
Of course these are also given no mass by~$\sigma$, so we restrict our attention without loss of generality to cases where $I$ contains all such flavours.

Integrate out all the chirals $X_j$ for $j\not\in I$ since we expect them to be massive.  The effective twisted superpotential is
\begin{equation}
  \tW_{\text{eff}} = - t_{\text{ren}}(\mu)\cdot\sigma - \sum_{j\not\in I} (Q^j\cdot \sigma) \biggl( \log\biggl(\frac{Q^j\cdot i\sigma}{\mu}\biggr) - 1 \biggr) ,
\end{equation}
and we search for solutions of \eqref{strangeGLSMeq1}, \eqref{strangeGLSMeq2}, \eqref{strangeGLSMeq3} for the resulting gauge theory.
We are only interested in solutions for which $\sigma\in\lie{q}_\CC^\perp$ so that the remaining chirals $X_j$ for $j\in I$ are not given a mass,
and for which all $X_j=0$ for $j\not\in I$ since they should be massive.

Focus now on components of \eqref{strangeGLSMeq2} and~\eqref{strangeGLSMeq3} along~$\lie{q}_\CC$.
They give
\begin{equation}
  \frac{\partial \tW}{\partial\sigma} = - t_{\text{ren}}(\mu) - \sum_{j\not\in I} Q^j \log\biggl(\frac{Q^j\cdot i\sigma}{\mu}\biggr) \in \lie{q}_\CC + 2\pi i \ZZ^r .
\end{equation}
Once these equations are solved for $\sigma\in\lie{q}_\CC^\perp$, one must check that masses are large ($\abs{Q^j\cdot\sigma} \gg \mu$) for $j\not\in I$.
These are precisely the condition for quantum Coulomb branch vacua of a sub-theory, with smaller gauge Lie algebra $\lie{q}^\perp\subset\lie{g}$ and one chiral multiplet of charge $(Q^j\bmod{\lie{q}})$ for each $j\not\in I$ (one can without any effect include gauge-neutral chirals for~$j\in I$).
Indeed, the Coulomb branch equation for this sub-theory is
\begin{equation}\label{subtheoryCoulomb}
  (t_{\text{ren}}(\mu)\bmod{\lie{q}_\CC}) + \sum_{j\not\in I} (Q^j\bmod{\lie{q}}) \log\biggl(\frac{(Q^j\bmod{\lie{q}})\cdot i\sigma}{\mu}\biggr) = (0 \bmod{\lie{q}})
\end{equation}
and the condition that $X_j$, $j\not\in I$ be massive reads
$\abs*{(Q^j\bmod{\lie{q}})\cdot\sigma} = \abs{Q^j\cdot\sigma} \gg \mu$.
Solutions $\sigma\in\lie{q}^\perp_{\CC}$ are generically isolated.
In addition, scaling $t\to\infty$ in a fixed direction, $\sigma$~can involve multiple scales: for a given solution the various $Q^j\cdot\sigma$ may behave as exponentials $\exp(t\cdot\lambda^j)$ with different parameters~$\lambda^j$.

Then, for each solution $\sigma\in\lie{q}^\perp_\CC$, we solve the D-term equations $\sum_{j\in I} Q^j \abs{X_j}^2 = \zeta_{\text{eff}}$ with
\begin{equation}\label{zetaeffquotienttheory}
  \zeta_{\text{eff}}
  = - \Re\biggl(\frac{\partial \tW}{\partial\sigma}\biggr)
  = \zeta_{\text{ren}}(\mu) + \sum_{j\not\in I} Q^j \log\abs*{\frac{Q^j\cdot\sigma}{\mu}} \in \lie{q} ,
\end{equation}
modulo gauge transformations.  Infinitesimal gauge transformations along~$\lie{q}^\perp$ act trivially, so
this is the Higgs branch of a ``quotient'' theory with gauge Lie algebra $\lie{q}^*=\lie{g}/\lie{q}^\perp$ and one chiral multiplet of charge~$Q^j$ for each $j\in I$.
The Higgs branch is empty whenever $\zeta_{\text{eff}}$ cannot be written as a positive linear combination of the $Q^j$ for $j\in I$.

The branch we found is a Coulomb branch if all $X$~vanish, a mixed branch if both chiral and vector multiplet scalars have non-zero vev, and a Higgs branch if $\sigma$~vanishes.
Importantly, the dimension of the branch is
\begin{equation}\label{dimensionofbranch}
  \abs{I} - \dim\lie{q} = \bigl\{j\bigm|Q^j\in\lie{q}\bigr\} - \dim\lie{q} .
\end{equation}
If the $Q^j$ that lie in $\lie{q}$ are linearly independent, this implies that the branch is an isolated vacuum.
In all models we study later in the paper, only the Higgs branch (for which $\lie{q}=\lie{g}$) has positive dimension.  The remaining branches are isolated vacua, and it is not important to work out whether they are mixed or Coulomb branches.

There is an interesting phase structure upon varying~$t$.
The set of solutions to the Coulomb branch equation~\eqref{subtheoryCoulomb} can change when the solution~$\sigma$ is such that one mass $\abs{Q^j\cdot\sigma}$ becomes of order~$\mu$ for $j\not\in I$.
Then one chiral becomes massless and the vacua are described by some larger choice of~$I$ and~$\lie{q}$.
On the other hand, the Higgs branch of the quotient theory changes geometry when $\zeta_{\text{eff}}$ given in~\eqref{zetaeffquotienttheory} changes sign.  The location of this phase transition depends on all components of the FI parameter (but not the theta angle), and we find in examples that the transition can take place along a wall subdividing classical phases\footnote{An interesting example is the $U(1)^2$ theory with four chirals of charges $Q^1=Q^2=(1,0)$, $Q^3=(0,1)$, $Q^4=(\alpha,1)$.  Its mixed branch with $\lie{q}=\Span((1,0))$ consists of two copies of $\mathbb{CP}^1$ with
$i\sigma=\bigl(0,\pm\mu\exp(-t_2/2)\bigr)$ and $\abs{X_1}^2+\abs{X_2}^2=\zeta_1-\alpha\zeta_2/2$ modulo $U(1)$.  It only exists when $\zeta_2\ll 0$ and $\zeta_1>\alpha\zeta_2/2$.  The $\zeta_1=\alpha\zeta_2/2$ boundary is a subphase transition across which each $\mathbb{CP}^1$ becomes a pair of Coulomb branch vacua.} (see \autoref{ssec:CoulGeom}).

\section{\label{sec:HJmodel}Hirzebruch-Jung model}

Our main example in this paper is an abelian GLSM considered in~\cite{Martinec:2002wg}, whose classical Higgs branch in different phases is a certain orbifold of~$\mC^2$ and its (partial) resolutions.
More precisely we consider the orbifold $\mC^2/\mZ_{n(p)}$ in which the generator $\omega=\exp(2\pi i/n)$ of $\mZ_n$ acts on $\mC^2$ by $(z_1,z_2)\mapsto(\omega z_1,\omega^pz_2)$ for some $0<p<n$ such that $\gcd(p,n)=1$.\footnote{The action $(z_1,z_2)\mapsto(\omega^jz_1,\omega^kz_2)$ of $\mZ_n$ on $\mC^2$ is faithful if $\gcd(j,k,n)=1$, so one may ask about more general quotients~$\mC^2/\mZ_{n(j,k)}$.  If $\gcd(j,n)=1=\gcd(k,n)$ then $\mC^2/\mZ_{n(j,k)}$ is $\mC^2/\mZ_{n(p)}$ where $pj=k\mod n$.  Otherwise $\mC^2/\mZ_{n(j,k)}$ is actually a quotient of $(\mC/\mZ_{\gcd(k,n)})\times(\mC/\mZ_{\gcd(j,n)})$, and by rescaling $z_1\mapsto z_1^{\gcd(k,n)}$ and $z_2\mapsto z_2^{\gcd(j,n)}$ before applying this argument we obtain a quotient $\mC^2/\mZ_{n'(p)}$ with $n'=n/\bigl(\gcd(j,n)\gcd(k,n)\bigr)$ and $p\frac{j}{\gcd(j,n)}=\frac{k}{\gcd(k,n)}\mod n'$.  The rescaling corresponds to blowing up each factor $\mC/\mZ_d$ to~$\mC$, which has the following effect.  Consider a $U(1)$ GLSM with two chiral multiplets of charges $1$ and $-d$.  Its Higgs branch in the two phases are $\mC/\mZ_d$ and~$\mC$ and the latter phase has $d-1$ additional Coulomb branch vacua.  It may be interesting to work out brane transport in detail for such quotients.}
In addition to the classical Higgs branch the model also has isolated quantum Coulomb/mixed branch vacua in many phases.

\subsection{Notation}

Before describing the GLSM let us introduce some notation.
Like every rational number in $(1,+\infty)$ the fraction $n/p$ has a unique continued fraction expansion
\begin{equation}
  \frac{n}{p} = [a_1,\ldots,a_r] \coloneqq a_1 - \frac{1}{a_2 - \frac{1}{\cdots - 1/a_r}}
\end{equation}
in terms of integers $a_\alpha\geq 2$.  This also defines~$r$.

\paragraph{Determinants.}

Then we consider the generalized Cartan matrix
\begin{equation}
  (C_{\alpha\beta})_{1\leq\alpha,\beta\leq r}
  \coloneqq
  \begin{pmatrix}
    a_1 & -1 & & 0\\
    -1 & a_2 & \ddots \\
    & \ddots & \ddots & -1 \\
    0 & & -1 & a_r
  \end{pmatrix}
\end{equation}
and its (diagonal) minors
\begin{equation}
  \begin{aligned}
    d_{ij} \coloneqq - d_{ji} & \coloneqq \det\bigl(C_{\alpha\beta}\bigr)_{i<\alpha,\beta<j}
    & & \text{for $0\leq i<j\leq r+1$,} \\
    d_{ii} & \coloneqq 0
    & & \text{for $0\leq i\leq r+1$.}
  \end{aligned}
\end{equation}
Here strict inequalities imply that the submatrix has size $j-i-1$, so $d_{i(i+1)}=1$ is the determinant of a $0\times 0$ matrix,
and we extended the notation to $i\geq j$ by antisymmetry for later convenience.
These determinants obey two recursion relations
\begin{equation}\label{drecursion}
  \begin{aligned}
    d_{i(j-1)} + d_{i(j+1)} & = a_j d_{ij} & & \text{for $0\leq i\leq r+1$ and $1\leq j\leq r$,} \\
    d_{(i-1)j} + d_{(i+1)j} & = a_i d_{ij} & & \text{for $1\leq i\leq r$ and $0\leq j\leq r+1$,}
  \end{aligned}
\end{equation}
and they can be related to partial continued fractions through
\begin{equation}
  \begin{aligned}
    [a_i,a_{i+1},\ldots,a_{j-1}] & = \frac{d_{(i-1)j}}{d_{ij}} \qquad \text{for $1\leq i<j\leq r+1$,} \\
    [a_j,a_{j-1},\ldots,a_{i+1}] & = \frac{d_{i(j+1)}}{d_{ij}} \qquad \text{for $0\leq i<j\leq r$.}
  \end{aligned}
\end{equation}
Note that since the continued fractions are all in $(1,+\infty)$ we learn that $d_{ij}<d_{(i-1)j}$ and $d_{ij}<d_{i(j+1)}$ for $i<j$ and these inequalities extend to all $i,j$ by antisymmetry.

\paragraph{Refined properties of determinants.}

We then define $p_j=d_{j(r+1)}$ and $q_j=d_{0j}$ for $0\leq j\leq r+1$, which can alternatively be defined as in \cite{Martinec:2002wg} by $p_{r+1}=q_0=0$ and $p_r=q_1=1$ and
\begin{equation}
  [a_j,a_{j+1},\ldots,a_r] = \frac{p_{j-1}}{p_j} \qquad \text{and} \qquad
  [a_j,a_{j-1},\ldots,a_1] = \frac{q_{j+1}}{q_j} .
\end{equation}
They obey $0=p_{r+1} < p_r < \cdots < p_1=p < p_0=n$ and $0=q_0 < q_1 < \cdots < q_{r+1}=n$.
The recursion relations~\eqref{drecursion} read $p_{j-1} + p_{j+1} = a_j p_j$ and $q_{j-1} + q_{j+1} = a_j q_j$ for $1\leq j\leq r$, from which we deduce by induction that
\begin{equation}\label{pqqpmodn}
  p_i q_j - p_j q_i = n d_{ij} \qquad \text{for $0\leq i,j\leq r+1$}
\end{equation}
because both quantities obey the same recursion relations~\eqref{drecursion} and agree for $0\leq i,j\leq 1$.
In fact this explicit formula for $d_{ij}$ implies (and is a special case of)
\begin{equation}\label{dddddd}
  d_{ij} d_{kl} - d_{ik} d_{jl} + d_{il} d_{jk} = 0 \qquad \text{for $0\leq i,j,k,l\leq r+1$.}
\end{equation}
This reproduces the recursion relations thanks to $d_{(i-1)(i+1)}=a_i$.
Taking $l=k\pm 1$ so that $d_{kl}=\pm 1$ we find $\gcd(d_{ik},d_{jk})|d_{ij}$.  Combining with permutations of $i,j,k$ we deduce
\begin{equation}\label{gcdgcdgcd}
  \gcd(d_{ij},d_{ik}) = \gcd(d_{ij},d_{jk}) = \gcd(d_{ik},d_{jk}) .
\end{equation}

Another consequence of the recursion relation is that $j\mapsto p_j$ is convex since its discrete Laplacian $p_{j-1}-2p_j+p_{j+1} = (a_j-2)p_j$ is non-negative (all $a_j\geq 2$), and likewise $j\mapsto q_j$ is convex.  Their sum is convex and $p_0+q_0 = p_{r+1}+q_{r+1}=n$ so $p_j+q_j\leq n$ for all $0\leq j\leq r+1$.  It is easy to check that equality only happens for $j=0$ or $j=r+1$ or when $p=n-1$ (all $a_j=2$).
Given the definitions of $p_j$ and $q_j$ the inequality reads $d_{j(r+1)}+d_{0j}\leq d_{0(r+1)}$.  It generalizes to $d_{ij}+d_{jk}\leq d_{ik}$ for $i<j<k$ with equality if and only if $a_\alpha=2$ for all $i<\alpha<k$.

Integer solutions $(x_0,\dots,x_{r+1})\in\ZZ^{r+2}$ of the recursion relation $x_{i-1}-a_ix_i+x_{i+1}=0$ for all $1\leq i\leq r$ appear in a few places in our work.  The lattice of solutions is a rank~$2$ sublattice of $\ZZ^{r+2}$, and any pair of solutions $(d_{ij})_{0\leq i\leq r+1}$ and $(d_{ik})_{0\leq i\leq r+1}$ spans an index $\abs{d_{jk}}$ sublattice inside it.

\paragraph{GLSM in one basis.}

The GLSM has gauge group $G=U(1)^r$ and $r+2$ chiral multiplets $X_0,\ldots,X_{r+1}$ (for convenience we label flavours starting at~$0$).
There are two convenient choices of bases for the Lie algebra, leading to two different charge matrices that are of course related to each other by a change of basis.

The factor $U(1)_\alpha$, namely the $\alpha$-th factor in $U(1)^r$, acts with charges $(1,-a_\alpha,1)$ on $(X_{\alpha-1},X_\alpha,X_{\alpha+1})$ and does not act on other~$X_\beta$.
In other words the charge matrix is
\begin{equation}\label{GLSMIcharges}
  \Bigl(Q_\alpha{}^i\Bigr)_{1\leq\alpha\leq r,0\leq i\leq r+1}
  =
  \begin{pmatrix}
    1 & -a_1 & 1 & 0 & \cdots & 0 \\
    0 & 1 & -a_2 & 1 & \ddots & \vdots \\
    \vdots & \ddots & \ddots & \ddots & \ddots & 0 \\
    0 & \cdots & 0 & 1 & -a_r & 1
  \end{pmatrix} ,
\end{equation}
and in particular charges of $X_1$, \dots, $X_r$ are minus the generalized Cartan matrix.
The action of~$G$ is faithful: if an element $(g_1,\dots,g_r)\in U(1)^r$ acts trivially then $g_1=1$ (because of the action on $X_0$), then $g_2=1$ (because of the action on $X_1$) and so on, so all $g_\alpha=1$.
We denote components of~$\zeta$ in this basis by~$\zeta_\alpha$.

\paragraph{GLSM in the second basis.}

We change basis by multiplying the charge matrix by $nC^{-1}$, whose components are integers $(nC^{-1})^{\alpha\beta} = p_{\max(\alpha,\beta)} q_{\min(\alpha,\beta)}$.
This yields
\begin{equation}\label{GLSMIIcharges}
  \biggl(\sum_{\beta=1}^r n(C^{-1})^{\alpha\beta}Q_\beta{}^i\biggr)_{1\leq\alpha\leq r,0\leq i\leq r+1}
  =
  \begin{pmatrix}
    p_1 & -n & 0 & \cdots & 0 & q_1 \\
    p_2 & 0 & -n & \ddots & \vdots & q_2 \\
    \vdots & \vdots & \ddots & \ddots & 0 & \vdots \\
    p_r & 0 & \cdots & 0 & -n & q_r
  \end{pmatrix} .
\end{equation}
Again, each factor acts on three chiral multiplets, but now these are $X_0$, $X_\alpha$ and $X_{r+1}$.
In this basis the gauge group takes the form $U(1)^r/(\mZ_n)^{r-1}$.
Indeed, all elements $(g^1,\dots,g^r)\in(\mZ_n)^r\subset U(1)^r$ such that $\prod_\alpha(g^\alpha)^{p_\alpha}=\prod_\alpha(g^\alpha)^{q_\alpha}=1$ act trivially, and in fact these two conditions are equivalent thanks to $p_\alpha=p_1 q_\alpha\mod n$, see~\eqref{pqqpmodn}.
In terms of the components~$\zeta_\alpha$ in the first basis, components of~$\zeta$ in this basis are $\zeta'_\alpha=\sum_{\beta=1}^r\bigl(n(C^{-1})^{\alpha\beta}\zeta_\beta\bigr)$ for $1\leq\alpha\leq r$.

\paragraph{Further comments.}

The case $a_1=\cdots=a_r=2$ is interesting because charge vectors sum to zero precisely in that case.
The generalized Cartan matrix is the Cartan matrix of the Lie algebra~$A_r$, and we compute $d_{ij} = j-i$.
In particular, $n=r+1$ and $p=r$, namely $\mZ_n$ acts by multiplication by $(\omega,\omega^{n-1})$, hence acts on $\mC^2$ as a subgroup of $SU(2)$.
The Higgs branch is in this case a Calabi-Yau manifold or orbifold depending on the phase.
The model then flows to a superconformal field theory.

Returning to general~$a_\alpha$, consider the phase where all of these sums are negative.
The classical vacuum equations imply $X_\alpha\neq 0$ for $1\leq\alpha\leq r$, and these fields are fixed up to a phase in terms of $\zeta$, $X_0$ and~$X_{r+1}$.
The phase is absorbed by a gauge transformation, and the gauge group is Higgsed down to the discrete subgroup $(\mZ_n)^r/(\mZ_n)^{r-1}\subset U(1)^r/(\mZ_n)^{r-1}$ that leaves $X_1,\ldots,X_r$ invariant.
Altogether the Higgs branch is spanned by $X_0$ and $X_{r+1}$, modulo the remaining $\mZ_n$ gauge transformations, which multiply $X_{r+1}$ and $X_0$ by powers of $(\omega,\omega^p)$ since $p_\alpha=pq_\alpha\mod n$.  As a complex orbifold, the Higgs branch is thus $\mC^2/\mZ_{n(p)}$ in this phase.
This phase is called the \textbf{orbifold phase}.
We discuss other phases in detail in \autoref{ssec:HiggsGeom}.

\subsection{\label{ssec:HiggsGeom}Higgs branch geometry}

We now turn to describing the geometry of the Higgs branch in each phase.
The Hirzebruch-Jung models also admit Coulomb and mixed branch vacua (see \autoref{ssec:CoulGeom}).
By a slight abuse of notation we denote the vev of the bottom component of a chiral multiplet by the same letter: the coordinate ring of the UV target space is thus $\mathcal{R} = \mathbb{C}[X_0,X_1,\dots,X_r,X_{r+1}]$.

\paragraph{Higgs branch in one phase.}

The IR Higgs branch~$X$ in a given phase admits a $(\CC^*)^2$ action obtained as the $(\CC^*)^{r+2}$ symmetry rotating individual $X_i$, quotiented by the $(\CC^*)^r$ gauge symmetry.
Orbits of the $(\CC^*)^2$ action are parametrized by the values of $\abs{P}^2$ and $\abs{Q}^2$ (or any other pair of chirals).
Orbits are typically $(\CC^*)^2$, but they reduce to $\CC^*$ in each locus $\{X_i=0\}$ and to a point at pairwise intersections thereof.

To describe the allowed values $(\abs{P}^2,\abs{Q}^2)$, consider the D-term equations for the GLSM written in the second basis~\eqref{GLSMIIcharges}:
under the $\alpha$-th $U(1)$ gauge factor the fields $P$, $X_\alpha$, $Q$ have charges $p_\alpha$, $-n$, $q_\alpha$ and other multiplets are neutral.
We denote by $\zeta'_\alpha=\sum_{\beta=1}^r\bigl(n(C^{-1})^{\alpha\beta}\zeta_\beta\bigr)$ the FI parameters in this basis (in terms of those in the first basis).
The D-term equations are then
\begin{equation}
  p_\alpha\abs{P}^2 + q_\alpha\abs{Q}^2 = \zeta'_\alpha + n \abs{X_\alpha}^2 , \quad \text{for } 1\leq \alpha\leq r .
\end{equation}
Up to the toric action, this fixes all $X_\alpha$ in terms of $(\abs{P}^2,\abs{Q}^2)$, provided that the linear inequalities $p_\alpha\abs{P}^2+q_\alpha\abs{Q}^2 \geq \zeta'_\alpha$ are obeyed.  The toric diagram of~$X$ depicted in \autoref{fig:HJtoric} thus consists of the subset~$S$ of the upper quadrant that lies above all lines
\begin{equation}\label{pqline}
  p_\alpha\abs{P}^2+q_\alpha\abs{Q}^2 = \zeta'_\alpha \,.
\end{equation}
Extending notation to include $\zeta'_0=\zeta'_{r+1}=0$ and $X_0=P$ and $X_{r+1}=Q$, the $\alpha=0$ and $\alpha=r+1$ lines are the two axes $\{\abs{P}^2=0\}$ and $\{\abs{Q}^2=0\}$.

\begin{figure}[t]
  \centering
  \begin{tikzpicture}
    \draw (0,0) -- (0,3);
    \draw (.7,-.5) -- (-.5,3.1) node [left] {\tiny $\{X_1=0\}$};
    \draw (1.5,-.5) -- (-.5,1.5) node [left] {\tiny $\{X_2=0\}$};
    \draw (2.8,-.5) -- (-.5,.6) node [left] {\tiny $\{X_3=0\}$};
    \draw (0,0) -- (3,0);
    \draw[very thick, {stealth}-{stealth}] (0,3) node [above] {$Q$} -- (0,1.6) -- (.3,.7) -- (.85,.15) -- (1.3,0) -- (3,0) node [right] {$P$};
    \node at (2,2) {$S$};
  \end{tikzpicture}
  \qquad
  \begin{tikzpicture}
    \draw (0,0) -- (0,3);
    \draw (.7,-.5) -- (-.5,3.1) node [left] {\tiny $\{X_1=0\}$};
    \draw (1.5,-.5) -- (-.5,1.5) node [left] {\tiny $\{X_2=0\}$};
    \draw (2.2,-.5) -- (-.5,.4) node [left] {\tiny $\{X_3=0\}$};
    \draw (0,0) -- (3,0);
    \draw[very thick, {stealth}-{stealth}] (0,3) node [above] {$Q$} -- (0,1.6) -- (.3,.7) -- (1,0) -- (3,0) node [right] {$P$};
    \node at (2,2) {$S$};
  \end{tikzpicture}
  \caption{\label{fig:HJtoric}Toric diagrams of two resolutions of $\mathbb{C}^2/\mathbb{Z}_{4(-1)}$.  These are Higgs branches in two different phases of the same GLSM of rank $r=3$.  On the left, all exceptional divisors are blown up, while on the right the third exceptional divisor is blown down.}
\end{figure}

In a given phase, the boundary of~$S$ consists of a collection of line segments along some of the lines~\eqref{pqline}.
Each such line segment corresponds to an exceptional divisor of the toric geometry, that can be blown down by varying from $\zeta'_\alpha\gg 0$ to $\zeta'_\alpha\ll 0$.
The $2^r$ phases of the GLSM are characterized by the set $A\subset \{1,\dots,r\}$ of divisors that are blown up, namely such that $\{X_\alpha=0\}$ is neither empty nor a point.
Since $q_\alpha/p_\alpha$ increases with $\alpha$ we learn that edges of the set~$S$ are segments of lines corresponding to $\{X_i=0\}$ for $i\in\{0\}\cup A\cup\{r+1\}$, in increasing order.
Later, we need the deleted set~$\Delta$.  It is the union of
the hyperplanes $\{X_\alpha=0\}$ for each $\alpha\in\{1,\dots,r\}\setminus A$,
and of the intersections $\{X_\alpha=X_\beta=0\}$ for $\alpha,\beta\in \{0\}\cup A\cup\{r+1\}$ that are not consecutive elements of this set (namely such that there exists $\gamma\in A$ with $\alpha<\gamma<\beta$).

Note that $X$ typically has orbifold singularities (see \autoref{ssec:local models} for details).
For instance in the phase $A=\emptyset$ the non-zero vevs of $X_1$, \dots, $X_r$ only break the gauge group down to $\mathbb{Z}_n$, which acts on $P$ and $Q$ with charges $p_1=p$ and $q_1=1$ (one could equally well choose charges $p_r=1$ and $q_r$ because $p_1 q_r = p_r q_1 = 1 \mod n$).
Thus, in that phase, $X=\mathbb{C}^2/\mathbb{Z}_{n(p)}$.

\paragraph{Description of phase boundaries.}

Phase boundaries occur when one of the lines~\eqref{pqline} touches~$S$ at a single point, namely when three of these lines intersect at a point that is above any other line~\eqref{pqline}.
The lines for $\alpha=i,j,k$ (with $0\leq i<j<k\leq r+1$) have such a common intersection when
\begin{equation}\label{pqzetaphaseboundary}
  \det
  \begin{pmatrix}
    p_i & q_i & \zeta'_i \\
    p_j & q_j & \zeta'_j \\
    p_k & q_k & \zeta'_k
  \end{pmatrix}
  = 0 ,
  \quad \text{and, for any $0\leq\ell\leq r+1$,} \quad
  \det
  \begin{pmatrix}
    p_i & q_i & \zeta'_i \\
    p_\ell & q_\ell & \zeta'_\ell \\
    p_k & q_k & \zeta'_k
  \end{pmatrix}
  \geq 0
\end{equation}
where we recall $\zeta'_0=\zeta'_{r+1}$.
The same linear condition on FI parameters can also be derived in the first basis as follows by solving D-term equations together with $X_i=X_j=X_k=0$.
The D-term equations for $i<\alpha<j$ together with $X_i=X_j=0$ give a unique solution for $\abs{X_{i+1}}^2,\dots,\abs{X_{j-1}}^2$ in terms of $\zeta_{i+1},\dots,\zeta_{j-1}$.
Likewise $X_j=X_k=0$ and the D-term equations for $i<\alpha<j$ give $\abs{X_\alpha}^2$ for $j<\alpha<k$ in terms of $\zeta_\alpha$ for the same range of~$\alpha$.
The $j$-th D-term equation $\abs{X_{j-1}}^2+\abs{X_{j+1}}^2=\zeta_j$ then provides a linear constraint on $\zeta_\alpha$ for $i<\alpha<k$.
This linear equation must be combined with the inequalities in~\eqref{pqzetaphaseboundary} (converted to basis~I).

The more conceptual point of view is to consider a $U(1)\subset U(1)^r$ that acts trivially on all chiral multiplets except $X_i$, $X_j$, $X_k$, as we do in~\eqref{localdivisorchargesU1Zm}.  Its FI parameter $\zeta_{\text{loc}}$ is given in~\eqref{localmodeltval} as a linear combination of $\zeta_\alpha$ for $i<\alpha<k$.  The corresponding D-term equation writes $\zeta_{\text{loc}}$ as a linear combination of $\abs{X_i}^2$, $\abs{X_j}^2$, $\abs{X_k}^2$, so the point $X_i=X_j=X_k=0$ characterizing wall-crossing occurs at $\zeta_{\text{loc}} = 0$.  In the phase with the exceptional divisor~$E_j$ blown up, its volume is controlled by $\abs{\zeta_{\text{loc}}}$, and is independent of $\zeta_i$ and~$\zeta_k$.  In particular in the fully resolved phase we have
\begin{equation}
  \Vol(E_j) \sim \zeta_j .
\end{equation}

\paragraph{Calabi--Yau walls.}

Besides the (classical) position of walls, it is interesting to determine which ones are Calabi-Yau walls, because these are true codimension~$1$ singularities around which it makes sense to study monodromies in the K\"ahler moduli space.
The wall at which $X_i$, $X_j$, $X_k$ can vanish at the same point is a cone of the charges~$Q^{\neq i,j,k}$.
It is Calabi-Yau if $Q^{\text{tot}}=\sum_\ell Q^\ell$ can be written as a linear combination of these.
We prove now that the condition is that $a_{i+1}=\dots=a_{k+1}=2$.

Consider first the case $i=0$, $k=r+1$, namely a wall-crossing from the orbifold phase.
In basis II, $Q^{\text{tot}}$ has components $p_\alpha+q_\alpha-n$ for $1\leq\alpha\leq r$ and
we want to write it as a linear combination of the charges $Q^{\neq i,j,k}$, which in basis II are $-n$ times each basis vector except the $j$-th one.
This exactly requires a vanishing $j$-th component $p_j+q_j-n=0$.
We proved below~\eqref{gcdgcdgcd} that this only happens in the Calabi-Yau case (all $a_\alpha=2$).
Of course, since bases are equivalent, we could have obtained the same conclusion in basis I, with more work.

Now consider the general case and work in basis I of the GLSM\@.
Start from $Q^{\text{tot}}$.  By subtracting a linear combination of the $Q_{<i}$ we can cancel components $1,\dots,i$ of~$Q^{\text{tot}}$.
Likewise subtracting a linear combination of the $Q_{>k}$ cancels components $k,\dots,r$.
In this way the problem reduces to the case $i=0$, $k=r+1$ which we have analysed, so we learn that
the wall is Calabi–Yau if and only if $a_{i+1}=\dots=a_{k-1}=2$.

An alternative point of view is to look at the local $U(1)\times\ZZ_m$ model of~\eqref{localdivisorchargesU1Zm} that describes the wall-crossing.
The wall is a true singularity if and only that local model has a true singularity, namely is Calabi-Yau.  We compute that the sum of charges in the local model is $d_{ik}-d_{ij}-d_{jk}$ up to a scaling.
We proved below~\eqref{gcdgcdgcd} that this vanishes if and only if $a_{i+1}=\dots=a_{k-1}=2$.

When $a_{i+1}=\dots=a_{k-1}=2$, one can check that the RG flow (which shifts $\zeta$ in the $-Q^{\text{tot}}$ direction) pushes deeper into the wall defined by the inequalities in~\eqref{pqzetaphaseboundary}, so the wall is an IR Calabi-Yau wall.

\subsection{\label{ssec:local models}Local models}

Let $E_i=\{X_i=0\}$ for $0\leq i\leq r+1$.
The sets $E_0$ and $E_{r+1}$ are always non-compact, while the $E_\alpha$ for $1\leq\alpha\leq r$ are exceptional divisors or are empty depending on whether $\alpha\in A$ or not.
We now describe the geometry near intersections $E_i\cap E_j$ for $0\leq i<j\leq r+1$ in phases where they exist, then near $E_j$ for $0\leq j\leq r+1$.

\paragraph{Local model near an intersection.}

An intersection $E_i\cap E_j$ is non-empty only in phases such that $i$ and $j$ are elements of $\{0\}\cup A\cup\{r+1\}$ (to have $E_i\neq\emptyset$ and $E_j\neq\emptyset$) and such that no other element $\alpha\in A$ is between $i$ and~$j$.  The second condition ensures $E_i$ and $E_j$ intersect, as is manifest in the $(\abs{P}^2,\abs{Q}^2)$ plane.
Near the intersection point $E_i\cap E_j$, all chiral multiplets other than $X_i$ and~$X_j$ have a vev.
By the Higgs mechanism this vev breaks the gauge symmetry down to the subgroup of elements of $U(1)^r$ that fix all $X_l$ other than $X_i$ and~$X_j$.

\begin{figure}[t]
  \centering
  \begin{tikzpicture}
    \draw (0,0) -- (0,3);
    \draw (0,0) -- (3,0);
    \filldraw[gray, opacity=.3] (0,3) -- (0,0) -- (3,0) -- (3,3);
    \draw[very thick, {stealth}-{stealth}] (0,3) node [above] {$X_2$} -- (0,0) -- (3,0) node [right] {$X_1$};
  \end{tikzpicture}
  \qquad
  \begin{tikzpicture}
    \draw (0,0) -- (0,3);
    \draw (.7,-.5) -- (-.5,3.1) node [left] {\tiny $\{X_1=0\}$};
    \draw (1.5,-.5) -- (-.5,1.5) node [left] {\tiny $\{X_2=0\}$};
    \draw (2.8,-.5) -- (-.5,.6) node [left] {\tiny $\{X_3=0\}$};
    \draw (0,0) -- (3,0);
    \draw[{stealth}-{stealth}] (0,3) node [above] {$Q$} -- (0,1.6) -- (.3,.7) -- (.85,.15) -- (1.3,0) -- (3,0) node [right] {$P$};
    \filldraw[gray, opacity=.3] (0.05,3) -- (0.05,1.45) -- (.3,.7) -- (.8,.2) -- (1.25,0.05) -- (3,0.05) -- (3,3);
    \draw[very thick] (0.05,1.45) -- (.3,.7) -- (.8,.2);
  \end{tikzpicture}
  \caption{\label{fig:locmodel}Toric diagrams of the Higgs branch of: (a) a local model of an intersection $E_i\cap E_j$, and (b) the full GLSM\@.
    The region shaded in gray is the image of (a) inside (b) under the embedding of complex manifolds/orbifolds.
    The toric diagram does not depic the orbifold group~$\ZZ_{d_{ij}}$ since it acts purely on phases of chiral multiplets.}
\end{figure}

Let us work in basis~I of the GLSM\@, and let $g=(g_1,\dots,g_r)\in U(1)^r$ be such an element.
Fixing $X_0$ requires $g_1=1$, in which case fixing $X_1$ requires $g_2=1$ and so on, so $g_1,\dots,g_i=1$.
Likewise $g_j=\dots=g_r=1$.
Next, we consider in turn the constraints coming from the fact that $g$ fixes $X_\alpha$ for $\alpha=j-1,j-2,\dots,i+1$.
Each step~$\alpha$ gives one component $g_{\alpha-1}=(g_\alpha)^{a_\alpha} (g_{\alpha+1})^{-1}$, and an explicit expression is
\begin{equation}\label{localmodelintersectiongalpha}
  g_\alpha = (g_{j-1})^{d_{\alpha j}} \qquad\text{for $i\leq\alpha\leq j$.}
\end{equation}
This relies on the initial cases $d_{jj}=0$ and $d_{(j-1)j}=1$ and the recursion relation~\eqref{drecursion} $d_{(\alpha+1)j}+d_{(\alpha-1)j}=a_\alpha d_{\alpha j}$, applied for $i<\alpha<j$.  Then, $g_i=1$ forces $g_{j-1}$ to be a $d_{ij}$-th root of unity.
Altogether we are left with a gauge group~$\ZZ_{d_{ij}}$.
The two remaining chirals $X_i$ and~$X_j$ have charges~$d_{(i+1)j}$ and~$1$ as summarized in the following table
\begin{equation}\label{localmodelintersection}
  \begin{tabular}{*{7}{>{$}c<{$}}}
    \toprule
           & X_i & X_j \\
    \midrule
      \ZZ_{d_{ij}} & d_{(i+1)j} & 1 \\
    \bottomrule
  \end{tabular}
  \qquad \text{or in another basis} \qquad
  \begin{tabular}{*{7}{>{$}c<{$}}}
    \toprule
           & X_i & X_j \\
    \midrule
      \ZZ_{d_{ij}} & 1 & d_{i(j-1)} \\
    \bottomrule
  \end{tabular}
\end{equation}
where we used $d_{i(j-1)}d_{(i+1)j}-d_{ij}d_{(i+1)(j-1)}=d_{i(i+1)}d_{(j-1)j}=1$ to invert $d_{(i+1)j}$ modulo~$d_{ij}$.
We conclude that near $E_i\cap E_j$ the Higgs branch is close to the orbifold $\CC^2/\ZZ_{d_{ij}(d_{(i+1)j})}$.

\paragraph{Consistency checks for the intersection model.}

This reproduces our earlier conclusions for the orbifold phase $A=\emptyset$, which is the only phase in which $E_0\cap E_{r+1}$ is non-empty.
We had found that the Higgs branch is $\CC^2/\ZZ_{n(p)}$ in that phase and indeed $d_{0(r+1)}=n$ and $d_{1(r+1)}=p_1=p$.
In fact all cases reduce to this one by noting that vevs of $X_0,\dots,X_{i-1}$ and $X_{j+1},\dots,X_{r+1}$ break completely the gauge factors with indices $1\leq\alpha\leq i$ and $j\leq\alpha\leq r$ (see above), thus reducing the problem to the orbifold phase of a $U(1)^{j-i-1}$ Hirzebruch-Jung model with $j-i+1$ chirals.
From this point of view, the orbifold singularities for $i+2\leq j$ are due to the presence of blown-down exceptional divisors $E_\alpha$ for $i<\alpha<j$.
In contrast, the residual gauge group trivial for $i+1=j$ because $d_{ij}=d_{i(i+1)}=1$, so intersections $E_i\cap E_{i+1}$ are smooth.

Recall that as a complex orbifold the Higgs branch of an abelian GLSM is a quotient by the complexified gauge group~$G_{\CC}$ of $V\setminus\Delta$, the space of chiral multiplets minus some deleted set consisting of coordinate subspaces.
This construction enables us to embed the Higgs branch of the local model~\eqref{localmodelintersection}, as a complex orbifold, into that of the full model in the given phase.
A point $(X_i,X_j)\in\CC^2/\ZZ_{d_{ij}(d_{(i+1)j})}$ is mapped to $(1,\dots,1,X_i,1,\dots,1,X_j,1,\dots,1)\in V=\CC^{r+2}$, up to $(\CC^*)^r$ gauge transformations, where the non-trivial entries are in positions $0\leq i<j\leq r+1$.
Our earlier analysis of what subgroup of $U(1)^r$ leaves the non-zero vevs of $X_l$ for $l\neq i,j$ extends trivially to the complexified setting and shows that the map is well-defined.
Its image consists of all $(X_0,\dots,X_{r+1})\in V$ that can be gauge-fixed to have $X_l=1$ for $l\neq i,j$, namely to the set $V\setminus\bigcup_{l\neq i,j} E_l$.
The toric diagram is depicted in \autoref{fig:locmodel}.

\paragraph{Local model near an exceptional divisor.}

We repeat the same analysis for an exceptional divisor~$E_j$ for $1\leq j\leq r$ in some phase.
Let $E_i$ and $E_k$ be non-empty and intersect $E_j$ at one point each, namely the indices should obey $i<j<k$ and be successive elements in the set $\{0\}\cup A\cup\{r+1\}$.
From the $(\abs{P}^2,\abs{Q}^2)$ toric diagrams we know that $E_j$ is topologically a two-sphere and it has a $U(1)$ isometry with two fixed points: $E_i\cap E_j$ and $E_j\cap E_k$.
In $E_j$, hence on a neighborhood thereof, all chiral multiplets except $X_i$, $X_j$, $X_k$ get a vev.
The Higgs mechanism breaks the gauge group to the subgroup of elements $g=(g_1,\dots,g_r)\in U(1)^r$ that leave all the vevs invariant.

Again we work in basis I and find that fixing $X_0,\dots,X_{i-1}$ and $X_{k+1},\dots,X_{r+1}$ forces $g_1=\dots=g_i=1$ and $g_k=\dots=g_r=1$.  Next, the calculations near~\eqref{localmodelintersectiongalpha} give $g_\alpha = (g_{i+1})^{d_{i\alpha}}$ for $i\leq\alpha\leq j$, and $g_\alpha = (g_{k-1})^{d_{\alpha k}}$ for $j\leq\alpha\leq k$.
The compatibility of these two expressions of $g_j$ means that the gauge group is parametrized by solutions of $g_{i+1}^{d_{ij}}=g_{k-1}^{d_{jk}}$, namely
\begin{gather}
  g_{i+1}=h^{d_{jk}/m}\omega^u \quad \text{and} \quad g_{k-1}=h^{d_{ij}/m}\omega^v \quad\text{for}\quad (h,\omega)\in U(1)\times\ZZ_m , \quad \text{so}
  \\
  g_\alpha = \begin{cases}
    1 & \text{if } 1\leq\alpha\leq i \text{ or } k\leq\alpha\leq r, \\
    h^{d_{i\alpha}d_{jk}/m}\omega^{ud_{i\alpha}} & \text{if } i\leq\alpha\leq j, \\
    h^{d_{ij}d_{\alpha k}/m}\omega^{vd_{\alpha k}} & \text{if } j\leq\alpha\leq k,
  \end{cases}
\end{gather}
where $m\in\ZZ_{\geq 1}$ and $u,v\in\ZZ$ are chosen to obey
\begin{equation}
  m=\gcd(d_{ij},d_{jk})=ud_{ij}-vd_{jk} .
\end{equation}
Different choices of $(u,v)$ amount to different choices of basis for~$\ZZ_m$.
The GLSM can be expressed in various choices of basis, related by automorphisms of $U(1)\times\ZZ_m$.
Besides conjugation that changes signs of all charges, one can add to the $\ZZ_m$ charges any multiple of the $U(1)$ charges, and multiply the $\ZZ_m$ charges by any invertible element of~$\ZZ_m$.

From how the $U(1)$ factor of the gauge group of the local model embeds into the $U(1)^r$ gauge group of the full GLSM we work out the FI-theta parameter of the local model,
\begin{equation}\label{localmodeltval}
  t_{\text{loc}} = \frac{1}{m} \sum_{\alpha=i+1}^{k-1} d_{i\min(\alpha,j)}d_{\max(\alpha,j)k} \, t_\alpha .
\end{equation}

The chiral multiplet $X_i$ transforms by $g_{i-1}g_i^{-a_i}g_{i+1}=h^{d_{jk}/m}\omega^u$ and
$X_k$ by $g_{k-1}g_k^{-a_k}g_{k+1}=h^{d_{ij}/m}\omega^v$
while $X_j$~transforms by $g_{j-1}g_j^{-a_j}g_{j+1}= h^{\ell/m}\omega^s$ with
\begin{equation}
  \begin{aligned}
    \ell & = d_{i(j-1)}d_{jk}-a_jd_{ij}d_{jk}+d_{ij}d_{(j+1)k}
    = -d_{i(j+1)}d_{jk}+d_{ij}d_{(j+1)k}
    = -d_{ik}
    \\
    s & \equiv -d_{i(j+1)} u + d_{(j+1)k} v \equiv d_{i(j-1)} u - d_{(j-1)k} v \mod{m}
  \end{aligned}
\end{equation}
where we used the recursion relation and~\eqref{dddddd} and $d_{j(j+1)}=1$ to simplify $\ell$ and to give two equally complicated expressions for~$s$.
Charges are summarized in the following table, with $m$, $u$, $s$, $v$ given above:
\begin{equation}\label{localdivisorchargesU1Zm}
  \begin{array}{cccc}
    \toprule
           & X_i & X_j & X_k \\
    \midrule
      U(1) & d_{jk}/m & -d_{ik}/m & d_{ij}/m \\
      \ZZ_m & u & s & v \\
    \bottomrule
  \end{array}
\end{equation}
It will be useful later that $X_i^{d_{i(j+1)}}X_jX_k^{-d_{(j+1)k}}$ and $X_i^{d_{ij}}X_k^{-d_{jk}}$ and $X_i^{-d_{i(j-1)}}X_jX_k^{d_{(j-1)k}}$ are gauge neutral.

\paragraph{Consistency checks for the exceptional divisor model.}

In this analysis we first reduced the model to a $U(1)^{k-i-1}$ Hirzebruch-Jung model with $k-i+1$ chirals.
Up to relabeling, this can be mapped to the case $i=0$ and $k=r+1$ (with a different value of~$r$), in which a single exceptional divisor, $E_j$, is blown up.
In that case, $m=\gcd(p_j,q_j)$ and the local model has $U(1)$ charges $(p_j,-n,q_j)/m$ and $\ZZ_m$ charges given above.
It is instructive to reproduce some of these results in basis II\@.
Elements $g=(g^1,\dots,g^r)\in U(1)^r/\ZZ_n^{r-1}$ that only acts on $X_0$, $X_j$, $X_{r+1}$ are those for which all $g^\alpha\in\ZZ_n$ except $g^j\in U(1)$.
The residual gauge group is thus $(U(1)_j\times\ZZ_n^{r-1})/\ZZ_n^{r-1}$ and the question is how the $\ZZ_n^{r-1}$ quotient is taken.
The $U(1)_j$ factor acts with charges $(p_j,-n,q_j)$, hence its $\ZZ_m$ subgroup acts trivially.
By a volume argument the residual gauge group must be $U(1)_j/\ZZ_m$ times a discrete abelian group of order~$m$.

Another consistency check is to determine which $(h,\omega)\in U(1)\times\ZZ_m$ fix~$X_i$.
Write $\omega=\exp(2\pi ia/m)$ and $h=\exp(-2\pi i b/d_{jk})$ with $a\in\ZZ$ by construction.
The condition is that $au\equiv b \bmod{m}$.
It is solved exactly by the $d_{jk}$ powers of $(h,\omega)=\bigl(\exp(-2\pi i/m),\exp(2\pi i u/d_{jk})\bigr)$.
The residual gauge group is $\ZZ_{d_{jk}}$, and it is easy to check that it acts on $X_j$ and $X_k$ with charges $d_{(j+1)k}$ and~$1$.
As expected from \eqref{localmodelintersection} we find the orbifold~$\CC^2/\ZZ_{d_{jk}(d_{(j+1)k})}$.
Near the other pole $E_i\cap E_j$ we similarly find $\CC^2/\ZZ_{d_{ij}(d_{i(j-1)})}\simeq \CC^2/\ZZ_{d_{ij}(d_{(i+1)j})}$.
The local model embeds as a complex orbifold into the full Higgs branch, and its image is the region in which no chiral vanishes except $X_i$, $X_j$, $X_k$.

Does the local model approximate well the metric on the divisor~$E_j$ in the full Higgs branch?  Not always.
Integrating out chirals that are nonzero near~$E_j$ and removing broken gauge fields is an approximation that is valid, provided that vevs of the remaining chiral multiplets are much less than those that we integrate out.
However, the vev of $X_i$ near the intersection $E_j\cap E_k$ may be bigger than some other chiral multiplets~$X_\alpha$ for $j<\alpha<k$, especially when FI parameters are taken close to a wall that corresponds to blowing up~$E_\alpha$ (see right side of \autoref{fig:HJtoric} for instance).
It may be interesting to make quantitative comparisons between the $U(1)$-invariant metrics on~$E_j$ for different models.
When discussing the metric we will assume that the regime of FI parameters is such that all chirals other than $X_i$, $X_j$, $X_k$ have large vevs in the neighborhood of $E_j$ that we are considering.

Very close to $E_i\cap E_j$ we know from \eqref{localmodelintersection} that the metric is that of~$\CC^2/\ZZ_{d_{ij}(d_{i(j-1)})}$, parametrized by $X_i$ and~$X_j$.
The submanifold $E_j=\{X_j=0\}$ has the same deficit angle at $E_i\cap E_j$ as $\CC/\ZZ_{d_{ij}}$.
The exceptional divisor is thus a topological two-sphere with $U(1)$ isometry and two conical singularities.
As a complex manifold/orbifold it is simply $\CP^1$, with projective coordinates $(X_i^{d_{ij}}:X_k^{d_{jk}})$, for the same reason that $\CC/\ZZ_n\simeq\CC$ under the map $X\to X^n$.
As a K\"ahler manifold the divisor $\{X_j=0\}$ of the local model~\eqref{localdivisorchargesU1Zm} could be called $\mathbb{WCP}^1_{b,a}$ with $b=d_{jk}$ and $a=d_{ij}$.
When $m=\gcd(d_{ij},d_{jk})=1$ the local model is a $U(1)$ GLSM\@, and the K\"ahler quotient construction of its Higgs branch coincides with a standard construction of weighted projective spaces (in any dimensions).
When $m>1$ the same $U(1)$ construction would simply construct $\mathbb{WCP}^1_{b/m,a/m}$ and one needs a further $\ZZ_m$ orbifold to obtain the correct conical singularities $\CC/\ZZ_a$ and~$\CC/\ZZ_b$.

\paragraph{Line bundles on \texorpdfstring{$\mathbb{WCP}^1$}{the weighted projective line}.}

In the resolved phase of the local model~\eqref{localdivisorchargesU1Zm}, the Higgs branch is the total space of the normal line bundle of the exceptional divisor~$E_j$.
(Away from $E_j$ the metric receives strong corrections.)
Let us determine what line bundle it is and discuss more general line bundles, as this is essential for our study of B-branes on Hirzebruch-Jung models in \autoref{ssec:BbranesHJ}.
For brevity we denote $a=d_{ij}$ and $b=d_{jk}$ so that the exceptional divisor is $\mathbb{WCP}^1_{b,a}$ with conical singularities $\CC/\ZZ_a$ and $\CC/\ZZ_b$.
Let $m=\gcd(a,b)$ and $\omega_p=\exp(2\pi i/p)$ for all $p>0$.

\paragraph{Geometric point of view on line bundles.}

We discussed above how $\mathbb{WCP}^1_{b,a}$ is constructed by gluing the cones $\CC/\ZZ_a$ with coordinate~$x$ and $\CC/\ZZ_b$ with coordinate~$y$.
We choose these coordinates to be single-valued before quotienting so the well-defined expressions on the orbifold are $x^a$ and~$y^b$.
The change of coordinates between the two is then $x^a=y^{-b}$.

A line bundle on $\mathbb{WCP}^1_{b,a}$ is built by gluing an orbifold (or equivariant) line bundle on $\CC/\ZZ_a$ and one on $\CC/\ZZ_b$ through a transition map.
The orbifold line bundles are characterized by their charge under the orbifold group, which in view of later identifications we denote respectively by $-\bar{\gamma}\in\ZZ_a$ under~$\ZZ_a$ and $\bar{\delta}\in\ZZ_b$ under~$\ZZ_b$.
Sections of the $\CC/\ZZ_a$ bundle are $f_{\text{N}}\colon\CC^*\to\CC$ such that $f_{\text{N}}(x)=\omega_a^{-\bar{\gamma}} f_{\text{N}}(\omega_a x)$.
The transition map must map that to a section $f_{\text{S}}\colon\CC^*\to\CC$ such that $f_{\text{S}}(y)=\omega_b^{\bar{\delta}} f_{\text{S}}(\omega_b y)$ of the other orbifold bundle,
by a relation of the form $f_{\text{N}}(x) = ({\cdots}) f_{\text{S}}(y)$ for $x^a=y^{-b}$.
Since neither $x$ nor $y$ determines the other uniquely in general, and since the orbifold line bundles may have non-zero charges, the coefficient $({\cdots})$ defining the transition map typically depends on both $x$ and~$y$, subject to the relation $x^a=y^{-b}$.
The transition map is thus
\begin{equation}\label{WCP1bundlegammadelta}
  f_{\text{N}}(x) = x^\gamma y^\delta f_{\text{S}}(y) .
\end{equation}
The transition map should reproduce the orbifold group actions on $f_{\text{N}}$ and $f_{\text{S}}$ when one keeps $y$ or~$x$ fixed, respectively.
This implies $-\bar{\gamma}=-\gamma\bmod{a}$ and $\bar{\delta}=\delta\bmod{b}$.
Altogether the line bundle is characterized by
\begin{equation}
  (\gamma,\delta)\in\ZZ^2/\bigl((a,-b)\ZZ\bigr) .
\end{equation}
This group is isomorphic to $\ZZ\times\ZZ_m$ with $m=\gcd(a,b)$.
In particular, on weighted projective spaces $\mathbb{WCP}^1_{b,a}$ with $\gcd(a,b)=1$, all line bundles are tensor powers of one line bundle that we call~$\cO(1)$.
In terms of the $U(1)$ GLSM discussed next, that line bundle is parametrized by a scalar of $U(1)$ charge~$1$.

\paragraph{GLSM point of view on line bundles.}

The $\mathbb{WCP}^1_{b,a}$ exceptional divisor is the Higgs branch of the following $U(1)\times\ZZ_m$ GLSM,
obtained from~\eqref{localdivisorchargesU1Zm} by dropping the chiral multiplet~$X_j$:
\begin{equation}
  \begin{array}{cccc}
    \toprule
           & X_i & X_k \\
    \midrule
      U(1) & b/m & a/m \\
      \ZZ_m & u & v \\
    \bottomrule
  \end{array}
\end{equation}
where $u,v\in \ZZ$ obey $(a/m)u-(b/m)v=1$, which implies for instance that $a/m$ and $v$ are coprime.
As a complex orbifold, $\mathbb{WCP}^1_{b,a}$ is parametrized by homogeneous coordinates $(x_i:x_k)\in\CC^2\setminus\{(0,0)\}$ with the identification
$(x_i:x_k)\sim(h^{b/m}\omega^ux_i:h^{a/m}\omega^vx_k)$ for all $(h,\omega)\in\CC^*\times\ZZ_m$, the complexified gauge group.
The coordinates $x$ and $y$ of the gluing description are obtained from $(x_i:x_k)$ by gauge-fixing $x_k=1$ or $x_i=1$ so $(x_i:x_k)\sim(x:1)\sim(1:y)$.

While not strictly necessary it is instructive to check that $x$ and $y$ are subject to $\ZZ_a$ and $\ZZ_b$ orbifold identifications.
Let us gauge-fix $x_k=1$.
The elements $(h,\omega)$ that leave $x_k$ fixed are those such that $h^{a/m}\omega^v=1$.
For each $\omega\in\ZZ_m$ there are $a/m$ possible~$h$, so in total there are $m(a/m)=a$ solutions.
On the other hand, $\ell\mapsto(\omega_a^{-\ell v},\omega_m^{\ell})$ defines an injective group morphism from $\ZZ_a$ to the space of solutions:
its kernel consists of $\ell$ such that $\ell=0\bmod{m}$ and $\ell v=0\bmod{a}$,
hence $\ell=m\ell'$ and $\ell'v=0\bmod{a/m}$,
hence (because $a/m$ and $v$ are coprime) $\ell'=0\bmod{a/m}$ and finally $\ell=0\bmod{a}$.
The residual gauge group is thus $\ZZ_a$ consisting of all $(\omega_a^{-\ell v},\omega_m^{\ell})\in U(1)\times\ZZ_m$.
The group $\ZZ_a$ acts by $(x:1)\mapsto (\omega_a^{-\ell vb/m+\ell ua/m} x:1)=(\omega_a^\ell x:1)$, namely a standard orbifold~$\CC/\ZZ_a$.
The situation is the same for the other pole.

The fiber of a line bundle is parametrized by a scalar with some charges $(\alpha,\beta)$ under $U(1)\times\ZZ_m$.
A section of that line bundle is then (a meromorphic function) $f\colon\CC^2\setminus\{(0,0)\}\to\CC$ such that
\begin{equation}\label{sectionofWCP1}
  h^\alpha\omega^\beta\,f\bigl(h^{b/m}\omega^u x_i: h^{a/m}\omega^v x_k\bigr) = f(x_i:x_k) ,
  \quad \text{for all } (h,\omega)\in \CC^*\times\ZZ_m .
\end{equation}
In the $\CC/\ZZ_a$ patch, \eqref{sectionofWCP1} becomes
\begin{equation}
  \omega_a^{\ell(-v\alpha+\beta a/m)}\,f\bigl(\omega_a^{\ell} x_i: 1\bigr) = f(x_i:1) ,
\end{equation}
which describes an equivariant line bundle with charge $-v\alpha+\beta a/m\bmod{a}$ on $\CC/\ZZ_a$.
The combination $-v\alpha+\beta a/m$ can be found more directly:
the charge vector $(\alpha,\beta)$ is an integer linear combination of those of $x_i$ and~$x_k$,
\begin{equation}
  \begin{pmatrix}\alpha\\\beta\end{pmatrix}
  = -\gamma\begin{pmatrix}b/m\\u\end{pmatrix}
  + \delta\begin{pmatrix}a/m\\v\end{pmatrix} ,
  \text{ where } \gamma=v\alpha-\beta a/m, \text{ and } \delta=u\alpha-\beta b/m
\end{equation}
are defined up to shifting $(\gamma,\delta)$ by multiples of $(a,b)$.

A gauge transformation that maps $(x_i:x_k)\mapsto(\lambda x_i:\mu x_k)$ acts on the section as $f\mapsto \lambda^{-\gamma} \mu^\delta f$.
We deduce that the equivariant line bundles from which our line bundle is built have charges $-\gamma\bmod{a}$ and $\delta\bmod{b}$, respectively.
We also deduce the transition map by converting~\eqref{sectionofWCP1} to $\gamma$ and~$\delta$ and imposing $x_k=1$ and $h^{b/m}\omega^u x_i=1$:
\begin{equation}
  f(x:1) = x^\gamma y^\delta f(1:y) .
\end{equation}
This is exactly~\eqref{WCP1bundlegammadelta} since $f_{\text{N}}(x)=f(x:1)$ and $f_{\text{S}}(y)=f(1:y)$.

\paragraph{The normal bundle.}

The normal bundle of $E_j$ in the Higgs branch of the Hirzebruch-Jung model is the same line bundle as in the local model~\eqref{localdivisorchargesU1Zm}.
It is thus parametrized by a scalar~$X_j$ with charges $(\alpha,\beta)=(-d_{ik}/m,s)$, with $s=d_{i(j-1)}u-d_{(j-1)k}v$, under the $U(1)\times\ZZ_m$ gauge group.
As observed below~\eqref{localdivisorchargesU1Zm} these are the same charges as $X_i^{-d_{i(j+1)}}X_k^{d_{(j+1)k}}$ and also the same charges as $X_i^{d_{i(j-1)}} X_k^{-d_{(j-1)k}}$.
In the notation above, $(\gamma,\delta)=(d_{i(j+1)},d_{(j+1)k})$.

\subsection{\label{ssec:CoulGeom}Coulomb and mixed branches}

The Hirzebruch-Jung models we consider also admit Coulomb and mixed branch vacua.
As explained in \autoref{ssec:mixedbranches}, these are found by searching, for each subspace of the chirals, some vacua in which these chirals (and no others) are not given a mass by~$\sigma$.
The dimension of the branch is~\eqref{dimensionofbranch}, which vanishes unless the set of charges of these chiral multiplets obey linear relations.
In our models, any $r-1$ of the charge vectors $Q^i$ are linearly independent, so the only branch of positive dimension is the one in which all chiral multiplets can get vevs, namely the Higgs branch.  Other branches only consist of isolated vacua, and the distinction between Coulomb and mixed vacua is unimportant.\footnote{In more complicated models, mixed branches can have positive dimension.}

In this section it is convenient to study the theory at some energy scale~$\mu$, and correspondingly introduce a notation for the dimensionless $\hat{\sigma}\coloneqq\sigma/\mu$.  We recall the renormalized $t_{\text{ren}}=\zeta_{\text{ren}}-i\theta_{\text{ren}}$ given in~\eqref{FIrenorm}:
\begin{equation}
  t_{\text{ren}}(\mu) = t_{\text{bare}} + Q^{\text{tot}} \log\Bigl(\frac{\mu}{\Lambda}\Bigr) .
\end{equation}

Coulomb branch vacua for example are found by extremizing the effective twisted superpotential
\begin{equation}\label{WeffCoulGeom}
  \mu^{-1} \tW_{\text{eff}} = - t_{\text{ren}}\cdot\hat{\sigma} - \sum_j (Q^j\cdot\hat{\sigma}) \bigl(\log\bigl(Q^j\cdot i\hat{\sigma}\bigr) - 1\bigr)
\end{equation}
where components $t_{\text{ren},\alpha}$~are only defined up to multiples of~$2\pi i$, hence the equation is in fact $\frac{\partial \tW_{\text{eff}}}{\partial\sigma^\alpha}\in 2\pi i\mathbb{Z}$.
Exponentiating it gives
\begin{equation}
  \prod_j \bigl(Q^j\cdot i\hat{\sigma}\bigr)^{Q^j_\alpha} = e^{-t_{\text{ren},\alpha}} .
\end{equation}

Let us consider solutions of these equations for some instructive examples.

\paragraph{One parameter model.}

For the $r=1$ Hirzebruch-Jung models, $n/p = a_1$ namely $p=1$ and $a_1=n$.
We get one equation $i\hat{\sigma} (-a_1 i\hat{\sigma})^{-a_1} i\hat{\sigma}= e^{-t_{\text{ren}}}$ namely
\begin{equation}\label{oneparameterCBeqs}
  (i\hat{\sigma})^{a_1-2}=e^{t_{\text{ren}}} (-a_1)^{-a_1} .
\end{equation}
This has $a_1-2$ solutions.  In the phase $\zeta_{\text{ren}}\ll 0$ they have $\abs{\sigma}\gg\mu$, namely the approximation makes sense.
The picture that emerges is that the one-parameter model has two phases
\begin{itemize}
\item $\zeta_{\text{ren}}\ll 0$ with Higgs branch $\mathbb{C}^2/\mathbb{Z}_{n(1)}$ and no Coulomb branch;
\item $\zeta_{\text{ren}}\gg 0$ with Higgs branch the total space of $\cO(-n)\to\mathbb{CP}^1$ and $n-2$ Coulomb branch vacua.
\end{itemize}
In the Calabi-Yau case, both phases are pure-Higgs.  A non-compact Coulomb branch (arbitrary $\hat{\sigma}$) opens up at $t=a_1\log(-a_1)\bmod{2\pi i}$ and the theory is singular.

\paragraph{Two-parameter models with \texorpdfstring{$p=2$}{p=2}: the equations.}
Next we consider $\mathbb{C}^2/\mathbb{Z}_{n(2)}$ (for $n=2k-1$), namely $r=2$, $a_1=k$ and $a_2=2$.
We recall the charge matrix for convenience:
\begin{equation}\label{twoparameter}
  \begin{array}{cccccc}
    \toprule
           & X_0 & X_1 & X_2 & X_3 \\
    \midrule
      U(1)_1 & 1 & -k & 1 & 0 \\
      U(1)_2 & 0 & 1 & -2 & 1 \\
    \bottomrule
  \end{array}
\end{equation}
Critical points of the effective twisted superpotential, namely solutions of
\begin{equation}\label{C2Zn2critical}
  \begin{aligned}
    i\hat{\sigma}_1 (i\hat{\sigma}_1-2i\hat{\sigma}_2) & = e^{-t^{\text{ren}}_1} (-k i\hat{\sigma}_1+i\hat{\sigma}_2)^{k} \\
    (-ki\hat{\sigma}_1+i\hat{\sigma}_2) i\hat{\sigma}_2 & = e^{-t^{\text{ren}}_2}(i\hat{\sigma}_1-2i\hat{\sigma}_2)^2
  \end{aligned}
\end{equation}
can be found explicitly by solving the second equation then the first:
\begin{equation}\label{C2Zn2criticalsols}
  \begin{aligned}
    \hat{\sigma}_1 & = v_{\pm} \hat{\sigma}_2 , \qquad
    (i\hat{\sigma}_2)^{k-2} = \frac{e^{t^{\text{ren}}_1} v_{\pm} (v_{\pm}-2)}{(1-k v_{\pm})^{k}} ,
    \quad \text{with} \\
    v_{\pm} & = 2-ke^{t^{\text{ren}}_2}/2 \pm \sqrt{(1-2k)e^{t^{\text{ren}}_2}+k^2e^{2t^{\text{ren}}_2}/4} .
  \end{aligned}
\end{equation}
Altogether we get $2(k-2)=n-3$ solutions.
In the Calabi-Yau case $\mathbb{C}^2/\mathbb{Z}_{3(2)}$, namely $k=2$, there are no Coulomb branch vacua for generic $t^{\text{ren}}_1$, $t^{\text{ren}}_2$.  The theory has a codimension~$1$ singular locus where a non-compact one-dimensional Coulomb branch ($\hat{\sigma}_1=v_{\pm}\hat{\sigma}_2$) opens up.  This singular locus asymptotes to (shifted) classical walls.
We henceforth consider the non-Calabi-Yau case $k>2$.

\newcommand{\phasesofGLSM}{%
  \begin{tikzpicture}[scale=1]
    \draw[double,-{stealth}=.3] (0,0) -- (1,0) node [right=-3pt] {\scriptsize $(1,0)$};
    \draw[thick,-{stealth}=.3] (0,0) -- (0,1) node [above=-3pt] {\scriptsize $(0,1)$};
    \draw[thick,-{stealth}=.3] (0,0) -- (-3,1) node [left=-3pt] {\scriptsize $(-k,1)$};
    \draw[thick,-{stealth}=.3] (0,0) -- (1,-2) node [below=-3pt] {\scriptsize $(1,-2)$};
    \draw[thick,densely dotted] (0,0) -- (-2.4,1.2);
    \draw[thick,densely dotted] (0,0) -- (1.5,-1.5);
    \node at (1.5,.75) {\scriptsize $A=\{1,2\}$};
    \node[fill=white,inner sep=0pt] at (-1,.75) {\scriptsize $A=\{2\}$};
    \node at (-1,-1) {\scriptsize $A=\emptyset$};
    \node[fill=white,inner sep=0pt] at (1.5,-1) {\scriptsize $A=\{1\}$};
  \end{tikzpicture}
  \qquad
  \begin{tikzpicture}[scale=.5]
    \draw[double,-{stealth}=.3] (0,0) -- (2,1) node [right] {\scriptsize $(2,1)$};
    \draw[thick,-{stealth}=.3] (0,0) -- (1,3) node [left] {\scriptsize $(1,k)$};
    \draw[thick,-{stealth}=.3] (0,0) -- (-5,0) node [below] {\scriptsize $(-n,0)$};
    \draw[thick,-{stealth}=.3] (0,0) -- (0,-5) node [above left] {\scriptsize $(0,-n)$};
    \draw[thick,densely dotted] (0,0) -- (2,-4);
    \draw[thick,densely dotted] (0,0) -- (-4.5,1.5);
    \node at (3,1.9) {\scriptsize $n{-}3$ Coul.};
    \node at (-2,1.5) {\scriptsize $k{-}2$ Coul.};
    \node at (-4,.5) {\scriptsize $k{-}2$ mixed};
    \node at (3,-2.5) {\scriptsize $n{-}3$}; \node at (3,-3) {\scriptsize Coul.};
    \node at (1,-3.5) {\scriptsize $n{-}3$}; \node at (1,-4) {\scriptsize mixed};
  \end{tikzpicture}
}
\begin{figure}[tbp]\centering
  \phasesofGLSM
  \caption{\label{fig:phasesC2Zn2}Phase structure for $\mathbb{C}^2/\mathbb{Z}_{n(2)}$ with $n=2k-1$ and $k>2$.
    The FI parameter runs towards the IR in the direction given by the double arrow, which in this model is parallel to a wall.
    Dotted lines denote transitions between sub-phases in which isolated massive vacua are Coulomb branch vacua or mixed branch vacua.
    To avoid clutter, we only indicate the sets~$A$ in one diagram and the numbers of mixed and Coulomb branch vacua in the other diagram (note that $n-3=2(k-2)$).
    Left: basis~I\@.  Right: basis~II\@.}
\end{figure}

\paragraph{Two-parameter models with \texorpdfstring{$p=2$}{p=2}: four phases.}

Besides being solutions of~\eqref{C2Zn2criticalsols}, Coulomb branch vacua must also be such that $\abs{\hat{\sigma}_1}$, $\abs{\hat{\sigma}_2}$, $\abs{\hat{\sigma}_1-2\hat{\sigma}_2}$, $\abs{-k\hat{\sigma}_1+\hat{\sigma}_2}$ are all large so as to make all chiral multiplets massive.
Depending on the phase, only some of these $n-3$ values of~$\hat{\sigma}$ are genuine Coulomb branch vacua.
We now consider in turn each of the $2^r=4$ phases we found when analysing the Higgs branch.
These are recapitulated in \autoref{fig:phasesC2Zn2}.
Recall that they are classified by the set of exceptional divisors that are blown up.
\begin{itemize}
\item \textbf{No divisor blown up} ($A=\emptyset$): $\zeta'^{\text{ren}}_1,\zeta'^{\text{ren}}_2\ll 0$, that is, $2\zeta^{\text{ren}}_1+\zeta^{\text{ren}}_2\ll 0$ and $\zeta^{\text{ren}}_1+k\zeta^{\text{ren}}_2\ll 0$.
  The equations~\eqref{C2Zn2critical} characterizing Coulomb branch vacua can be usefully combined into
  \begin{equation}
    \begin{aligned}
      (-k i\hat{\sigma}_1+i\hat{\sigma}_2)^{2k-1} & = e^{2t^{\text{ren}}_1+t^{\text{ren}}_2} (i\hat{\sigma}_1)^2 i\hat{\sigma}_2 , \\
      (i\hat{\sigma}_1-2i\hat{\sigma}_2)^{2k-1} & = e^{t^{\text{ren}}_1+kt^{\text{ren}}_2} i\hat{\sigma}_1 (i\hat{\sigma}_2)^k ,
    \end{aligned}
  \end{equation}
  and in addition $\abs{\hat{\sigma}_1}$, $\abs{\hat{\sigma}_2}$, $\abs{\hat{\sigma}_1-2\hat{\sigma}_2}$, $\abs{-k\hat{\sigma}_1+\hat{\sigma}_2}$ must all be large.
  Dividing the equations by $(\abs{\hat{\sigma}_1}+\abs{\hat{\sigma}_2})^{2k-1}$ and using that the exponentials are small in this phase,
  we learn that both $\abs{-k \hat{\sigma}_1+\hat{\sigma}_2}$ and $\abs{\hat{\sigma}_1-2\hat{\sigma}_2}$ must be parametrically smaller than $\abs{\hat{\sigma}_1}+\abs{\hat{\sigma}_2}$.
  This is impossible since $k\neq 1/2$.  There are no Coulomb branch vacua in this phase.

\item \textbf{First divisor blown up} ($A=\{1\}$): $2\zeta^{\text{ren}}_1+\zeta^{\text{ren}}_2\gg 0$ and $\zeta^{\text{ren}}_2\ll 0$.
  For this phase we work out the $t^{\text{ren}}_2\to-\infty$ asymptotics of~\eqref{C2Zn2criticalsols} to be
  \begin{equation}\label{t2tominfty}
    \hat{\sigma}_1 \sim \hat{\sigma}_2 \sim (-k\hat{\sigma}_1+\hat{\sigma}_2) \sim e^{(2t^{\text{ren}}_1+t^{\text{ren}}_2)/(2k-4)} , \qquad
    (\hat{\sigma}_1-2\hat{\sigma}_2) \sim e^{(2t^{\text{ren}}_1+(k-1)t^{\text{ren}}_2)/(2k-4)} .
  \end{equation}
  While the first three combinations are large throughout the phase, the last one is large only in a sub-phase $2\zeta^{\text{ren}}_1+(k-1)\zeta^{\text{ren}}_2\gg 0$.
  In that sub-phase we get $n-3=2k-4$ Coulomb branch vacua.

  In the other sub-phase, the mass $\abs{\hat{\sigma}_1-2\hat{\sigma}_2}$ of the chiral multiplet~$X_2$ becomes small, so that a better approximation is to only integrate out $X_0$, $X_1$, $X_3$ and get an effective twisted superpotential for a vector multiplet scalar constrained to have $\hat{\sigma}_1-2\hat{\sigma}_2=0$.
  The asymptotics~\eqref{t2tominfty} remain correct for $\hat{\sigma}_1$, $\hat{\sigma}_2$, $(-k\hat{\sigma}_1+\hat{\sigma}_2)$.
  Up to unimportant constant shifts, the non-trivial D-term equation is the following relation between two-component vectors:
  \begin{equation}
    \!\begin{pmatrix}1\\-2\end{pmatrix} \abs{X_2}^2 \!=\! \begin{pmatrix}\zeta^{\text{ren}}_1\\\zeta^{\text{ren}}_2\end{pmatrix}+\biggl(\!\!\begin{pmatrix}1\\0\end{pmatrix}+\begin{pmatrix}-k\\1\end{pmatrix}+\begin{pmatrix}0\\1\end{pmatrix}\!\!\biggr)\frac{2\zeta^{\text{ren}}_1+\zeta^{\text{ren}}_2}{2k-4}
    =
    -\frac{2\zeta^{\text{ren}}_1+(k-1)\zeta^{\text{ren}}_2}{2k-4} \!\begin{pmatrix} 1 \\ -2 \end{pmatrix} \!.
  \end{equation}
  There are solutions, hence mixed branch vacua, when $2\zeta^{\text{ren}}_1+(k-1)\zeta^{\text{ren}}_2\ll 0$.
  The number of solutions is $2k-4$ because that is the number of possible overall phases for~$\hat{\sigma}$, just like in the sub-phase we already analyzed.
  Upon crossing the wall, the $2k-4$ vacua remain well-separated in the direction transverse to $\hat{\sigma}_1-2\hat{\sigma}_2\simeq 0$,
  hence each vacuum gets deformed continuously to a vacuum on the other side.
  There is no phase transition: the isolated massive vacua simply correspond to different combinations (Coulomb versus mixed) of the UV fields.

\item \textbf{Second divisor blown up} ($A=\{2\}$): $\zeta^{\text{ren}}_1+k\zeta^{\text{ren}}_2\gg 0$ and $\zeta^{\text{ren}}_1\ll 0$.
  In this phase we rewrite~\eqref{C2Zn2critical} as
  \begin{equation}
    \begin{aligned}
      i\hat{\sigma}_1 (i\hat{\sigma}_1-2i\hat{\sigma}_2) & = e^{-t^{\text{ren}}_1} (-k i\hat{\sigma}_1+i\hat{\sigma}_2)^k , \\
      (i\hat{\sigma}_1-2i\hat{\sigma}_2)^{2k-1} & = e^{t^{\text{ren}}_1+kt^{\text{ren}}_2} i\hat{\sigma}_1 (i\hat{\sigma}_2)^k .
    \end{aligned}
  \end{equation}
  with $\abs{\hat{\sigma}_1}$, $\abs{\hat{\sigma}_2}$, $\abs{\hat{\sigma}_1-2\hat{\sigma}_2}$, $\abs{-k\hat{\sigma}_1+\hat{\sigma}_2}$ all big.
  Since $\zeta^{\text{ren}}_1\ll 0$, the first equation requires $\abs{-k\hat{\sigma}_1+\hat{\sigma}_2}$ to be parametrically smaller than $\abs{\hat{\sigma}_1}+\abs{\hat{\sigma}_2}$.
  Plugging $\hat{\sigma}_2\simeq k\hat{\sigma}_1$ in the second equation gives
  \begin{equation}\label{asymptsigma1phase2}
    (i\hat{\sigma}_1)^{k-2} \simeq e^{t^{\text{ren}}_1+kt^{\text{ren}}_2} k^k (1-2k)^{1-2k} ,
  \end{equation}
  which has $k-2$ solutions.  From the first equation we then deduce
  \begin{equation}
    (-k i\hat{\sigma}_1+i\hat{\sigma}_2) \simeq e^{(t^{\text{ren}}_1+2t^{\text{ren}}_2)/(k-2)} k^{2/(k-2)} (1-2k)^{-3/(k-2)}
  \end{equation}
  which is large or not depending on the sign of $\zeta^{\text{ren}}_1+2\zeta^{\text{ren}}_2$.
  The line $\zeta^{\text{ren}}_1+2\zeta^{\text{ren}}_2=0$ splits the phase into two parts.
  In the sub-phase $\zeta^{\text{ren}}_1+2\zeta^{\text{ren}}_2\gg 0$ we have $k-2$ Coulomb branch vacua, and in the other sub-phase, none.

  The next step is to look for mixed branch vacua for which the mass $\abs{-k\hat{\sigma}_1+\hat{\sigma}_2}$ of~$X_1$ is small.
  The asymptotics~\eqref{asymptsigma1phase2} are unchanged and the D-term equation reads essentially
  \begin{equation}
    \begin{pmatrix}-k\\1\end{pmatrix}\abs{X_1}^2
    = \begin{pmatrix}\zeta^{\text{ren}}_1\\\zeta^{\text{ren}}_2\end{pmatrix}
    + \biggl(\!\begin{pmatrix}1\\0\end{pmatrix}+\begin{pmatrix}1\\-2\end{pmatrix}+\begin{pmatrix}0\\1\end{pmatrix}\!\biggr)\frac{\zeta^{\text{ren}}_1+k\zeta^{\text{ren}}_2}{k-2}
    = -\frac{\zeta^{\text{ren}}_1+2\zeta^{\text{ren}}_2}{k-2} \begin{pmatrix}-k\\1\end{pmatrix} .
  \end{equation}
  There are solutions, hence mixed branch vacua, when $\zeta^{\text{ren}}_1+2\zeta^{\text{ren}}_2\ll 0$.
  Each of the $k-2$ Coulomb branch vacua of one subphase gets deformed continuously to a mixed branch vacuum in the other subphase.

\item \textbf{Both divisors blown up} ($A=\{1,2\}$): $\zeta^{\text{ren}}_1\gg 0$, $\zeta^{\text{ren}}_2\gg 0$.
  The $t^{\text{ren}}_2\to+\infty$ asymptotics of~\eqref{C2Zn2criticalsols} are different for the $v_+$ and the $v_-$ solutions.  For~$v_+$,
  \begin{equation}
    \hat{\sigma}_2 \sim \hat{\sigma}_1 \sim (\hat{\sigma}_1-2\hat{\sigma}_2) \sim e^{(t^{\text{ren}}_1+kt^{\text{ren}}_2)/(k-2)} , \qquad (-k\hat{\sigma}_1+\hat{\sigma}_2) \sim e^{(t^{\text{ren}}_1+2t^{\text{ren}}_2)/(k-2)}
  \end{equation}
  and for~$v_-$,
  \begin{equation}
    \hat{\sigma}_1 \sim (\hat{\sigma}_1-2\hat{\sigma}_2) \sim (-k\hat{\sigma}_1+\hat{\sigma}_2) \sim e^{t^{\text{ren}}_1/(k-2)} , \qquad
    \hat{\sigma}_2 \sim e^{t^{\text{ren}}_1/(k-2)} e^{-t^{\text{ren}}_2} .
  \end{equation}
  All of these $n-3=2k-4$ solutions are genuine Coulomb branch vacua.
\end{itemize}

\paragraph{General rank.}

In general Hirzebruch-Jung models it is difficult to determine Coulomb branch and mixed branch vacua explicitly, but we can count them.
The idea is to turn on generic twisted masses so that the Higgs branch reduces to isolated massive vacua too.
Since one can smoothly vary between any two phases, by turning on a theta angle to avoid singularities in FI-theta parameter space, the number of vacua in all phases must be the same.
In particular, in the orbifold phase there are $n$ Higgs branch vacua due to twisted sectors, and no Coulomb/mixed branch vacua.

The question then boils down to finding the number of vacua on the Higgs branch when generic twisted masses are turned on.
We reiterate that the Higgs branch only changes at classical phase boundaries, shifted according to the discussion below~\eqref{tmusol}.

The effect of twisted masses~$m_i$ is to change the mass equation from $(Q^i\cdot\sigma)X_i=0$ to $(Q^i\cdot\sigma+m_i)X_i=0$ for all~$i$.
For generic~$m_i$, at most $r$ of the masses $Q^i\cdot\sigma+m_i$ can vanish at the same time, so at most $r$ of the $X_i$ may be non-zero.
In other words, at least two of the $X_i$~must vanish.
The Higgs branch thus reduces to the intersections $E_i\cap E_j=\{X_i=X_j=0\}$.
At each such non-empty intersection there are $d_{ij}\geq 1$ vacua, due to the twisted sectors for the $\ZZ_{d_{ij}}$~orbifold group.
Altogether there are
\begin{equation}\label{countCoulombmixedvacua}
  \begin{aligned}
    d_{0\alpha_1} + d_{\alpha_1\alpha_2} + \dots + d_{\alpha_{\ell-1}\alpha_\ell} + d_{\alpha_\ell (r+1)}
    & \quad \text{Higgs branch vacua, hence} \\
    n - \bigl(d_{0\alpha_1} + d_{\alpha_1\alpha_2} + \dots + d_{\alpha_{\ell-1}\alpha_\ell} + d_{\alpha_\ell (r+1)}\bigr)
    & \quad \text{Coulomb/mixed vacua,}
  \end{aligned}
\end{equation}
where we denoted $A=\{\alpha_1,\dots,\alpha_\ell\}$ the set of blown up exceptional divisors.
When we turn off twisted masses the isolated Coulomb/mixed branch vacua are unaffected while Higgs branch vacua spread onto the whole Higgs branch.

In \autoref{ssec:wallcrossingHiggsing} we explain how the number of Coulomb/mixed branch vacua jumps when crossing a wall,
by restricting to a local model of the wall with gauge group $U(1)\times\Gamma$ for $\Gamma$ a discrete group.
In Hirzebruch-Jung models, crossing a wall means blowing up an exceptional divisor~$E_j$.
The local model~\eqref{localdivisorchargesU1Zm} for this transition has gauge group $U(1)\times\ZZ_m$ with $m=\gcd(d_{ij},d_{jk})$ where $E_i$ and $E_k$ are the two divisor intersecting~$E_j$.
Then the number of Coulomb branch vacua should increase by $m$ times\footnote{The gauge group is abelian so $\ZZ_m$ acts trivially on the non-zero vevs of~$\sigma$ in Coulomb branch vacua.  Then $\ZZ_m$ converts each vacuum of the $U(1)$ theory to $m$ due to twisted sectors.} the sum of $U(1)$ charges, so
\begin{equation}
  m\biggl(\frac{d_{ik}}{m} - \frac{d_{ij}}{m} - \frac{d_{jk}}{m}\biggr) \geq 0 .
\end{equation}
This is consistent with~\eqref{countCoulombmixedvacua}.
The jump $d_{ik}-d_{ij}-d_{jk}$ vanishes if and only if the local model is Calabi-Yau, namely $a_{i+1}=\dots=a_{k-1}=2$.

\section{\label{sec:hpf}Hemisphere partition function and geometry of B-branes}

B-branes are a special class of boundary states in $\mathcal{N}=(2,2)$ 2d SCFTs that preserve a particular subalgebra of the superconformal algebra.
In this work we mainly deal with theories that do not flow to an SCFT in the IR\@.
More precisely we work with $\mathcal{N}=(2,2)$ supersymmetric UV descriptions with an anomalous axial R-symmetry.
B-branes can still be defined in these theories as boundary states that preserve a subalgebra $\mathbf{2}_{B}\subset (2,2)$ \cite{Hori:2000ck,Aspinwall:2009isa,Witten:1992fb}.

In this section we focus on the Higgs branch.
Deep in a phase different branches of the theory are well-separated,
hence a B-brane decomposes into one part for each branch
in a way that we explore in \autoref{sec:Band restriction rule}.
For some class of B-branes, called band-restricted (or grade-restricted), the Higgs branch part has a simple geometric construction.
\textbf{We ignore for now this subtle issue of band restriction, also explained in \autoref{sec:Band restriction rule}.}

We review how to compute the central charge of a B-brane, defined directly in the UV via supersymmetric localization \cite{Hori:2013ika,Honda:2013uca,Sugishita:2013jca}.
This provides a powerful tool to analyze the behaviour of B-branes along the RG flow or in the gauge-decoupling limit.
The localization formula is singular in our non-compact setting and we determine how to regularize it using R-charges.
Among several classes of branes in abelian GLSM we find that compactly-supported branes have finite central charge.
We also specialize to B-branes on Higgs branches of the Hirzebruch-Jung models and compute their (zero-instanton) central charges,
which we compare in \autoref{sec:Ktheoreticaspects} with a geometric definition in terms of (the untwisted sector of) K-theory.\footnote{The low-lying instanton corrections should capture contributions from twisted sectors.}

\subsection{\label{ssec:BbraneGLSM}Field theory description of B-branes in GLSMs [review]}

Consider a GLSM on a half space $\mathcal{H}=\mathbb{R}\times \mathbb{R}_{\leq 0}$.
We denote left and right supercharges by $(\mathbf{Q}_{\pm},\overline{\mathbf{Q}}_{\pm})$.
Then the $\mathbf{2}_{B}\subset (2,2)$ supersymmetry we want to preserve is the subalgebra generated by $\mathbf{Q}_{+}+\mathbf{Q}_{-}$ and its conjugate $\overline{\mathbf{Q}}_{+}+\overline{\mathbf{Q}}_{-}$.

\paragraph{Algebraic data.}

The boundary term
\begin{equation}
  \Tr_\rho\biggl(\Pexp \int_{\partial\mathcal{H}} A\biggr) , \qquad
  A = (\Re(\sigma)+iv_{\tau}) \mathrm{d}\tau
\end{equation}
is invariant under $\mathbf{2}_{B}$ supersymmetry and gauge-invariant for any representation $\rho$ of~$G$.
When the superpotential~$W$ is nonzero we also need to add an extra term to the action in order to preserve $\mathbf{2}_{B}$.
This term is also a holonomy,
\begin{equation}
  \Tr_{M}\biggl(\Pexp \int_{\partial\mathcal{H}} \mathcal{A}(\mathbf{T})\biggr), \qquad
  \mathcal{A}(\mathbf{T})\coloneqq \frac{1}{2}\psi^{i}\partial_{i}\mathbf{T}-\frac{1}{2}\bar{\psi}^{\bar{i}}\overline{\partial}_{\bar{i}}\mathbf{T}^{\dag}-\frac{1}{2}\{\mathbf{T},\mathbf{T}^{\dag}\} ,
\end{equation}
where $M$ is introduced below and $\mathbf{T}$ is a matrix factorization of $W$ (the tachyon profile) \cite{Herbst:2008jq,Lazaroiu:2003zi}.
The most general boundary action preserving $\mathbf{2}_{B}$~supersymmetry that we will consider is given by
\begin{equation}
  \Tr_{M}\biggl(\Pexp \int_{\partial\mathcal{H}}\Bigl(A+ \mathcal{A}(\mathbf{T})\Bigr)\biggr).
\end{equation}
Altogether, specifying the B-brane requires the following \textbf{algebraic data} $\mathcal{B}=(M,\rho,\mathbf{r}_{*},\mathbf{T})$.
\begin{itemize}
\item A $\mathbb{Z}_{2}$-graded, finite dimensional free $\Sym(V^*)$ module (\textbf{Chan-Paton vector space}) denoted by $M=M_{0}\oplus M_{1}$. For the cases when $W\neq 0$, we will need $\rank(M_{0})=\rank(M_{1})$.
\item Two representations, $\rho\colon G\rightarrow GL(M)$ and $\mathbf{r}_{*}\colon\lie{u}(1)_{V}\rightarrow \lie{gl}(M)$.
\item A \textbf{matrix factorization} $\mathbf{T} \in \End_{\Sym(V^{*})}(M)$ of the superpotential $W\in\Sym(V^{*})$, i.e., a $\ZZ_2$~odd endomorphism such that $\mathbf{T}^{2}=iW\cdot\id_{M}$
  and such that the group actions $\rho$ and $\mathbf{r}_*$ are compatible with the action of $G$ and $U(1)_{V}$ on the chiral matter $X\in V$:
  for all $\lambda\in U(1)_{V}$ and $g\in G$,
  \begin{equation}\label{rhodef}
    \begin{aligned}
      \lambda^{\mathbf{r}_{*}}\mathbf{T}(\lambda^{R}X)\lambda^{-\mathbf{r}_{*}} & = \lambda \mathbf{T}(X) , \\
      \rho(g)^{-1}\mathbf{T}(\rho_{\text{m}}(g)\cdot X)\rho(g) & = \mathbf{T}(X) .
    \end{aligned}
  \end{equation}
\end{itemize}

\paragraph{Choice of contour.}

However, there is still a piece of data that we need to fix to fully define a B-brane on a GLSM\@: a profile for the vector multiplet scalar~$\sigma$.
This data consists of a gauge-invariant middle-dimensional subvariety of~$\lie{g}_{\CC}$, or equivalently its intersection $L\subset\lie{t}_{\CC}$ with the Cartan algebra, which we refer to as the contour.
An \textbf{admissible contour} is a gauge invariant, middle-dimensional~$L$ that is a continuous deformation of the real contour $L_{\RR}\coloneqq\{\Im\sigma=0\}$
such that the boundary effective twisted superpotential
\begin{equation}
  \tW_{\text{eff},\rho}(\sigma)\coloneqq \Biggl(\sum_{\alpha>0}\pm i\pi\,\alpha\cdot\sigma\Biggr)-\Biggl(\sum_j (Q^j\cdot\sigma)\biggl(\log\biggl(\frac{iQ^j\cdot\sigma}{\Lambda}\biggr)-1\biggr)\Biggr)-t_{\text{bare}}\cdot\sigma+2\pi i\,\rho\cdot\sigma
\end{equation}
approaches $+\infty$ in all asymptotic directions of~$L$.
Signs in the sum over positive roots~$\alpha$ of~$G$ depend on the Weyl chamber in which $\Re\sigma$ lies; this sum is absent in abelian GLSMs.
The full B-brane is then given by $(\mathcal{B},L)$.

\paragraph{Hemisphere partition function.}

The brane's central charge is given by~\cite{Hori:2013ika}\footnote{We give the result for a non-abelian gauge group but only use it in abelian cases.}
\begin{equation}
  \label{ZD2loc}
  Z_{D^2}(\cB) = C(\mathfrak{r}\Lambda)^{\hat{c}/2} \int_{L} \mathrm{d}^{l_G}\hat{\sigma}
  \biggl( \prod_{\alpha > 0} (\alpha\cdot\hat{\sigma}) \sinh\bigl(\pi\alpha\cdot\hat{\sigma}\bigr) \biggr)
  \prod_j \Gamma \biggl( iQ^j\cdot\hat{\sigma} + \frac{R_j}{2} \biggr) e^{it_{\text{ren}}\cdot\hat{\sigma}} f_{\cB}(\hat{\sigma})
\end{equation}
where $R_j$ is the R-charge of the $j$-th chiral multiplet.
Here $\mathfrak{r}$ is the radius of the disk $D^{2}$ and $\Lambda$ the UV energy scale. So we identify $\mu=\mathfrak{r}^{-1}$ and then the renormalized FI-theta parameter is
\begin{equation}
t_{\text{ren},\alpha}(\mu)=\zeta_{\text{bare},\alpha}-i\theta_{\text{bare},\alpha}-\biggl(\sum_{j}Q^{j}_{\alpha}\biggr)\log \mathfrak{r}\Lambda
\end{equation}

 Note that the only dependence of the partition function on the choice of brane is through the \textbf{brane factor}
\begin{equation}
  f_{\cB}(\hat{\sigma}) \coloneqq \Tr_{M}\Bigl(e^{i\pi\mathbf{r}_{\ast}}e^{2\pi\rho(\hat{\sigma})}\Bigr) ,
\end{equation}
which itself does not depend on the matrix factorization~$\mathbf{T}$. However, when coupling the GLSM to the curvature of $D^{2}$, one of the couplings enters $\mathcal{A}(\mathbf{T})$ as $\mathbf{r}_{\ast}/\mathfrak{r}$, explaining the factor $e^{i\pi\mathbf{r}_{\ast}}$ in $f_{\cB}(\hat{\sigma})$.

In the following we focus on abelian GLSMs with zero superpotentials.\footnote{Then $\mathbf{T}$ squares to zero, namely $M_0\xrightarrow{\mathbf{T}}M_1\xrightarrow{\mathbf{T}}M_0$ is a $\ZZ_2$-graded complex.}
The partition function is then
\begin{equation}
  \label{ZD2locab}
  Z_{D^2}(\cB) = C(\mathfrak{r}\Lambda)^{\hat{c}/2} \int_{L} \mathrm{d}^{r}\hat{\sigma}  \prod_j \Gamma\biggl( iQ^j\cdot\hat{\sigma} + \frac{R_j}{2} \biggr) e^{it_{\text{ren}}\cdot\hat{\sigma}} f_{\cB}(\hat{\sigma}).
\end{equation}
To be precise, the localization formula was only derived when R-charges of all chiral multiplets obey $0<R_j<2$.
This ensures that none of the poles of one-loop determinants $\Gamma(iQ^j\cdot\hat{\sigma}+R_j/2)$ lie on the contour~$L_{\RR}$ that $L$~is a deformation of.  The contour $L_{\RR}$ must then be deformed, without crossing any poles, into a contour~$L$ that ensures convergence at large~$\abs{\hat{\sigma}}$.

\paragraph{Calabi-Yau abelian GLSM.}

Consider first the Calabi-Yau case.
For each phase, the contour integral can be closed and expressed as a sum of residues.
Each Gamma function has poles along an infinite family of parallel hyperplanes $Q^j\cdot\hat{\sigma} = i (R_j/2 + k)$ for $k\geq 0$.
For generic~$R_j$, hyperplanes have at most $r$-fold intersections, which are solutions $\hat{\sigma}_{J,k}$ of $Q^j\cdot\hat{\sigma} = i (R_j/2 + k_j)$ for $j\in J$, where $J$ is a set of $r$~flavours and $k_j\geq 0$ are integers that physically count world-sheet instantons (vortices).

When closing the contours one picks up a residue from some of these families of points.
The choice of what~$J$ to pick up depends on the direction in which contours are closed, which depends on the asymptotics of the integrand and in particular on the phase in which the FI parameter $\zeta=\Re t$ lies.\footnote{Since $t_{\text{ren}}=t_{\text{bare}}$ we omit these subscripts.}
A convenient shortcut is to use that the sum of residues should converge: the factor $e^{it\cdot\hat{\sigma}}$ must go to zero as any of the $k_j\to\infty$.
This occurs precisely when $\zeta$ is a positive linear combination of the~$Q^j$, namely when $\zeta\in\Cone_J$:
\begin{equation}\label{ZD2locabHiggs}
  Z_{D^2,\text{residue}}(\cB) = \frac{(\mathfrak{r}\Lambda)^{\hat{c}/2}}{(-2\pi i)^{\dim V-r}} \!\! \sum_{J\mid\zeta\in\Cone_J} \sum_{k:J\to\ZZ_{\geq 0}} \!\!
  \pm \res_{i\hat{\sigma}=i\hat{\sigma}_{J,k}} \Biggl( \prod_j \Gamma\biggl( iQ^j\cdot\hat{\sigma} + \frac{R_j}{2} \biggr) e^{it\cdot\hat{\sigma}} f_{\cB}(\hat{\sigma}) \Biggr) ,
\end{equation}
where we fixed the normalization constant~$C$ for later convenience.
The shortcut does not fix the sign with which residues should be summed.\footnote{We take the residue of the integrand as a function of $i\hat{\sigma}$ rather than $\hat{\sigma}$ to cancel some factors of~$i$.  For instance $\res_{\hat{\sigma}=0}\Gamma(i\hat{\sigma})=-i$ while $\res_{i\hat{\sigma}=0}\Gamma(i\hat{\sigma})=1$.}
A more precise analysis gives that the sign $\pm$ in this formula is $\sign\bigl(\det(Q^j)_{j\in J}\bigr)$.
Incidentally, the sign and the poles that contribute coincide with those selected by the Jeffrey-Kirwan (JK) prescription with JK parameter~$\zeta$.

While in the Calabi-Yau case $Z_{D^2}(\cB)=Z_{D^2,\text{residue}}(\cB)$, in general this sum of residues is only part of the complete hemisphere partition function.

\paragraph{Non-Calabi-Yau abelian GLSMs.}

We turn to theories with anomalous $U(1)_A$.
In \autoref{sec:Band restriction rule} we discuss the shape of the contour~$L$ and the behaviour of the integrand at large~$\abs{\hat{\sigma}}$, which depends on the brane factor.
In phases with only a Higgs branch, namely such that $Q^{\text{tot}}=\sum_j Q^j$ belongs to the closure of the phase (see \autopageref{para:pureHiggs}), the contour can be closed and $Z_{D^2}(\cB)$ gives the sum of residues~\eqref{ZD2locabHiggs}.

In other phases, the series over~$k$ that appear in~\eqref{ZD2locabHiggs} are asymptotic series, as can be seen by working out the large~$k$ behaviour of the Gamma functions that multiply~$e^{it_{\text{ren}}\cdot\hat{\sigma}}$.
Finitely deep in the phase, the contour integral is evaluated by deforming~$L$ to pass finitely many poles.
This expresses $Z_{D^2}(\cB)$ as a sum of finitely many residues in the asymptotic series, plus a remaining contour integral.
Provided the brane is ``band-restricted'', namely obeys a bound~\eqref{bandsanomalous} on the charge vectors of the representation~$\rho$ hence on the degrees of~$f_{\cB}$ as a multivariate polynomial in the $\exp 2\pi\hat{\sigma}^\alpha$,
the remaining contour integral is computed using a saddle-point approximation that identifies it with a contribution due to mixed or Coulomb branches.
For such band-restricted branes, the Higgs branch contribution is exclusively due to the asymptotic series of residues.

For now, we focus on resolving the singularity at $\hat{\sigma}=0$ that occurs in our models because of vanishing R-charges, namely because of a non-compact target space.
We denote by $Z_{D^2,\text{residue}}^{\text{0-instanton}}(\cB)$ this zero-instanton term, which is the term $k=0$ in~\eqref{ZD2locabHiggs}.

\subsection{\label{ssec:Bbraneabelian}B-brane category and geometric projection to the Higgs branch}

Branes preserving B-type supersymmetry naturally form a category whose morphisms are given by quantizing string states between them.
For a sigma model to a target $X$ this category is equivalent \cite{Aspinwall:2001pu} to the derived category $D^b(X)$ whose objects are given by bounded chain complexes of vector bundles\footnote{A useful fact is that it is equivalent to work with complexes of vector bundles when coherent sheaves admit locally free resolutions, which is the case in all the cases considered in this work.} and whose morphisms are given by chain complexes modulo chain homotopy with quasi-isomorphisms formally inverted.

\paragraph{RG flow, gauge-decoupling and geometric functor.}

For targets~$X$ that can be realized as the Higgs branch in some phase of a GLSM with $W=0$, it is useful to start with the brane category of the (UV) GLSM\@, denoted by $D^b(V,G)$.
This is a $G$-equivariant\footnote{Our notation leaves implicit the action of $G$ on the vector space $V$ in which chiral multiplets take values.} version of $D^b(V)$;
for instance for $G=U(1)^r$, objects in this category are equivalent to complexes of Wilson lines (equivariant line bundles) $\cW(q)$ for $q \in \Hom(G,U(1))$.
Any vector bundle on $V$ is of course trivial, but $G$ may act nontrivially.
Then one can flow from the UV GLSM down to the IR\@, or as discussed below~\eqref{FIrenorm} to some intermediate energy scale well-described by the phase of interest.
Each brane in $D^b(V,G)$ reduces to a brane in this phase, which typically has contributions from all branches of the theory.

To conveniently reach an arbitrary phase at some intermediate energy scale, we fix the energy scale~$\mu$ and FI parameter at that scale, and consider the gauge-decoupling limit $g\to+\infty$.  This amounts to an RG flow while simulteneously varying the bare FI parameter to keep the renormalized coupling $t_{\text{ren}}(\mu)$ fixed.
The image of a GLSM B-brane~$\cB\in D^b(V,G)$ under this limit is denoted by $F_{g\to\infty}(\cB)$.
In particular its Higgs branch part lies in $D^b(X)$, giving a functor
\begin{equation}
F_{g\to\infty,\text{Higgs}}\colon D^b(V,G)\to D^b(X) .
\end{equation}
On the other hand, the restriction and projection from $V$ to $X=(V\setminus\Delta)/G$ provides another functor $F_{\text{geom}}\colon D^b(V,G)\to D^b(X)$.

In \autoref{sec:Band restriction rule} we learn that these two functors are equal in phases with only a Higgs branch (e.g.\@, all phases in Calabi-Yau models)
\begin{equation}\label{gdHiggseqgeom}
  F_{g\to\infty,\text{Higgs}} = F_{\text{geom}} \quad \text{in pure-Higgs phases,}
\end{equation}
and that they otherwise agree (only) for B-branes that are band-restricted, namely that are built from Wilson lines with charges in some range.
In other words, for these GLSM branes, naive geometric considerations give the correct Higgs branch image.
We also explain in \autoref{sec:Band restriction rule} how every GLSM brane is equivalent to one that is suitably band-restricted.
Together this allows to determine the Higgs branch image of every GLSM brane.

\paragraph{Focusing on $F_{\text{geom}}$ throughout \autoref{sec:hpf}.}
In this section we work with the simpler~$F_{\text{geom}}$.
This meshes well with the fact that we work with $Z_{D^2,\text{residue}}$ rather than the full~$Z_{D^2}$.
Indeed, branes with the same image under~$F_{\text{geom}}$ have the same value of $Z_{D^2,\text{residue}}$, as we will explain near~\eqref{polezerocancellation}, just as branes with the same image under the gauge-decoupling functor~$F_{g\to\infty}$ (including all branches) have the same~$Z_{D^2}$.

This has an interesting consequence if one wants to compute the hemisphere partition function of a Higgs branch brane $\cB_{\text{Higgs}}\in D^b(X)$, where $X$ is the Higgs branch in any phase.
Consider a band-restricted brane $\cB\in D^b(V,G)$ that reduces to it (plus mixed or Coulomb parts), namely with $F_{g\to\infty,\text{Higgs}}(\cB)=\cB_{\text{Higgs}}$.
We want to compute the Higgs part of $Z_{D^2}(\cB)$, which as discussed at the end of \autoref{ssec:BbraneGLSM} is simply $Z_{D^2,\text{residue}}(\cB)$.
On the other hand, by the discussion near~\eqref{gdHiggseqgeom} based on \autoref{sec:Band restriction rule}, $F_{\text{geom}}(\cB)=F_{g\to\infty,\text{Higgs}}(\cB)$.
Now, since $Z_{D^2,\text{residue}}$ only depends on the geometric image of a brane, we conclude that
\begin{equation}\label{computingZofHiggsbranch}
  Z_{D^2}(\cB_{\text{Higgs}}) = Z_{D^2,\text{residue}}(\cB')
\end{equation}
for any brane $\cB'$ with $F_{\text{geom}}(\cB')=\cB_{\text{Higgs}}$.  This condition is much easier to check than finding a brane whose Higgs branch image is $\cB_{\text{Higgs}}$.  Of course, since the brane $\cB'$ need not be grade-restricted, its image under the gauge-decoupling limit may have nothing to do with $\cB_{\text{Higgs}}$, just as $Z_{D^2,\text{residue}}(\cB')$ may have nothing to do with the actual Higgs branch part of $Z_{D^2}(\cB')$, because the remaining contour integral discussed at the end of \autoref{ssec:BbraneGLSM} contributes to the Higgs branch part.

\paragraph{Derived category of~$X$.}

We return to describing the derived category of our toric variety or orbifold~$X$.
It is generated by the line bundles~\cite{Aspinwall:2008jk,Herbst:2008jq}
\begin{equation}
  \cO(q)\coloneqq F_{\text{geom}}\bigl(\cW(q)\bigr) .
\end{equation}
We also introduce the notation $\cF(q)\coloneqq \cF\otimes\cO(q)$ for any sheaf~$\cF$.
From the GIT viewpoint, the deleted set in a given phase is $\Delta=\bigcup_J\Delta_J$ for a collection of linear subspaces $\Delta_J=\{X_i=0\;\allowbreak\forall i\in J\}$.
Each structure sheaf $\cO_{\Delta_J}\in D^b(V,G)$ is mapped by $F_{\text{geom}}$ to a trivial brane in $D^b(X)$.
This sheaf admits a Koszul resolution as a complex of equivariant line bundles, mapped by $F_{\text{geom}}$ to a complex of line bundles~$\cO(q)$.
The resulting complex is trivial.
This explains why the derived category $D^b(X)$ is generated by the $\cO(q)$ subject to one relation for each deleted set~$\Delta_J$.
One interpretation is that we start in the UV with a ``free'' category consisting of all possible bound states of Wilson lines and then in different phases we impose different sets of relations to construct the B-brane categories $D^b(X)$; these relations in turn encode the geometry of~$X$.

\paragraph{Koszul resolutions.}

Let us describe the Koszul resolution $\cK_{\{j\}}\in D^b(V,G)$ of the structure sheaf $\cO_{\{X_j=0\}}\in D^b(V,G)$ of a hyperplane.
For some explicit calculations it is helpful to treat coherent sheaves as modules.
Then we wish to resolve the $R=\mathbb{C}[X_1,\dots,X_N]$-module $R/(X_j)$ through the exact sequence
\begin{equation}
  0 \to R \xrightarrow{X_j} R \twoheadrightarrow R/(X_j) \to 0 , \quad \text{that is,} \quad
  0 \to \cW(-Q^j) \xrightarrow{X_j} \cW(0) \twoheadrightarrow \cO_{\{X_j=0\}} \to 0
\end{equation}
involving multiplication by~$X_j$.
The Koszul resolution is then a two-term complex of line bundles
\begin{equation}\label{hyperplaneKoszul}
  \cK_{\{j\}} \coloneqq \bigl(\cW(-Q^j) \xrightarrow{X_j} \cW(0)\bigr) .
\end{equation}
It is important to keep up with the gauge charges at each stage of the resolution: different gauge charges lead to different bundles on the quotient~$X$.  To resolve the module $R/(X_j)$ the two~$R$ above should be taken to have gauge charges $-Q^j$ and~$0$, respectively, where $Q^j$ is the charge vector of~$X_j$.

The Koszul resolution~$\cK_J$ of $\cO_{\Delta_J}$ is then obtained by using that $\cO_{\Delta_J} = \bigotimes_{j\in J}\cO_{\{X_j=0\}}$.
Denoting by $m$ the number of elements of~$J$ we get the resolution
\begin{equation}\label{KoszulResolutionOfSheaf}
  \cK_J = \bigotimes_{j\in J} \cK_{\{j\}} = \bigl( \cW \to \cW^{\oplus m} \to \dots \to
  \cW^{\oplus\binom{m}{k}} \to \dots \to \cW^{\oplus m} \to \cW \bigr)
\end{equation}
where the $2^m=\sum_{k=0}^m\binom{m}{k}$ Wilson lines in the resolution are labeled by subsets $I\subset J$, and the non-zero maps are as follows.
From the Wilson line labelled $I\sqcup\{j\}$ to that labelled~$I$ the map is multiplication by $\pm X_j$.  Signs are chosen to make~\eqref{KoszulResolutionOfSheaf} exact except at the right-most point.
More generally, the resolution $\cK_J(q)$ of $\cO_{\Delta_J}(q)$ is this complex in which
the Wilson line labelled~$I$ has gauge charge $q-\sum_{j\in I}Q^j$ and R-charge $-\sum_{j\in I} R_j/2$.
We deduce that $\cO_{\Delta_J}(q)$ has brane factor
\begin{equation}\label{Koszulbranefactor}
  f_J(\hat{\sigma}) = e^{2\pi q\cdot\hat{\sigma}} \prod_{j\in J} (1 - e^{-2\pi Q^j\cdot\hat{\sigma}+i\pi R_j}) .
\end{equation}

\paragraph{Geometrically empty branes.}

Consider $\Delta_J$~lying in the deleted set~\eqref{DeletedSetAbelianGLSM}, namely $\zeta\not\in\Cone_{^{\complement}J}$, which means that $\zeta$ cannot be written as a positive linear combination of charge vectors without using at least one of the $Q^j$, $j\in J$.
Then by construction of the deleted set $\cO_{\Delta_J}\in D^b(V,G)$ is mapped by $F_{\text{geom}}$ to an empty brane in~$D^b(X)$, and likewise for the Koszul resolution, $F_{\text{geom}}(\cK_J)=F_{\text{geom}}(\cO_{\Delta_J})=0\in D^b(X)$.

Interestingly, the brane factor~\eqref{Koszulbranefactor} has zeros at every pole of one-loop determinants of~$X_j$ for $j\in J$.
This is the key observation to show near~\eqref{polezerocancellation} that $Z_{D^2,\text{residue}}=0$, consistently with the fact that $F_{\text{geom}}(\cO_{\Delta_J})=0$.

In a \emph{pure-Higgs phase}, as discussed in~\eqref{gdHiggseqgeom}, the Higgs branch image of a complex of Wilson lines is correctly captured by the functor~$F_{\text{geom}}$, and the same holds in other phases for \emph{band-restricted branes}, only.
In such cases, the complex of Wilson lines~\eqref{KoszulResolutionOfSheaf} reduces to some brane whose Higgs branch part is empty but whose mixed and Coulomb parts are typically non-trivial,
\begin{equation}\label{geom-empty-claim}
  F_{g\to\infty,\text{Higgs}}(\cK_J) = 0 \qquad \text{for pure-Higgs phase or band-restricted brane.}
\end{equation}

\subsection{\label{ssec:regZU1}Examples of partition function regularization in a \texorpdfstring{$U(1)$}{U(1)} GLSM}

Before coming back to general considerations for multiparameter models in the next subsection,
let us do some calculations in one-parameter ($r=1$) Hirzebruch-Jung models.
These have gauge group $U(1)$ and three chiral multiplets $X_0$, $X_1$, $X_2$ of charges $1$, $-n$, $1$.
The orbifold phase $\zeta\ll 0$ has Higgs branch $\CC^2/\ZZ_{n(1)}$.
The resolved phase $\zeta\gg 0$ has Higgs branch the total space of $\cO(-n)\to\CP^1$, and has $n-2$ massive vacua.
As explained in \autoref{sec:Band restriction rule}, there is a band restriction rule in the resolved phase:
only branes built from Wilson lines $\cW(q)$ within the window $\abs{\theta/(2\pi)+q}<n/2$ are such that
the sum of residues gives the correct Higgs branch contribution to the hemisphere partition function.
Nevertheless, we work out in the next subsections that the quantity relevant to the Higgs branch geometry is $Z_{D^2,\text{residue}}(\cB)$, whose 0-instanton part we focus on now.

\paragraph{Preliminaries on R-charges and contour pinching.}

The localization formula for the hemisphere partition function~\eqref{ZD2locab} is only valid when R-charges are all in the interval $(0,2)$.
We thus turn on positive R-charges $R_0$, $R_1$, $R_2$ for $X_0$, $X_1$, $X_2$.
The localization result
\begin{equation}\label{welldefinedrank1ZD2B}
  Z_{D^2}(\cB) = \frac{(\mathfrak{r}\Lambda)^{\hat{c}/2}}{(-2\pi i)^2} \int_L \frac{\mathrm{d}\hat{\sigma}}{2\pi} \, \Gamma( i\hat{\sigma} + R_0/2) \Gamma(-in\hat{\sigma} + R_1/2) \Gamma( i\hat{\sigma} + R_2/2) e^{it_{\text{ren}}\hat{\sigma}} f_{\cB}(\hat{\sigma})
\end{equation}
is then well-defined, and poles due to positively and negatively charged chiral multiplets are on different sides of the contour:
\begin{alignat}{5}
  i\hat{\sigma} & = -R_j/2-k<0 \qquad && \text{for } j\in\{0,2\} \text{ and } k\geq 0 ,
  \\
  i\hat{\sigma} & = \frac{1}{n}\bigl(R_1/2+k\bigr)>0 && \text{for } k\geq 0 .
\end{alignat}
By mixing R-symmetry with the gauge symmetry, we take $R_1=0$ and deform~$L$ slightly to keep all poles of each Gamma function on the same side of~$L$ as for positive R-charges.
Contrarily to compact models, the remaining R-charges $R_0$ and~$R_2$ are needed for regularization and the localization result depends non-trivially on them: there are divergences as $R_0,R_2\to 0$.

With $R_1=0$, consider the limit $R_0\to 0$; then the contour gets pinched between a pair of poles at negative and zero~$i\hat{\sigma}$.
(A slightly more complicated pinching occurs as both $R_0,R_2\to 0$.)
On very general grounds such a pinching makes the integral blow up like the inverse distance between the poles.
For any function~$f$ that is holomorphic in $x$ in a neighborhood of the contour,
\begin{equation}\label{generalgroundspinching}
  \int_{\RR} \frac{f(x)\,dx}{(x-i\varepsilon)(x+i\varepsilon)}
  = 2\pi i \frac{f(i\varepsilon)}{2i\varepsilon} +
  \int_{\tikz{
      \draw[-{stealth}](0,0)--(.2,0);
      \draw[-{stealth}](.2,0)--(.3,0)arc(180:0:.1 and .15)--(.7,0);
      \draw(.7,0)--(.8,0);
      \fill (.4,.05) circle (.02);
      \fill (.4,-.05) circle (.02);
    }} \frac{f(x) \, dx}{(x-i\varepsilon)(x+i\varepsilon)}
  = \frac{2\pi i f(0)}{2i\varepsilon} + O(1)
\end{equation}
as $\varepsilon\to 0$, where the contour is a slight deformation of~$\RR$ that goes above both poles.
We will not need the $\varepsilon^0$ term but a quick calculation shows that it is given by a principal value prescription
$\lim_{\varepsilon\to 0}\int_{\abs{x}>\varepsilon} (f(x)-f(0)) dx/x^2$.
Observe also that the singular term can be computed directly as $f(0) = \res_{x\to 0} \bigl( x \lim_{\varepsilon\to 0} (\text{integrand}) \bigr)$, without explicitly decomposing the integrand into two singular factors $(x\pm i\varepsilon)^2$ and a holomorphic function~$f$.

Coming back to our $U(1)$ model with charges $1$, $-n$, $1$, we discuss five instructive cases in the resolved phase.

\paragraph{The image $\cO(q)$ of a Wilson line.}
As our first example of brane we specialize~\eqref{welldefinedrank1ZD2B} to a Wilson line of charge~$q$, whose brane factor is~$e^{2\pi q\hat{\sigma}}$.
Like in~\eqref{generalgroundspinching} we can shift the contour through the pole at $i\hat{\sigma}=0$ and the remaining integral is smooth as $R_j\to 0$, so the only singular contribution as $R_0,R_2\to 0$ is the residue
\begin{equation}\label{ZD2r1Wq}
  Z_{D^2,\text{residue}}^{\text{0-instanton}}(\cO(q)) \simeq \frac{(\mathfrak{r}\Lambda)^{\hat{c}/2}}{n} \frac{\Gamma(R_0/2)}{-2\pi i} \frac{\Gamma(R_2/2)}{-2\pi i} + O(1)
  = - \frac{(\mathfrak{r}\Lambda)^{\hat{c}/2}}{n\pi^2R_0R_2} + O\biggl(\frac{1}{R_0}\biggr) + O\biggl(\frac{1}{R_2}\biggr) .
\end{equation}
This quadratic divergence is not an artifact of how we regularized.
Up to a factor it is the $U(1)^2$ equivariant volume of an orbifold of~$\CC^2$, the $\ZZ_n$ orbifold group being responsible for the $1/n$ factor.
More precisely, the divergent terms are ($1/n$ times) the partition function of a pair of free chiral multiplets of R-charges $R_0$ and~$R_2$,
which parametrize the two non-compact directions in the support of the brane.
None of the divergent terms depends on the charge~$q$.

\paragraph{The structure sheaf $\cO_{E_1}$ of the exceptional divisor.}
Next, consider the structure sheaf of the exceptional divisor.
The structure sheaf has Koszul resolution $\cO(n)\xrightarrow{X_1}\cO$ hence brane factor $f_1(\hat{\sigma}) = 1 - e^{2\pi n\hat{\sigma}}$, that is, a difference of Wilson line brane factors with $q=0$ and $q=n$.
Taking the difference cancels all $R_j\to 0$ divergences of~\eqref{ZD2r1Wq}.
Euler's formula $\Gamma(x)\sin\pi x=\pi/\Gamma(1-x)$ tells us that
\begin{equation}
  \Gamma(-in\hat{\sigma}) f_1(\hat{\sigma})
  = \frac{-2\pi i e^{\pi n\hat{\sigma}}}{\Gamma(1+in\hat{\sigma})}
\end{equation}
has no pole.
Then the localization formula does not exhibit contour pinching since all poles on one side have been cancelled.
We compute, through a contour integral or directly through~\eqref{ZD2locabHiggs},
\begin{equation}\label{ZD2P1divisor}
  \begin{aligned}
    Z_{D^2,\text{residue}}^{\text{0-instanton}}\Bigl(\cW(n)\xrightarrow{X_1} \cW(0)\Bigr) & = \frac{(\mathfrak{r}\Lambda)^{\hat{c}/2}}{(-2\pi i)^2} \int_0 \frac{\mathrm{d}\hat{\sigma}}{2\pi} \, \frac{-2\pi i e^{\pi n\hat{\sigma}}}{\Gamma(1+in\hat{\sigma})} \Gamma(i\hat{\sigma})^2 e^{it_{\text{ren}}\hat{\sigma}} \\
    & = (\mathfrak{r}\Lambda)^{\hat{c}/2} \frac{t_{\text{ren}} - i\pi n + (n-2)\gamma}{-2\pi i} ,
  \end{aligned}
\end{equation}
where $\int_0$ denotes an integral around the pole at $i\hat{\sigma}=0$.
The key aspect of this brane that leads to a finite partition function is that poles of the one-loop determinant of~$X_1$ are cancelled by the brane factor.
If we had given a positive R-charge $R_1$ to $X=X_1$ too, it would appear in the brane factor in exactly the correct way to cancel the pole of the one-loop determinant,
namely through $-in\hat{\sigma}\to -in\hat{\sigma}+R_1/2$.
Note that this brane is not band-restricted.

\paragraph{Other branes with compact support~$E_1$.}
Any brane that is supported on the exceptional divisor~$E_1$ has a brane factor that is a multiple of~$f_1$, before introducing R-charges~$R_j$.
For branes that do not respect the flavour symmetries (isometries of the Higgs branch) there is no preferred way to include R-charges in their brane factors.
However, it is natural to impose that the brane factor is still a multiple of~$f_1$.
The brane factor then cancels poles of $\Gamma(-in\hat{\sigma})$, which avoids contour pinching.
The regularized partition function is then finite as $R_j\to 0$, and unambiguous since adding any $O(R_j)$ terms (times~$f_1$) to the brane factor simply shifts the regularized partition function by~$O(R_j)$.
For instance the twist $\cO_E(q)$ with resolution $\cO(q+n)\xrightarrow{X_1}\cO(q)$ has brane factor $e^{2\pi q\hat{\sigma}} f_1(\hat{\sigma})$ which gives
\begin{equation}
  Z_{D^2,\text{residue}}^{\text{0-instanton}} \Bigl(\cW(q+n)\xrightarrow{X_1} \cW(q)\Bigr) = (\mathfrak{r}\Lambda)^{\hat{c}/2} \frac{t_{\text{ren}} - i\pi (n+2q) + (n-2)\gamma}{-2\pi i} ,
\end{equation}
found by shifting $t\to t-2\pi i q$.
Another example is the brane with resolution $\cO(kn)\xrightarrow{X_1^k}\cO(0)$, which results in
$(\mathfrak{r}\Lambda)^{\hat{c}/2} \, k\bigl(t_{\text{ren}}- i\pi kn + (n-2)\gamma\bigr)/(-2\pi i)$.  Again this is finite, consistent with the fact that this brane's support is compact.

\paragraph{Noncompact branes with the same brane factor.}
We now illustrate that non-compact branes with the same brane factor at $R_j=0$ can have different, geometrically meaningful, regularized partition functions.
Consider a brane~$\cB_{k_0,k_1,k_2}$ with resolution $\cO(0)\xrightarrow{X_0^{k_0}X_1^{k_1}X_2^{k_2}}\cO(k_0-nk_1+k_2)$ for some $k_i\geq 0$, where the arrow denotes multiplication by the monomial~$X_0^{k_0}X_1^{k_1}X_2^{k_2}$.
This brane is supported on the base $\CP^1$ and two noncompact fibers: $\{X_0=0\}\cup\{X_1=0\}\cup\{X_2=0\}$.
The brane factor, including R-charges, is then
\begin{equation}
  f_{\cB_{k_0,k_1,k_2}}(\hat{\sigma}) = 1 - e^{2\pi(k_0-nk_1+k_2)\hat{\sigma}-i\pi(k_0R_0+k_2R_2)}
\end{equation}
and its $R_j\to 0$ limit only depends on $k_0-nk_1+k_2$.
For instance when $k_0+k_2=nk_1$ the brane factor is zero, as for an empty brane.
The partition function computed using~\eqref{ZD2r1Wq} is in general divergent:
\begin{equation}
  \begin{aligned}
    Z_{D^2,\text{residue}}^{\text{0-instanton}}\bigl(\cB_{k,l}\bigr)
    & = (\mathfrak{r}\Lambda)^{\hat{c}/2}\frac{1-e^{-i\pi(k_0R_0+k_2R_2)}}{(-2\pi i)^2} \biggl( \frac{1}{n} \Gamma\biggl(\frac{R_0}{2}\biggr) \Gamma\biggl(\frac{R_2}{2}\biggr) + O(1) \biggr) \\
    & = (\mathfrak{r}\Lambda)^{\hat{c}/2}\frac{1}{n(2\pi i)} \biggl(k_0\Gamma\biggl(\frac{R_2}{2}\biggr) + k_2\Gamma\biggl(\frac{R_0}{2}\biggr)\biggr) + \dots
  \end{aligned}
\end{equation}
Ignoring the factor $2\pi i$, the two terms have a geometric interpretation as contributions of the non-compact supports $\{X_0=0\}$ and $\{X_2=0\}$ of the brane, which have multiplicity $k_0$ and~$k_2$ respectively.

\paragraph{Noncompact branes respecting only part of the Higgs branch isometry.}
A last instructive case is a brane $\cO(0)\xrightarrow{G_k(X_0,X_2)}\cO(k)$ for $G_k$ a homogenous polynomial of degree~$k$.
Roots of~$G_k$ define points with homogenous coordinates $(X_0:X_2)$ on~$\CP^1$.
In the resolved phase the brane is supported on the corresponding fibers of the total space of $\cO(-n)\to\CP^1$.
Unless $G_k$ is a monomial, it does not transform in a definite way under the $U(1)^2$ isometries acting on $X_0$ and $X_2$ that we used to introduce R-charges $R_0$ and~$R_2$.
Thus one cannot deform the brane factor to introduce $R_0$ and~$R_2$.
However, $G_k$ has definite charge under the diagonal subgroup $U(1)\subset U(1)^2$, so that it makes sense to turn on $R_0=R_2>0$.
Such an R-charge is enough to avoid contour pinching and regularize the partition function.

To summarize, the hemisphere partition function can be regularized for branes with support on~$E_1$, and for non-compact branes that preserve an isometry of the Higgs branch.
Contour pinching as $R_j\to 0$ reflects the existence of non-compact branes for which the correct partition function is infinite.

\subsection{\label{ssec:regZgeneral}Regularization for compact branes in abelian GLSMs}

We are interested in models with non-compact Higgs branch.
In Calabi-Yau models the superconformal algebra of the IR limit contains a $U(1)$ R-symmetry that splits into left-moving and right-moving parts.
Such a left/right split symmetry must act trivially on non-compact directions.\footnote{More precisely, in a nonlinear sigma model, the $U(1)$ isometry given by a Killing vector $\xi_I$ (namely such that $\nabla_{(I}\xi_{J)}=0$) splits if $\nabla_{[I}\xi_{J]}=0$ too.  A $U(1)$ rotation of a cylinder splits, but not a $U(1)$ rotation of a cone or plane.}
This IR R-symmetry is typically visible in the UV\@ (it could also be emergent),
in which case the relevant localization calculation is the one involving that R-symmetry.
As we just argued it must assign R-charge~$0$ to gauge-invariant polynomials in chiral multiplets that span the non-compact directions.
The natural generalization to non-compact models that are not Calabi-Yau is to apply the localization result in which all non-compact directions have R-charge~$0$.

At face value the localization result is singular whenever any R-charge vanishes, because the contour passes through a pole at $\hat{\sigma}=0$ (other poles are not problematic).
The obvious regularization is to turn on a small positive R-charge~$R_j$ for each chiral multiplet~$X_j$, that is, mix the R-symmetry with gauge and flavour symmetries.
Geometrically, the mixing with flavour symmetries amounts to working equivariantly with respect to isometries of the Higgs branch.
In principle such a regularization is only adapted for branes that preserve an isometry of the Higgs branch,
but we find in examples that the regularization can be extended to some other branes with compact support.

\paragraph{Compact models (non-zero superpotential).}
In compact models it is typically possible to mix the R-symmetry with $\varepsilon$ times a gauge symmetry so as to shift all R-charges into $(0,2)$.
This mixing with gauge charges amounts to a shift of the integration variable~$\hat{\sigma}$.
The localization result thus only depends on~$\varepsilon$ through an overall factor~$e^{t_{\text{ren}}\varepsilon/2}$, independent of the brane, and which disappears as $\varepsilon\to 0$.
Since the resulting integral is regular, any $O(\varepsilon)$ correction to the brane factor drops out as $\varepsilon\to 0$.
Therefore the result is finite for any brane, and is insensitive to the precise regularisation.
As an example, consider the quintic hypersurface GLSM\@, a $U(1)$ model with chiral multiplets $P$ and $X_i$, $1\leq i\leq 5$ of charges $-5$ and~$1$ and with a superpotential $W=PG_5(X)$ for $G_5$ a generic degree~$5$ homogenous polynomial;
instead of the usual R-charges $2$ and $0$ for $P$ and $X$ one uses $2-5\varepsilon$ and~$\varepsilon$ for $\varepsilon\in(0,2/5)$.
An alternative point of view, rather than shifting~$\hat{\sigma}$, is that the contour $L$ is not $\RR$ but a shift (more generally a deformation) thereof such that all poles of $\Gamma(iQ^j\hat{\sigma})$ for $Q^j>0$ are on one side of~$L$, and those with $Q^j<0$ on the other side.

\paragraph{Branes in non-compact models.}

In non-compact models, such as the quintic GLSM above without its superpotential, the non-compact directions are spanned by some gauge invariants with R-charge~$0$.
Mixing R-symmetry with gauge symmetry does not affect their R-charge, thus it cannot make all chiral multiplets have positive R-charge.

Let us discuss the case of branes supported on distinguished subsets
\begin{equation}
  E_K=\{X_j=0\;\forall j\in K\}
\end{equation}
of the Higgs branch (in Hirzebruch-Jung models these include the exceptional divisors).
Depending on~$K$ this may be empty, compact, or noncompact.
The set $E_K$ is defined as solutions to D-term equations with the further constraint $X_j=0$ for $j\in K$, so
\begin{equation}
  \sum_{i\not\in K} Q^i \abs{X_i}^2 = \zeta_{\text{ren}} ,
\end{equation}
modulo gauge transformations.
This has solutions ($E_K\neq\emptyset$) if and only if $\zeta_{\text{ren}}\in\Cone_{({}^\complement K)}$.
Under this condition, let us prove that $E_K$~is compact if and only if there exists $\hat{s}$ such that $Q^i\cdot\hat{s}>0$ for all $i\not\in K$.
If there exists such an~$\hat{s}$ then the norm of points in~$E_K$ is bounded:
$\lVert X\rVert^2\leq(\zeta_{\text{ren}}\cdot\hat{s})/\min_{i\not\in K}(Q^i\cdot\hat{s})$.
Conversely, if there exists no such~$\hat{s}$ then the polygonal cone $\Cone_{({}^\complement K)}$ (which has finitely many edges) does not lie in any half-space, which implies that the cone contains a line through~$0$.
In turn this implies that there exists a vanishing linear combination $\sum_{i\not\in K}\lambda_i Q^i=0$ with positive coefficients $\lambda_i>0$.
From any $X\in E_K$ we can then build arbitrarily large solutions by shifting each $\abs{X_i}^2$ by the same multiple of~$\lambda_i$, thus $E_K$ is noncompact.
We discuss each of these cases in turn:
\begin{itemize}
\item $E_K$~empty hence $\zeta_{\text{ren}}\not\in\Cone_{({}^\complement K)}$,
\item $E_K$~compact hence $\zeta_{\text{ren}}\in\Cone_{({}^\complement K)}$ and there exists $\hat{s}$ such that $Q^i\cdot\hat{s}>0$ for all $i\not\in K$,
\item $E_K$~noncompact hence $\zeta_{\text{ren}}\in\Cone_{({}^\complement K)}$ and there is no such~$\hat{s}$.
\end{itemize}

For branes supported in a union of sets $E_K$ that is compact we argue that the regularized partition function has an unambiguous finite $R_j\to 0$ limit, which we will compute for Hirzebruch-Jung models to match it with a geometric calculation in \autoref{sec:Ktheoreticaspects}.
In contrast, branes that are non-compact only have a meaningful regularized partition function if they respect enough isometries of the Higgs branch.
The result typically diverges as $R_j\to 0$ and different branes that have the same brane factor at $R_j=0$ may give different regularized partition functions.

\paragraph{Empty branes.}

Let us begin with GLSM branes whose geometric image (image under $F_{\text{geom}}$) in the Higgs branch category~$D^b(X)$ is empty.
Besides showing in a simple case that noncompactness of the Higgs branch is not an issue, the main purpose is to show that the residue contribution $Z_{D^2,\text{residue}}$ vanishes for a brane whose geometric image is empty, hence $Z_{D^2,\text{residue}}$ only depends on the geometric image.  We used this result to get~\eqref{computingZofHiggsbranch}.

Consider the brane $\cO_{\Delta_K}$ for $\Delta_K=\{X_j=0\;\allowbreak\forall j\in K\}$ an irreducible component of the deleted set $\Delta\subset V$.
Given the second description of~$\Delta$ in~\eqref{DeletedSetAbelianGLSM}, the possible~$K$ are characterized by the fact that $\zeta_{\text{ren}}\not\in\Cone_{({}^\complement K)}$, namely the FI parameter cannot be written as a linear combination of $\{Q^j\mid j\not\in K\}$.
Recall now that the Higgs branch hemisphere partition function~\eqref{ZD2locabHiggs} picks up residues labeled by a set of $r$~flavours $J$ such that $\zeta_{\text{ren}}\in\Cone_J$.
Together this implies that $J\not\subset {}^{\complement}K$ namely $J\cap K\neq \emptyset$.
Each pole that contributes obeys $Q^j\cdot\hat{\sigma}\in i(R_j/2+\ZZ_{\geq 0})$ for all $j\in J$, hence for at least one $j\in K$.
However, the brane factor~\eqref{Koszulbranefactor} has zeros whenever $Q^j\cdot\hat{\sigma}\in i(R_j/2+\ZZ_{\geq 0})$ for any $j\in K$.
To reiterate, the brane factor cancels all poles of one-loop determinants of the chirals $X_j$, $j\in K$, because Euler's formula $\Gamma(x)\sin\pi x=\pi/\Gamma(1-x)$ yields
\begin{equation}\label{polezerocancellation}
  \begin{aligned}
    \bigl(1 - e^{-2\pi Q^j\cdot\hat{\sigma}+i\pi R_j}\bigr) \Gamma\biggl(iQ^j\cdot\hat{\sigma} + \frac{R_j}{2}\biggr)
    & = -2\pi i\,e^{-\pi Q^j\cdot\hat{\sigma}+i\pi R_j/2} \biggm/\Gamma\biggl(1-iQ^j\cdot\hat{\sigma} - \frac{R_j}{2}\biggr)
  \end{aligned}
\end{equation}
which has no pole.
All residue contributions in that phase are thus eliminated, namely $Z_{D^2,\text{residue}}=0$ for that brane.

In our calculations it was crucial that R-charges appear in the brane factor~\eqref{Koszulbranefactor} in the same way as in chiral multiplet one-loop determinants, so that the cancellation~\eqref{polezerocancellation} took place.
We deduced the brane factor from the Koszul complex~\eqref{KoszulResolutionOfSheaf} whose morphisms are multiplications by chiral multiplets with well-defined R-charges.
An arbitrary complex may in general fail to have well-defined R-charges; our regularization of the hemisphere partition function can then fail to be defined.

\paragraph{Compact branes.}

The structure sheaf of~$E_K$ has a Koszul resolution~\eqref{KoszulResolutionOfSheaf} with brane factor~\eqref{Koszulbranefactor}.\footnote{In case $E_K$ is contained in an orbifold singularity, its structure sheaf is actually a fractional brane rather than a usual D-brane wrapping~$E_K$.}
As explained in~\eqref{polezerocancellation} this brane factor cancels all poles of the one-loop determinant of $X_j$ for $j\in K$:
explicitly their product gives a factor $-2\pi i\,e^{\cdots}/\Gamma(1-\cdots)$ with no pole.
Recall now that the residue part of the hemisphere partition function~\eqref{ZD2locabHiggs} is a sum, over sets~$J$ of $r$~flavours such that $\zeta_{\text{ren}}\in\Cone_J$, of residues at common poles of the chiral multiplets $X_j$, $j\in J$.
Any such residue with $J\cap K\neq\emptyset$ vanishes due to the brane factor.
The sum is thus restricted to $J\subset{}^\complement K$.  Altogether,
\begin{equation}\label{ZD2HiggsEK}
  Z_{D^2,\text{residue}} \simeq (\mathfrak{r}\Lambda)^{\hat{c}/2} \!\!\!\!\!\!\!\!\! \sum_{\substack{J\subset {}^\complement K\mid\zeta_{\text{ren}}\in\Cone_J\\k:J\to\ZZ_{\geq 0}}} \!\!\!\!\!\!\!\!\! \pm \! \res_{i\hat{\sigma}=i\hat{\sigma}_{J,k}} \!\! \biggl(
  \frac{ e^{it_{\text{ren}}\cdot\hat{\sigma}} e^{\sum_{j\in K}(-\pi Q^j\cdot\hat{\sigma}+i\pi R_j/2)}\prod_{j\not\in K} \Gamma\bigl( iQ^j\cdot\hat{\sigma} + R_j/2 \bigr)}{(-2\pi i)^{\#({}^\complement K\setminus J)} \prod_{j\in K} \Gamma\bigl( 1- iQ^j\cdot\hat{\sigma} - R_j/2 \bigr)} \biggr) , \!
\end{equation}
where the sign $\pm$ is $\sign(\det(Q^\ell)_{\ell\in J})$.

Consider the case of a compact~$E_K$, such that there exists $\hat{s}$ with $Q^i\cdot\hat{s}>0$ for all $i\not\in K$.
In the contour formula for the hemisphere partition function, shift the integration variable $\hat{\sigma}$ to $\hat{\sigma}-i\varepsilon\hat{s}$ for some small $\varepsilon>0$.
This shifts the argument of all numerator Gamma functions (those with $j\not\in K$) by a positive amount, just like R-charges, thus none of their poles intersect the contour as all $R_j\to 0$.
We have no control on the signs of $Q^i\cdot\hat{s}$ for $i\in K$, which are shifts of arguments of Gamma functions in the denominator, but these factors do not contribute any pole.
Altogether the contour integral remains finite as $R_j\to 0$.
Just as in one parameter examples, the brane factor of any brane supported on~$E_K$ should be a multiple of the brane factor of the structure sheaf of~$E_K$.
That brane factor cancels poles from all chiral multiplets with $j\in K$, hence
the regularized partition function remains finite as $R_j\to 0$ too.
This generalizes readily to branes supported on the union of all compact~$E_K$:
their brane factor is a sum of brane factors supported on each~$E_K$.
These compact branes exhibit no contour pinching.

\paragraph{Non-compact branes.}

Finally, for a non-compact~$E_K$ we expect the regularized partition function of its structure sheaf to have singular contributions at $R_j\to 0$.
We worked them out in one-parameter examples in the previous subsection.
It would be very interesting to relate these singular contribution to an equivariant integral on the support of the brane, but we postpone such an investigation to later work.

\subsection{\label{ssec:BbranesHJ}B-brane category of Hirzebruch-Jung models}

We apply here the general considerations of the previous subsections to Hirzebruch-Jung models of arbitrary rank.
We describe the derived category $D^b(X)$ of coherent sheaves on the Higgs branch~$X$ in terms of generators and relations.
We determine the pull-back of each generator to local models of the orbifold points and of exceptional divisors.
Finally, we calculate the regularized hemisphere partition function for some compact branes that we compare with geometry in \autoref{sec:Ktheoreticaspects}.

\paragraph{Generators and relations.}

Fix a phase specified by the collection~$A$ of blown up divisors.
Recall that $D^b(X)$ is generated by the line bundles $\cO(b_1,\dots,b_r)$ on~$X$, defined to be the images (under $F_{\text{geom}}$) of the Wilson line branes $\cW(b_1,\dots,b_r)$ with charges $b$ under the $U(1)^r$ gauge group of the GLSM\@.
The tensor product $\cO(b_1,\dots,b_r)\otimes\cO(c_1,\dots,c_r) = \cO(b_1+c_1,\dots,b_r+c_r)$
means we could restrict our attention to branes with a single non-zero $b_i=1$, but it will be clearer to keep all~$b_i$.

Sections of $\cO(0,\dots,0)$ are just $G$-invariant functions on $V\setminus\Delta$, hence are functions on the Higgs branch $(V\setminus\Delta)/G$, so this is the structure sheaf of the Higgs branch, $\cO=\cO(0,\dots,0)$.
Multiplication by $X_j$ maps from $\cO$ to $\cO(\dots,0,1,-a_j,1,0,\dots)$, so the latter sheaf is $\cO$~twisted by the divisor $E_j=\{X_j=0\}$.  Explicitly,
\begin{equation}
  \begin{aligned}
    \cO(E_0) & = \cO(1,0,\dots) , \qquad \cO(E_1) = \cO(-a_1,1,0,\dots) , \\
    \cO(E_\alpha) & = \cO(\dots,0,1,-a_\alpha,1,0,\dots) \text{ for $1<\alpha<r$}, \\
    \cO(E_r) & = \cO(\dots,0,1,-a_r) , \qquad \cO(E_{r+1}) = \cO(\dots,0,1) ,
  \end{aligned}
\end{equation}
Since the Cartan matrix of charges has determinant~$n$ rather than~$1$, tensor products of the line bundles $\cO(E_\alpha)$ for $1\leq\alpha\leq r$ do not give all $\cO(b_1,\dots,b_r)$.
On the other hand the line bundles $\cO(E_j)$ for $0\leq j\leq r+1$ do.

Any B-brane of the GLSM whose support is in the deleted set~$\Delta$ gives a trivial brane in the Higgs branch theory (we study Coulomb/mixed parts of the brane in \autoref{sec:Band restriction rule}).
Therefore, the line bundles $\cO(b_1,\dots,b_r)$ are subject to one relation for each irreducible component of the deleted set~$\Delta$.
This gives two types of relations.
\begin{itemize}
\item For each divisor that is not blown up (each $\alpha\in\{1,\dots,r\}\setminus A$),
  $\Delta$ contains the hyperplane $\{X_\alpha=0\}$.
  Its structure sheaf has Koszul resolution~\eqref{hyperplaneKoszul} $\cW\xrightarrow{X_\alpha}\cW$ by line bundles,
  thus the complex $\cO(-E_\alpha)\xrightarrow{X_\alpha}\cO$ is trivial in~$D^b(X)$.
\item For each pair of non-consecutive blown-up divisor (each $\alpha,\beta\in A$ such that no $\gamma\in A$ obeys $\alpha<\gamma<\beta$), $\Delta$ contains the intersection $\{X_\alpha=X_\beta=0\}$.
  Its structure sheaf has Koszul resolution~\eqref{KoszulResolutionOfSheaf}, hence the following complex on~$X$ is trivial:
  \begin{equation}
    \cO(-E_\alpha-E_\beta) \xrightarrow[(X_\alpha,X_\beta)]{} \cO(-E_\beta) \oplus \cO(-E_\alpha) \xrightarrow[(X_\beta,-X_\alpha)]{} \cO.
  \end{equation}
\end{itemize}

\paragraph{Pull-backs.}

Our goal now is to clarify what the generators $\cO(b_1,\dots,b_r)$ are by determining their pullbacks to Higgs branches of local models discussed in \autoref{ssec:local models}.
Recall that these local models were found by determining that the $U(1)^r$ gauge group is Higgsed down to some subgroup~$H$ when some chiral multiplets have a non-zero vev.
The Wilson line $\cW(b_1,\dots,b_r)$ can be realized by the insertion of a 1d Fermi multiplet with charges $b\colon U(1)^r\to U(1)$.
After Higgsing its charge under $H$ is deduced from $H\subset U(1)^r\xrightarrow{b} U(1)$.
The resulting Wilson line in the local model has a clear geometric meaning.

\paragraph{Pull-back to an intersection.}

Consider first the local model~\eqref{localmodelintersection} for an intersection point $E_{\{i,j\}}=E_i\cap E_j$ of two divisors, with $0\leq i<j\leq r+1$.
The residual gauge group $H=\ZZ_{d_{ij}}$ embeds into $U(1)^r$ (in basis~I) as
\begin{equation}
  \ZZ_{d_{ij}} \ni 1 \mapsto (1,\dots,1,\omega^{d_{(i+1)j}},\omega^{d_{(i+2)j}},\dots,\omega^{d_{(j-1)j}},1,\dots,1)
\end{equation}
where $\omega=\exp(2\pi i/d_{ij})$ and the entries in positions $\alpha\in\{i,\dots,j\}$ are~$\omega^{d_{\alpha j}}$.  Note that $d_{(j-1)j}=1$.
The Higgs branch image of a Wilson line with charges $(b_1,\dots,b_r)$ in basis~I therefore has the following equivariant line bundle as its pullback to the neighborhood of $E_i\cap E_j$:
\begin{equation}
  \cW(q) \text{ on } \CC^2/\ZZ_{d_{ij}(d_{(i+1)j})}
  \text{ with } \ZZ_{d_{ij}} \text{-charge } q = \sum_{\alpha=i+1}^{j-1} d_{\alpha j} b_\alpha .
\end{equation}
While the expression is asymmetric between $i$ and~$j$ one can change basis in~$\ZZ_{d_{ij}}$ by multiplying all charges by $d_{i(j-1)}$.  Using $d_{i(j-1)}d_{\alpha j} = d_{ij} d_{\alpha(j-1)} + d_{i\alpha} d_{(j-1)j} \equiv d_{i\alpha} \bmod{d_{ij}}$, we find
\begin{equation}
  \cW(q) \text{ on } \CC^2/\ZZ_{d_{ij}(d_{i(j-1)})}
  \text{ with } \ZZ_{d_{ij}} \text{-charge } q = \sum_{\alpha=i+1}^{j-1} d_{i\alpha} b_\alpha .
\end{equation}
In both bases, the charge~$q$ only involves charges $b_\alpha$ for indices~$\alpha$ such that $E_\alpha$ is not blown up.

\paragraph{Pull-back to an exceptional divisor.}

Consider next the local model~\eqref{localdivisorchargesU1Zm} of an exceptional divisor $E_j$, in a phase where that divisor intersects $E_i$ and~$E_k$ for $0\leq i<j<k\leq r+1$.
The gauge group is $H=U(1)\times\ZZ_m$ with $m=\gcd(d_{ij},d_{jk})$,
and $(h,\omega)\in U(1)\times\ZZ_m$ is mapped to $g\in U(1)^r$ with the following coordinates in basis~I\@: $g_\alpha=1$ for $\alpha\leq i$ or $\alpha\geq k$, while
\begin{equation}
  \begin{aligned}
    g_\alpha & = \bigl(h^{d_{jk}/m}\omega^u\bigr)^{d_{i\alpha}} \text{ for } i\leq\alpha\leq j, \\
    g_\alpha & = \bigl(h^{d_{ij}/m}\omega^v\bigr)^{d_{\alpha k}} \text{ for } j\leq\alpha\leq k.
  \end{aligned}
\end{equation}
Note that $d_{ii}=d_{kk}=0$ hence these formulas are compatible.
A Wilson line with charges $(b_1,\dots,b_r)$ in basis~I thus maps to a Wilson line with charges
\begin{equation}\label{DbHJpullback}
  \begin{aligned}
    \frac{d_{jk}}{m} \sum_{\alpha=i+1}^{j-1} d_{i\alpha} b_\alpha
    + \frac{d_{ij} d_{jk} b_j}{m}
    + \frac{d_{ij}}{m} \sum_{\alpha=j+1}^{k-1} d_{\alpha k} b_\alpha & \text{ under } U(1) , \\
    u \sum_{\alpha=i+1}^{j-1} d_{i\alpha} b_\alpha + v \sum_{\alpha=j+1}^{k-1} d_{\alpha k}b_\alpha & \text{ under } \ZZ_m .
  \end{aligned}
\end{equation}
For a Wilson line with a single non-zero~$b_\beta$ all of these charges vanish except one ($j=\beta$) if $E_\beta$ is blown up, and two otherwise.
In the first case, the Higgs branch image of the Wilson line is a non-trivial line bundle on the weighted projective space~$E_\beta$ but has trivial pullback near each orbifold point or any other exceptional divisor.
In the second case ($E_\beta$~not blown-up) the Higgs branch image has a non-trivial pullback, with charge $d_{\beta j} b_\beta$, near the orbifold point $E_i\cap E_j$ with $i<\beta<j$ and non-trivial pullbacks on $E_i$ and~$E_j$.

The case of the fully resolved phase is instructive: then all $d_{ij}$ and $m$ appearing above are equal to~$1$ and the Higgs branch image of the Wilson line $\cW(b_1,\dots,b_r)$ is a line bundle (on the Hirzebruch-Jung resolution) whose pull-back to each $\CP^1$ exceptional divisor~$E_j$ is~$\cO(b_j)$.
This is consistent with the fact that the gauge group of the local model near~$E_j$ is in that case the $j$-th $U(1)$ factor in $U(1)^r$ (in basis~I).

\subsection{\label{ssec:ZD2HJ}Central charges in Hirzebruch-Jung models}

\paragraph{Fractional D0 branes at intersections.}

We explain around~\eqref{polezerocancellation} why the residue part $Z_{D^2,\text{residue}}$ of the hemisphere partition function correctly captures the Higgs branch contribution of a GLSM brane that reduces to a given Higgs branch brane, regardless of band restriction.
Let us apply this to fractional D0 branes at the intersection $E_i\cap E_j$, which is a $\ZZ_{d_{ij}}$-orbifold point.\footnote{If $d_{ij}=1$ the brane is a regular D0 brane at a smooth point.}
The brane factor is
\begin{equation}
  f(\hat{\sigma}) = \bigl(1-e^{2\pi i (iQ^i\cdot\hat{\sigma}+R_i/2)}\bigr) \bigl(1-e^{2\pi i (iQ^j\cdot\hat{\sigma}+R_j/2)}\bigr) e^{2\pi\rho\cdot\hat{\sigma}}
\end{equation}
where the twist by a Wilson line $\cW(\rho)$ affects the $\ZZ_{d_{ij}}$ charge of the fractional brane.
Incidentally, this brane can often not be made band-restricted for any choice of~$\rho$.

The brane's support is compact, thus as explained in \autoref{ssec:regZgeneral} the brane factor cancels enough poles to avoid contour pinching.
Specifically, the brane factor cancels poles from one-loop determinants of $X_i$ and~$X_j$.
Notice that there are only $r$ chiral multiplets other than $X_i$ and~$X_j$, which is exactly the rank of the gauge group, so the $r$-fold integral picks up exactly one family of residues, labeled by $J={}^\complement\{i,j\}$ in the notation of~\eqref{ZD2HiggsEK}.
We find
\begin{equation}
  Z_{D^2,\text{residue}}^{\text{0-instanton}} = \pm (\mathfrak{r}\Lambda)^{\hat{c}/2} \!\!\!\!\res_{i\hat{\sigma}=i\hat{\sigma}_{\{i,j\}}} \! \biggl(
  \frac{e^{(it_{\text{ren}}+2\pi\rho)\cdot\hat{\sigma}}}{(-2\pi i)^2} \!\prod_{\ell=i,j} \frac{-2\pi i\,e^{-\pi Q^\ell\cdot\hat{\sigma}+i\pi R_\ell/2}}{\Gamma\bigl( 1- iQ^\ell\cdot\hat{\sigma} - R_\ell/2 \bigr)}
  \!\prod_{\ell\neq i,j} \Gamma\bigl( iQ^\ell\cdot\hat{\sigma} + R_\ell/2 \bigr) \biggr) ,
\end{equation}
where the sign is $\sign(\det(Q^\ell)_{\ell\neq i,j})$ and $i\hat{\sigma}_{\{i,j\}}$ is the solution of $iQ^\ell\cdot\hat{\sigma} + R_\ell/2=0$ for all $\ell\neq i,j$.
This solution is linear in the R-charges, and we do not need its explicit expression~\eqref{sigmaij} yet.

Computing the residue gives a factor $1/\det(Q^\ell)_{\ell\neq i,j}$, which combines with the sign to give an absolute value.
The matrix has a block form, so
\begin{equation}
  \det(Q^\ell)_{\ell\neq i,j}
  = \det \begin{pmatrix}
    U & N_1 & 0 \\
    0 & -C_{(ij)} & 0 \\
    0 & N_2 & L
  \end{pmatrix}
  = \det(-C_{(ij)}) = (-1)^{j-i} d_{ij} ,
\end{equation}
where $U$ and $L$ are upper and lower triangular matrices with $1$ on the diagonal, $N_1$ and $N_2$ are matrices with a single non-zero entry equal to~$1$ in the corner closest to the diagonal of the main matrix, and $C_{(ij)}$ consists of rows and columns from $(i+1)$ to $(j-1)$ included of the generalized Cartan matrix.
We deduce\footnote{Note that $\hat{\sigma}\to 0$ as $R\to 0$, as can be seen for instance from the explicit expression~\eqref{sigmaij}.}
\begin{equation}
  Z_{D^2,\text{residue}}^{\text{0-instanton}} = (\mathfrak{r}\Lambda)^{\hat{c}/2} \lim_{R\to 0} \frac{1}{d_{ij}} \biggl(
  \frac{e^{(it_{\text{ren}}+2\pi\rho)\cdot\hat{\sigma}}}{(-2\pi i)^2} \prod_{\ell=i,j} \frac{-2\pi i\,e^{-\pi Q^\ell\cdot\hat{\sigma}+i\pi R_\ell/2}}{\Gamma\bigl( 1- iQ^\ell\cdot\hat{\sigma} - R_\ell/2 \bigr)} \biggr)_{i\hat{\sigma}=i\hat{\sigma}_{\{i,j\}}}
  = \frac{(\mathfrak{r}\Lambda)^{\hat{c}/2}}{d_{ij}} .
\end{equation}
Recall that the intersection $E_i\cap E_j$ is a fixed point of the orbifold group $\ZZ_{d_{ij}}$.
This central charge does not depend on the $\ZZ_{d_{ij}}$-charge of the fractional brane (no $\rho$ dependence).
A collection of $d_{ij}$ fractional branes with all possible $\ZZ_{d_{ij}}$ charges gives a D0 brane that can move away from the orbifold point, which is consistent with the fact that this collection has central charge~$1$ independent of which orbifold point we start from.
For $d_{ij}=1$ (so $j=i+1$) we are simply discussing a usual D0 brane.
We chose the normalization of the hemisphere partition function to make this case very simple.

\paragraph{Structure sheaf of an exceptional divisor.}

We now turn to the structure sheaf of an exceptional divisor~$E_j$, $1\leq j\leq r$, in a phase in which it is blown up.
As usual we denote by $i<j<k$ the neighboring exceptional divisors.
Again, the brane's support $E_j$~is compact so the brane factor
\begin{equation}
  f(\hat{\sigma}) = \bigl(1-e^{2\pi i (iQ^j\cdot\hat{\sigma}+R_j/2)}\bigr) e^{2\pi\rho\cdot\hat{\sigma}}
\end{equation}
cancels enough poles to avoid contour pinching.

Specializing~\eqref{ZD2HiggsEK} to the present case, the residue part of the regularized hemisphere partition function is a sum over sets~$J$ of $r$~flavours with $j\not\in J$ and $\zeta\in\Cone_J$.
The D-term equation gives a criterion: $\zeta\in\Cone_J$ if and only if $E_{({}^\complement J)}=\{X_i=0\mid i\not\in J\}$ is a non-empty subset of the Higgs branch.
Given that $j\in {}^\complement J$ and ${}^\complement J$ has $r+2-r=2$ elements we find ${}^\complement J$ can be $\{i,j\}$ or $\{j,k\}$.

\paragraph{Structure sheaf: residue calculation.}

We first compute the term with $J={}^\complement\{i,j\}$.
The leading term of~\eqref{ZD2HiggsEK} is given by taking all vorticities (denoted $k_j\geq 0$ there) to be zero.  One of the residues is then taken at the unique solution of
$Q^\ell\cdot\hat{\sigma} = i R_\ell/2$ for all $\ell\neq i,j$, which we denote $\hat{\sigma}_{\{i,j\}}$.
Explicitly,
\begin{equation}\label{sigmaij}
  i\hat{\sigma}_{\{i,j\}}^\alpha
  = \begin{cases}
    - \sum_{\ell=0}^{\alpha-1} d_{\ell\alpha} \frac{R_\ell}{2}
    & \text{for } 1\leq\alpha\leq i, \\
    - \frac{1}{d_{ij}}\sum_{\ell=0}^{r+1} d_{i\min(\alpha,\ell)} d_{j\max(\alpha,\ell)} \frac{R_\ell}{2}
    & \text{for } i\leq\alpha\leq j , \\
    - \sum_{\ell=\alpha+1}^{r+1} d_{\alpha\ell} \frac{R_\ell}{2}
    & \text{for } j\leq\alpha\leq r .
  \end{cases}
\end{equation}
The two formulas for $\hat{\sigma}_{\{i,j\}}^i$ agree, as do the two formulas for~$\hat{\sigma}_{\{i,j\}}^j$.
Using the recursion relation $d_{\ell(i-1)}-a_id_{\ell i}+d_{\ell(i+1)}=0$
and $d_{ij}d_{\ell(i+1)} - d_{i\ell} d_{j(i+1)}=d_{i(i+1)}d_{\ell j}=d_{\ell j}$,
we work out
\begin{equation}
  \begin{aligned}
    iQ^i\cdot\hat{\sigma}_{\{i,j\}} + \frac{R_i}{2}
    = \sum_{\ell=0}^{r+1} \frac{d_{\ell j}}{d_{ij}} \frac{R_\ell}{2}
  \end{aligned}
\end{equation}
and likewise
\begin{equation}\label{iQjsigmaij}
  \begin{aligned}
    iQ^j\cdot\hat{\sigma}_{\{i,j\}} + \frac{R_j}{2}
    = \sum_{\ell=0}^{r+1} \frac{d_{i\ell}}{d_{ij}} \frac{R_\ell}{2} .
  \end{aligned}
\end{equation}
The residue that appears in the hemisphere partition function is then
\begin{equation}
  \begin{aligned}
    & \sign\bigl(\det(Q^\ell)_{\ell\neq i,j}\bigr)
    \res_{i\hat{\sigma}=i\hat{\sigma}_{\{i,j\}}} \biggl(
    \frac{e^{it_{\text{ren}}\cdot\hat{\sigma}-\pi Q^j\cdot\hat{\sigma}+i\pi R_j/2}\prod_{\ell\neq j} \Gamma\bigl( iQ^\ell\cdot\hat{\sigma} + R_\ell/2 \bigr)}{-2\pi i\,\Gamma\bigl( 1- iQ^j\cdot\hat{\sigma} - R_j/2 \bigr)} \biggr)
    \\
    & \quad =
    \frac{1}{d_{ij}}
    \frac{e^{it_{\text{ren}}\cdot\hat{\sigma}_{\{i,j\}}+i\pi\sum_{\ell=0}^{r+1} (d_{i\ell}/d_{ij})(R_\ell/2)}\Gamma\bigl(\sum_{\ell=0}^{r+1} \frac{d_{\ell j}}{d_{ij}} \frac{R_\ell}{2}\bigr)}{-2\pi i\, \Gamma\bigl( 1- \sum_{\ell=0}^{r+1} \frac{d_{i\ell}}{d_{ij}} \frac{R_\ell}{2} \bigr)}
    \\
    & \quad = \frac{i}{2\pi}\Biggl(
    \frac{2}{\sum_{\ell=0}^{r+1} d_{\ell j} R_\ell}
    + \biggl( \frac{2it_{\text{ren}}\cdot\hat{\sigma}_{\{i,j\}}}{\sum_{\ell=0}^{r+1} d_{\ell j} R_\ell}
    + \frac{(i\pi-\gamma)\sum_{\ell=0}^{r+1} d_{i\ell} R_\ell}{d_{ij}\sum_{\ell=0}^{r+1} d_{\ell j} R_\ell}
    - \frac{\gamma}{d_{ij}} \biggr)
    + O(R)\Biggr)
  \end{aligned}
\end{equation}
where the factor $1/d_{ij}$ comes from the determinant of the matrix of charges $Q^\ell$, $\ell\neq i,j$ when computing the residue, $t_{\text{ren}}\cdot\hat{\sigma}_{\{i,j\}}$ can be computed from~\eqref{sigmaij}, and we used $\Gamma(x)=\frac{1}{x}-\gamma+O(x)$ and $\Gamma(1-x) = 1 + \gamma x + O(x^2)$.

The same steps give the residue corresponding to $J={}^\complement\{j,k\}$.
In fact, most intermediate calculations can be skipped: for example $iQ^k\cdot\hat{\sigma}_{\{j,k\}}+R_k/2$ is immediately obtained from $iQ^j\cdot\hat{\sigma}_{\{i,j\}}+R_j/2$ by replacing $(i,j)\to(j,k)$ in~\eqref{iQjsigmaij}.
The residue that appears in the hemisphere partition function is then
\begin{equation}
  \begin{aligned}
    & \sign\bigl(\det(Q^\ell)_{\ell\neq j,k}\bigr)
    \res_{i\hat{\sigma}=i\hat{\sigma}_{\{j,k\}}} \biggl(
    \frac{e^{it_{\text{ren}}\cdot\hat{\sigma}-\pi Q^j\cdot\hat{\sigma}+i\pi R_j/2}\prod_{\ell\neq j} \Gamma\bigl( iQ^\ell\cdot\hat{\sigma} + R_\ell/2 \bigr)}{-2\pi i\, \Gamma\bigl( 1- iQ^j\cdot\hat{\sigma} - R_j/2 \bigr)} \biggr)
    \\
    & \quad =
    \frac{i}{2\pi} \Biggl( \frac{2}{\sum_{\ell=0}^{r+1} d_{j\ell} R_\ell}
    + \biggl( \frac{2it_{\text{ren}}\cdot\hat{\sigma}_{\{j,k\}}}{\sum_{\ell=0}^{r+1} d_{j\ell} R_\ell}
    + \frac{(i\pi-\gamma)\sum_{\ell=0}^{r+1} d_{\ell k} R_\ell}{d_{jk}\sum_{\ell=0}^{r+1} d_{j\ell} R_\ell}
    - \frac{\gamma}{d_{jk}} \biggr)
    + O(R) \Biggr) .
  \end{aligned}
\end{equation}

\paragraph{Structure sheaf: Higgs branch part of the hemisphere partition function.}

Summing the two residues, the $O(1/R)$ divergence cancels as expected.
Using
\begin{equation}
  i\hat{\sigma}_{\{i,j\}}^\alpha
  - i\hat{\sigma}_{\{j,k\}}^\alpha
  = \begin{cases}
    0
    & \text{for } 1\leq\alpha\leq i ,
    \\
    (d_{i\alpha}/d_{ij}) \sum_{\ell=0}^{r+1} d_{\ell j} R_\ell/2
    & \text{for } i\leq\alpha\leq j ,
    \\
    (d_{\alpha k}/d_{jk}) \sum_{\ell=0}^{r+1} d_{\ell j} R_\ell/2
    & \text{for } j\leq\alpha\leq k ,
    \\
    0
    & \text{for } k\leq\alpha\leq r ,
  \end{cases}
\end{equation}
and relations between the $d_{\beta\gamma}$, we finally get the finite $R\to 0$ limit
\begin{equation}\label{D2chargebigresult}
  Z_{D^2,\text{residue}}^{\text{0-instanton}}
  \! = \! (\mathfrak{r}\Lambda)^{\hat{c}/2}\frac{i}{2\pi}\Biggl(\!
  \Biggl(\sum_{\alpha=i+1}^{j-1} \frac{d_{i\alpha}}{d_{ij}} t^{\text{ren}}_\alpha\!\Biggr)\!
  + t^{\text{ren}}_j
  + \!\Biggl(\sum_{\alpha=j+1}^{k-1} \frac{d_{\alpha k}}{d_{jk}} t^{\text{ren}}_\alpha\!\Biggr)
  - i\pi \frac{d_{ik}}{d_{ij}d_{jk}}
  + \frac{d_{ik}-d_{jk}-d_{ij}}{d_{ij}d_{jk}} \gamma\!\Biggr)
\end{equation}
where $t^{\text{ren}}_j$ could be included in either of the two sums by extending the bounds to $\alpha=j$.

Whenever $i=j-1$ and $k=j+1$, in particular in the fully-resolved phase, this formula reduces to
\begin{equation}\label{kthrE}
  Z_{D^2,\text{residue}}^{\text{0-instanton}} = (\mathfrak{r}\Lambda)^{\hat{c}/2} \frac{i}{2\pi} \Bigl( t^{\text{ren}}_j-i\pi a_j+(a_j-2)\gamma \Bigr) ,
\end{equation}
which coincides with the result~\eqref{ZD2P1divisor} for the one-parameter model.
More generally, the central charge coincides with the central charge one can compute from the local model~\eqref{localdivisorchargesU1Zm}:
\begin{equation}
  Z_{D^2,\text{residue}}^{\text{0-instanton}}
  = (\mathfrak{r}\Lambda)^{\hat{c}/2} \frac{i}{2\pi} \frac{1}{m} \frac{t^{\text{ren}}_{\text{loc}} - i\pi d_{ik}/m + (d_{ik}/m - d_{jk}/m - d_{ij}/m)\gamma}{(d_{ij}/m)(d_{jk}/m)}
\end{equation}
where $t^{\text{ren}}_{\text{loc}}$ is given in~\eqref{localmodeltval},  $m=\gcd(d_{ij},d_{jk})$, and the $1/m$ factor is due to the orbifold.

\section{\label{sec:Ktheoreticaspects}Central charge of B-branes from K-theory}

In this section we study the central charge of B-branes on abelian GLSMs, when projected to their image in the Higgs branch. We saw in \autoref{ssec:HiggsGeom} that the geometry of the Higgs branch corresponds to a toric variety with at most abelian quotient singularities. Denote it by $X_{\zeta}$, where $\zeta$ is the real part of the FI parameter (we omit the subscript ``ren'': all FI parameters in this section are renormalized at some scale~$\mu$). The derived category $D(X_{\zeta})$ of such spaces is a well known mathematical object, as we previously reviewed in \autoref{ssec:Bbraneabelian}.

\paragraph{Geometric central charges in compact models.}

In models with compact target, the central charge $Z$ of a B-brane $\mathcal{V}\in D(X_{\zeta})$ is a map
\begin{equation}
  Z\colon D(X_{\zeta})\longrightarrow \mathbb{C}
\end{equation}
that is holomorphic in $\zeta$ and multivalued in the K\"ahler moduli space.
Moreover, $Z$~factors through $K_{0}(X_{\zeta})$, the Grothendieck group of $D(X_{\zeta})$ spanned by holomorphic vector bundles:
\begin{equation}
  Z\colon K_{0}(X_{\zeta})\longrightarrow \mathbb{C} .
\end{equation}
Specifically, the central charge is given by \cite{Iritani:2009ab}
\begin{equation}\label{ccleading}
  Z(\mathcal{V})= \int_{X_{\zeta}}e^{\tau}\hbar^{c_1(X_\zeta)/(2\pi i)}\widehat{\Gamma}(TX_{\zeta})\ch(\mathcal{V})+\dots
\end{equation}
Here, $\tau$~denotes the complexified K\"ahler class of~$X_{\zeta}$, so $\tau\in H^{2}(X_{\zeta},\mathbb{R}/\mathbb{Z})+i\mathcal{K}_{X_{\zeta}}$, with $\mathcal{K}_{X_{\zeta}}$ the K\"ahler cone of $X_{\zeta}$, and the ``$+\ldots$'' denote instanton corrections.
The meaning of the parameter $\hbar$ can be traced to a $\mathbb{C}^{*}$-equivariant cohomology on the worldsheet $\mathbb{P}^{1}$ (see e.g., \cite[section 10.2.3]{Cox:1999ab}). Finally, $\widehat{\Gamma}(TX_{\zeta})$ denote the gamma class of the tangent bundle. It is defined as
\begin{equation}\label{gamma-class}
  \widehat{\Gamma}(TX_{\zeta})
  \coloneqq \prod_{l=1}^{\dim X_{\zeta}}\Gamma\left(1-\frac{\lambda_{l}}{2\pi i}\right)
  = 1 + \frac{\gamma}{2\pi i} c_1(TX_{\zeta}) + \dots
\end{equation}
where $\gamma$ is the Euler-Mascheroni constant, $\lambda_{l}$ are the Chern roots of $TX_{\zeta}$ and $c_1$~its first Chern class.
The real part of $Z$ is related to the RR-charge of the B-brane $\mathcal{V}$ \cite{Green:1996dd,Minasian:1997mm}.
The phase of the central charge plays an important role on the stability of B-branes \cite{Douglas:2000ah,bridgeland2007stability}.
An exact expression for $Z$ (including all instanton corrections) in geometric phases of local and compact Calabi-Yau manifolds was proposed by Hosono in~\cite{Hosono:2004jp} and used to define an integral structure on the $A$-model compatible with mirror symmetry \cite{Iritani:2009ab,Katzarkov:2008hs}.

In the following we review the necessary mathematical framework to write~\eqref{ccleading} for branes with compact support on local toric geometries and compare it with localization calculations for the projection of GLSM branes $(\mathcal{B},L)$ into the Higgs branch. All our analysis will concern only the leading term of $Z$ and we ignore the instanton corrections, leaving them for future work.

\subsection{K-theory and cohomology of toric varieties [review]}

We need to define the K-group $K_{0}(X_{\zeta})$, the cohomology $H^{*}(X)$ and the Chern character $\ch\colon K_{0}(X_{\zeta})\otimes \mathbb{Q}\rightarrow H^{\text{even}}(X_{\zeta})$ when $X_{\zeta}$ is toric.\footnote{In most of this section we will be working with K-theory with $\mathbb{Q}$ or $\mathbb{C}$ coefficients, hence $\ch$ is an isomorphism. Of course, when discussing questions such as integrality structure of the central charge map, one needs to consider $\mathbb{Z}$-valued K-theory} Since we are dealing with $X_{\zeta}$ toric and noncompact, we must make a distinction between objects with compact support and the ones with non-compact support.
We will see that we can make sense of the central charge geometrically for B-branes with compact support.\footnote{It would also be interesting to define K-group, cohomology, and Chern character equivariant with respect to isometries of~$X_\zeta$, and define the central charge of some non-compact branes, to compare it with the hemisphere partition function regularized by the addition of R-charges.}
For this purpose we have to define in addition the compact K-group $K^{\text{c}}_{0}(X_{\zeta})$ and cohomology $H^{*}_{\text{c}}(X_{\zeta})$ which are modules of their noncompact counterparts, and the compact Chern character $\ch^{\text{c}}\colon K_{0}^{\text{c}}(X_{\zeta})\rightarrow H^{\text{even}}_{\text{c}}(X_{\zeta})$.

Our starting point is the data of a fan~$\Sigma$, namely a consistent collection of rational polyhedral cones, on a lattice $N$ of rank~$d$.
It defines a toric variety~$\mathbb{P}_{\Sigma}$.
For a review and conventions used here, the reader can consult~\cite{Hori:2003ic}.
We denote by $\Sigma(1)=\{v_{1},\ldots,v_{n}\}$ the rays of~$\Sigma$.
For any $J\subset I=\{1,\ldots, n\}$ we let $\sigma_J$ be the cone spanned by $\{v_j\mid j\in J\}$.
For $\sigma\in\Sigma$, we define $\Star(\sigma)=\{ \sigma' \in \Sigma \mid \sigma \subseteq \sigma'\}$.

\paragraph{Cohomology.}

The untwisted sector of the cohomology of~$\mathbb{P}_{\Sigma}$ is generated by classes $D_{i}$ of cohomological degree~$2$ that correspond to toric divisors of~$\mathbb{P}_{\Sigma}$, modulo relations:
\begin{equation}
  H^{*}_{0}(\mathbb{P}_{\Sigma})=\frac{\mathbb{C}[D_{i}]_{i\in I}}{\bigl<\{\sum_{i}m(v_{i})D_{i}\mid m\in N^{*}\},\mathcal{I}_{\text{SR}}\bigr>} ,
\end{equation}
where the Stanley-Reisner (SR) ideal $\mathcal{I}_{\text{SR}}$ is spanned by products $\prod_{i\in J}D_{i}$ for every $J\subset I$ that does not span a cone in~$\Sigma$.
There are also twisted sectors,\footnote{For example in partial resolutions of singularities, there are usually twisted sectors.} labelled by $\nu\in\operatorname{Box}(\Sigma)$, namely by lattice points $\nu\in N$ that can be written for some cone $\sigma_J\in\Sigma$ as $\sum_{i\in J}\nu_i v_i$ with all $\nu_i\in[0,1)$.
The twisted sector cohomology~$H^{*}_{\nu}$ is the cohomology of the toric substack described by the quotient fan $\Sigma_\nu\coloneqq\Sigma/\sigma(\nu)$, whose rays are the rays in $S_{\nu}\coloneqq\Star(\sigma(\nu))\setminus\sigma(\nu)$, where $\sigma(\nu)\in\Sigma$ is the minimal cone in~$\Sigma$ that contains~$\nu$.
It has the presentation
\begin{equation}
  H^*_{\nu}(\mathbb{P}_{\Sigma})= \frac{\mathbb{C}[\overline{D}_{i}]_{v_i\in S_{\nu}(1)}}{\bigl<\bigl\{\sum_{i\in S_{\nu}(1)}m(v_{i})\overline{D}_{i}\bigm| m\in \Ann(v_{i}\in\sigma(\nu))\bigr\},\mathcal{I}^{\nu}_{\text{SR}}\bigr>} .
\end{equation}
The SR ideal $\mathcal{I}^{\nu}_{\text{SR}}$ in the $\nu$ twisted sector is spanned by $\prod_{i\in J}\overline{D}_{i}$ for $\sigma_J$ not a cone in $\Star(\sigma(\nu))$.
The full cohomology is the direct sum of twisted sector cohomologies, where $\nu=0$ is the untwisted sector.

\paragraph{Compactly-supported cohomology.}

The compact cohomology is also a direct sum over $\nu\in\operatorname{Box}(\Sigma)$.
Its untwisted sector is generated as an $H^{*}_0(\mathbb{P}_{\Sigma})$-module by symbols~$F_{\{i\}}$ of cohomological degree~$2$ corresponding to \emph{compact} toric divisors of~$\mathbb{P}_{\Sigma}$.
Relations are simpler in terms of more general elements~$F_{J}$, where $J$ labels cones of~$\Sigma$ whose interior is in the interior of~$\Sigma$ (we denote the collection of such cone interiors by~$\Sigma^{\circ}$):
\begin{equation}
  H^{*}_{\text{c},0}(\mathbb{P}_{\Sigma}) = \frac{\bigoplus_{\sigma_{J}^{\circ}\in \Sigma^{\circ} } \bigl( H^{*}_0(\mathbb{P}_{\Sigma})\,F_{J} \bigr)}{\langle H_{1},H_{2}\rangle} ,
\end{equation}
where the two types of relations are given by
\begin{equation}
  \begin{alignedat}{4}
    H_{1}&=\bigl\{ D_{i}F_{J}=F_{J\cup \{i\}} & {} \bigm| i\not\in J \text{ and } \sigma_{J\cup \{i\}}^{\circ}\in \Sigma^{\circ}\bigr\} , \\
    H_{2}&=\bigl\{ D_{i}F_{J}=0 & {} \bigm| i\not\in J \text{ and } \sigma_{J\cup \{i\}}^{\circ}\not\in \Sigma^{\circ}\bigr\} .
  \end{alignedat}
\end{equation}
The twisted sectors $H^{*}_{\text{c},\nu}(\mathbb{P}_{\Sigma})$ have a similar description as for the untwisted case, just replacing the commutative ring by $H^{*}_{\nu}(\mathbb{P}_{\Sigma})$ and the fan by the quotient fan $\Sigma_{\nu}\coloneqq\Sigma/\sigma(\nu)$. Explicitly,
\begin{gather}
  H^{*}_{\text{c},\nu}(\mathbb{P}_{\Sigma})= \frac{\bigoplus_{\sigma_{J}^{\circ}\in\Sigma_{\nu}^{\circ}}\bigl(H^{*}_{\nu}(\mathbb{P}_{\Sigma})\,\overline{F}_{J}\bigr)}{\langle H^{\nu}_{1},H^{\nu}_{2}\rangle} ,
  \\
  \begin{alignedat}{4}
    H^{\nu}_{1} & = \bigl\{ \overline{D}_{i}\overline{F}_{J}=\overline{F}_{J\cup \{i\}} & {} \bigm| i\not\in J \text{ and } \sigma_{J\cup \{i\}}^{\circ}\in \Sigma_{\nu}^{\circ}\bigr\} , \\
    H^{\nu}_{2} & = \bigl\{ \overline{D}_{i}\overline{F}_{J}=0 & {} \bigm| i\not\in J \text{ and } \sigma_{J\cup \{i\}}^{\circ}\not\in \Sigma_{\nu}^{\circ}\bigr\} .
  \end{alignedat}
\end{gather}

\paragraph{K-theory and compactly-supported K-theory.}

Next, K-theory (with complex coefficients) is given by the following ring~\cite{borisov2005k,borisov2015applications,borisov2005orbifold}:\footnote{The elements $1$ and~$R_i$ are K-theory classes of the structure sheaf and of its twist by a toric divisor, so $1-R_i^{-1}$ is the K-theory class of the structure sheaf of that toric divisor.  The ideal~$\mathcal{I}_K$, like the SR ideal in cohomology, has one relation for each empty intersection of toric divisors.}
\begin{equation}\label{K0toric}
  K_{0}(\mathbb{P}_{\Sigma})= \frac{\mathbb{C}[R_i^{\pm 1}]_{i\in I}}{\Bigl<\Bigl\{ 1-\prod_{i\in I}R_{i}^{m(v_{i})}\Bigm|m\in N^{*}\Bigr\},\mathcal{I}_{K}\Bigr>}, \qquad
  \mathcal{I}_{K}=\biggl<\prod_{i\in J}(1-R_{i})\biggm| \sigma_J\not\in \Sigma \biggr> .
\end{equation}
The compact version $K^{\text{c}}_{0}(\mathbb{P}_{\Sigma})$ is the quotient of a free $K_{0}(\mathbb{P}_{\Sigma})$-module generated by the symbols~$G_{J}$ with $\sigma_{J}^{\circ}\in \Sigma^{\circ}$ by the relations
\begin{equation}
  \begin{alignedat}{4}
    & \bigl\{ (1-R_{i}^{-1})G_{J}=G_{J\cup \{i\}} & {} \bigm| i\not\in J \text{ and } \sigma_{J\cup \{i\}}^{\circ}\in \Sigma^{\circ}\bigr\} , \\
    & \bigl\{ (1-R_{i}^{-1})G_{J}=0 & {} \bigm| i\not\in J \text{ and } \sigma_{J\cup \{i\}}^{\circ} \not\in \Sigma^{\circ}\bigr\} .
  \end{alignedat}
\end{equation}
Note that cohomologies are sums over sectors $\nu\in\operatorname{Box}(\Sigma)$ while K-theories are not given in such a form.

\paragraph{Chern characters.}

Finally we have the Chern character maps, which give isomorphisms (here we are always considering K-theory and cohomology with complex coefficients):
\begin{equation}
  \ch\colon K_{0}(\mathbb{P}_{\Sigma})\rightarrow H^{*}(\mathbb{P}_{\Sigma}),\qquad
  \ch^{\text{c}}\colon K^{\text{c}}_{0}(\mathbb{P}_{\Sigma})\rightarrow H^{*}_{\text{c}}(\mathbb{P}_{\Sigma}) .
\end{equation}
Explicitly, the Chern character~$\ch$ is given by its projection $\ch_{\nu}$ to each sector~$\nu$:
\begin{equation}\label{chgamma}
  \ch_{\nu}(R_{i})=
  \begin{cases}
    1, & \sigma_{\{i\}}\not\in \Star(\sigma(\nu)) , \\
    e^{\overline{D}_{i}}, & \sigma_{\{i\}}\in S_{\nu} = \Star(\sigma(\nu)) \setminus \sigma(\nu) , \\
    e^{2\pi i \nu_{i}}\prod_{v_j\not\in \sigma(\nu)}\ch_{\nu}(R_{j})^{m_{i}(v_{j})}, & \sigma_{\{i\}}\in\sigma(\nu) ,
  \end{cases}
\end{equation}
where $\nu_i$ is the coefficient of~$v_i$ in the decomposition of $\nu$ as a linear combination of rays in~$\sigma(\nu)$, and
where $m_i\in N^*$ is such that $m_i(v_i)=-1$ and $m_i(v_j)=0$ for all other $v_j\in\sigma(\nu)$.
The compactly-supported Chern character is given by
\begin{equation}
  \begin{gathered}
    \ch^{\text{c}}_{\nu}\Bigl(\prod_{i}R_{i}^{k_{i}}G_{J}\Bigr)
    = \biggl( \prod_{i}\ch_{\nu}(R_{i})^{k_{i}} \biggr)  \ch^{\text{c}}_{\nu}(G_{J})
    \\
    \ch^{\text{c}}_{\nu}(G_{J}) =
    \begin{cases}
      0 & \text{for } \sigma_J\not\in \Star(\sigma(\nu)), \\
      \biggl(\prod\limits_{i\in J,v_i\not\in \sigma(\nu)}\frac{1-e^{-\overline{D}_{i}}}{\overline{D}_{i}}\biggr)
      \biggl(\prod\limits_{i\in J,v_i\in \sigma(\nu)}(1-\ch_{\nu}(R_{i})^{-1})\biggr)
      \overline{F}_{J\cap S_{\nu}}, & \text{for } \sigma_J \in \Star(\sigma(\nu)),
    \end{cases}
  \end{gathered}
\end{equation}
where $J\cap S_{\nu}$ denotes the intersection of $J$ with indices of rays in~$S_{\nu}$.
The next-to-last factor is defined by Taylor expanding $(1-e^{-x})/x=1-x/2+\dots$ then replacing $x$ by~$\overline{D}_{i}$.

\paragraph{Integration.}

One last thing we need is the integration map
\begin{equation}\label{integration-map}
  \int\colon H^{*}_{\text{c}}(\mathbb{P}_{\Sigma})\longrightarrow \mathbb{C} .
\end{equation}
It is a sum over sectors.  The contribution from a sector~$\nu$ is defined by
\begin{equation}
  \int \overline{F}_{J}=\frac{1}{\abs{\Vol(J)}},\qquad \text{for } |J|=\rank(N_{\nu})
\end{equation}
and zero otherwise,
where $\abs{\Vol(J)}$ is the index of the lattice spanned by $J$ inside the quotient lattice $N_{\nu}\coloneqq N/\Span(\sigma(\nu))$.

\subsection{K-theory and cohomology in Hirzebruch-Jung models}

In order to apply the machinery reviewed in the previous subsection, we first need to connect the fan description with the GLSM data. This a well known construction, with slight subtleties when the gauge group has a discrete factor.

\paragraph{Fan from GLSM.}

Part of the data of an abelian GLSM with \emph{connected} gauge group $U(1)^r$ is an $r\times d$ matrix of charges~$Q_{\alpha}^{j}$.
This matrix gives relations between $d$ vectors on the lattice $N$, of rank $d-r$.
This lattice can be constructed as a quotient
\begin{equation}
  N = \ZZ^d \bigm/ \Span_{\ZZ} \bigl( \{Q_\alpha\mid 1\leq\alpha\leq r\} \bigr) .
\end{equation}
Basis vectors in $\ZZ^d$ project to vectors $S\coloneqq\{v_{1},\ldots,v_{d}\}\subset N$ whose relations are given by the charge matrix as $\sum_{j=1}^d Q_\alpha^j v_j = 0$.
(For calculations, one can choose $d-r$ linearly independent vectors~$v_j$ as a basis, then express all others as rational combinations\footnote{It may not be possible to find a subset of $d-r$ vectors $v_i$ whose \emph{integer} span is~$N$.  For instance if $v_1=(1,0)$, $v_2=(2,2)$, $v_3=(0,5)$ inside $N=\ZZ^2$ then $\Span_{\ZZ}(v_1,v_2)=\ZZ\times 2\ZZ\subsetneq N$, $\Span_{\ZZ}(v_2,v_3)=2\ZZ\times\ZZ\subsetneq N$, $\Span_{\ZZ}(v_1,v_3)=\ZZ\times 5\ZZ\subsetneq N$.} thereof using the relations: this realizes~$N$ as a lattice in~$\QQ^{d-r}$.)
The dual lattice $N^*=\Hom_{\ZZ}(N,\ZZ)$ embeds naturally in $\ZZ^d$ through
\begin{equation}
  N^*\ni m \mapsto \bigl(m(v_1),\dots,m(v_d)\bigr)\in\ZZ^d .
\end{equation}
What characterizes $N^*$ inside $\ZZ^d$ is that $m(\sum_{j=1}^d Q_\alpha^j v_j)=0$.
In other words, an element $(a_1,\dots,a_d)\in\ZZ^d$ is in~$N^*$ if and only if $\sum_{j=1}^d Q_\alpha^j a_j = 0$ for all~$\alpha$.
Physically, this is the condition for $\prod_{i=1}^d X_i^{a_i}$ to be gauge-invariant.

The construction extends to \emph{arbitrary} abelian GLSMs by defining $N^*\subset\ZZ^d$ as the set of powers $(a_1,\dots,a_d)$ such that $\prod_{i=1}^d X_i^{a_i}$ is gauge-invariant, then defining $N\coloneqq N^{**}$ as the lattice dual to~$N^*$.
Then $N$ is the weight lattice of the flavour symmetry group, since each element of~$N$ maps $N^*$ (gauge-invariant operators) to $\ZZ$ (their charge under the flavour symmetry).
In that approach, the vectors $v_i\in N$ are defined to map $(a_1,\dots,a_d)\mapsto a_i$.
Their integer span is a sublattice of~$N$, not necessarily all of~$N$ like for a connected gauge group.\footnote{The gauge group is then $G=\Hom(\ZZ^d/N^*,U(1))$, the group of phase rotations of the $X_i$ that leaves invariant all gauge-invariant monomials.}

The other important piece of data is the D-term equations.
For a choice of $\zeta$ in the interior of a K\"ahler cone, the D-term equations have no solution when certain subsets of the chirals vanish simultaneously.
The fan $\Sigma_{\zeta}$ consists of the cones~$\sigma_J$, for $J\subset\{1,\dots,d\}$,
such that D-term equations have solutions for which $X_j=0$ for all $j\in J$ simultaneously.
In particular, rays of~$\Sigma_{\zeta}$ are labeled by chirals that may vanish somewhere on the Higgs branch.
The fan only depends on the K\"ahler cone in which $\zeta$~lies.

\paragraph{Fan for Hirzebruch-Jung models.}

We used two GLSMs to study the quantum geometry of the Hirzebruch-Jung resolutions. The two are simply related by a change of basis on the $U(1)^{r}$ weights of the chiral fields. Since this change of variables was characterized by a transformation matrix of determinant not~$1$, the gauge groups of the corresponding models differ by a nontrivial finite quotient. This is just and artifact of the presentation and since the two charge matrices $Q_{\alpha}^{j}$ and $n(C^{-1})_{\alpha}^{\beta}Q_{\beta}^{j}$ are related by an invertible linear transformation, the constraints on~$v_j$ are the same either way:
\begin{equation}
  \sum_j Q_\alpha^j v_j=0\quad\forall\alpha
  \iff
  \sum_j \biggl( \sum_\beta n(C^{-1})_\alpha^\beta Q_\beta^j\biggr) v_j=0\quad\forall\alpha .
\end{equation}
They take the simplest form in basis~II\@: $n v_\alpha = p_\alpha v_0 + q_\alpha v_{r+1}$ for $1\leq\alpha\leq r$.
This motivates us to choose coordinates where $v_0=(n,0)$ and $v_{r+1}=(0,n)$ so that in that basis,
\begin{equation}
  S=\{v_{j}\}_{j=0}^{r+1}\qquad v_{j}\coloneqq(p_{j},q_{j}) = (q_jp-d_{1j}n,q_j) = q_jv_1-d_{1j}v_0
\end{equation}
where we used $pq_j-p_j=p_1q_j-q_1p_j=nd_{1j}$.
Since the gauge group is connected, we conclude $N=\Span_{\ZZ}(\{v_j\})=\Span_{\ZZ}(v_0,v_1)$.
We easily compute the dual basis, hence the dual lattice, finding
\begin{equation}
  N=\Span_{\mathbb{Z}}\bigl((n,0),(p,1)\bigr) , \qquad
  N^{*}= \Span_{\mathbb{Z}}\Bigl( \Bigl(\frac{1}{n},\frac{-p}{n}\Bigr), (0,1)\Bigr).
\end{equation}
In particular, $N$ has index~$n$ in~$\ZZ^2$.
Note in addition that $(q_j/n,-p_j/n)\in N^*$ for all $0\leq j\leq r+1$ because
\begin{equation}
  \Bigl(\frac{q_j}{n},\frac{-p_j}{n}\Bigr)\cdot(n,0) = q_j\in\ZZ \quad\text{and}\quad
  \Bigl(\frac{q_j}{n},\frac{-p_j}{n}\Bigr)\cdot(p,1) = \frac{1}{n} (p_1 q_j-p_jq_1) = d_{1j}\in\ZZ .
\end{equation}

To find the fan, we are instructed by the previous discussion to look at the deleted sets imposed by the D-term equations.
As we saw in \autoref{ssec:HiggsGeom}, a phase of the Hirzebruch-Jung model is characterized by the set $A=\{\alpha_1,\dots,\alpha_\ell\}\subset\{1,\dots,r\}$ of blown up divisors.
Let $\alpha_0=0$ and $\alpha_{\ell+1}=r+1$ for simplicity.
The fan~$\Sigma_\zeta$ only depends on~$A$ so we denote it~$\Sigma_A$.
The deleted sets are hyperplanes $\{X_\alpha=0\}$ for each $\alpha\in\{1,\dots,r\}\setminus A$ and intersections $\{X_{\alpha_i}=X_{\alpha_j}=0\}$ for each $0\leq i,j\leq\ell+1$ such that $\abs{i-j}\geq 2$.
The rays of~$\Sigma_A$ are thus all $v_{\alpha_i}$, $0\leq i\leq\ell+1$, and its $2$-dimensional cones are given by $\{\alpha_i,\alpha_{i+1}\}$ for $0\leq i\leq\ell$.
Some examples of these fans are given in \autoref{fig:latticefan}.

\begin{figure}[t]
  \centering
  \begin{tikzpicture}[scale=.5]
    \fill[opacity=.5, gray] (0,0) -- (5,0) -- (6,3) -- (1,3);
    \fill[opacity=.5, gray] (0,0) -- (1,3) -- (1,8) -- (0,5);
    \draw (2,1) circle (.2);
    \draw (4,2) circle (.2);
    \draw (-.5,0) -- (8.5,0);
    \draw (0,-.5) -- (0,8.5);
    \draw[-{stealth}, very thick] (0,0) -- (5,0) node [below] {$\scriptstyle v_0$};
    \draw[-{stealth}, very thick] (0,0) -- (1,3) node [above right,inner sep=2pt] {$\scriptstyle v_2$};
    \draw[-{stealth}, very thick] (0,0) -- (0,5) node [left] {$\scriptstyle v_3$};
    \foreach \x in {0,...,8} {
      \foreach \y in {0,...,8} {
        \pgfmathtruncatemacro{\z}{mod(\x-2*\y,5)}
        \ifnum\z=0\relax
        \filldraw(\x,\y) circle (.1);
        \fi
      }
    }
  \end{tikzpicture}
  \quad
  \begin{tikzpicture}[scale=.5]
    \fill[opacity=.5, gray] (0,0) -- (5,0) -- (7,1) -- (2,1);
    \fill[opacity=.5, gray] (0,0) -- (2,1) -- (3,4) -- (1,3);
    \fill[opacity=.5, gray] (0,0) -- (1,3) -- (1,8) -- (0,5);
    \draw (-.5,0) -- (8.5,0);
    \draw (0,-.5) -- (0,8.5);
    \draw[-{stealth}, very thick] (0,0) -- (5,0) node [below] {$\scriptstyle v_0$};
    \draw[-{stealth}, very thick] (0,0) -- (2,1) node [above right,inner sep=1pt] {$\scriptstyle v_1$};
    \draw[-{stealth}, very thick] (0,0) -- (1,3) node [above] {$\scriptstyle \:\:\:\:v_2$};
    \draw[-{stealth}, very thick] (0,0) -- (0,5) node [left] {$\scriptstyle v_3$};
    \foreach \x in {0,...,8} {
      \foreach \y in {0,...,8} {
        \pgfmathtruncatemacro{\z}{mod(\x-2*\y,5)}
        \ifnum\z=0\relax
        \filldraw(\x,\y) circle (.1);
        \fi
      }
    }
  \end{tikzpicture}
  \caption{\label{fig:latticefan}Lattice~$N$ and fan~$\Sigma_A$ of two resolutions of $\CC^2/\ZZ_{5(2)}$.  The $\operatorname{Box}$ of each two-dimensional cone in~$\Sigma_A$ is shaded in gray.  Left: the resolution with divisor $E_2$ blown up has two twisted sectors (two circled points); these twisted sectors are at the orbifold point $E_2\cap E_0$.  Right: the full resolution has no twisted sector.}
\end{figure}

\paragraph{Cohomology for Hirzebruch-Jung models.}

Twisted sectors are labeled by $\nu\in\operatorname{Box}(\Sigma_A)$ so let us find $\operatorname{Box}(\sigma)$ for all cones.
First, consider one-dimensional cones: the element $(q_{\alpha+1}/n,-p_{\alpha+1}/n)\in N^*$ maps $v_\alpha$ to $(p_\alpha q_{\alpha+1}-q_\alpha p_{\alpha+1})/n = d_{\alpha(\alpha+1)} = 1$, so $v_\alpha$ is not a multiple of any other element of~$N$, hence $\operatorname{Box}(v_\alpha)=\{0\}$.
Next, for two-dimensional cones we count
\begin{equation}
  \abs[\big]{\operatorname{Box}(\{v_\alpha,v_\beta\})}
  = \abs*{\frac{N}{\Span_{\ZZ}(v_\alpha,v_\beta)}}
  = \frac{\abs{\ZZ^2/\Span_{\ZZ}(v_\alpha,v_\beta)}}{\abs{\ZZ^2/N}}
  = \frac{\det(v_\alpha,v_\beta)}{n} = d_{\alpha\beta} ,
\end{equation}
so each two-dimensional cone $\{v_{\alpha_i},v_{\alpha_{i+1}}\}$ contributes $d_{\alpha_i\alpha_{i+1}}-1$ twisted sectors.
(This counting is consistent with the orbifold group $\ZZ_{d_{\alpha_i\alpha_{i+1}}}$ that we found geometrically earlier.)
We deduce that for any non-zero $\nu\in\operatorname{Box}(\Sigma_A)$ the smallest cone $\sigma(\nu)$ containing $\nu$ is a two-dimensional cone.
The quotient fan $\Sigma_A/\sigma(\nu)$ is then trivial (it has no ray), and its cohomology is~$\CC$.
The full cohomology is thus
\begin{equation}\label{cohomology}
  H^{*}(\mathbb{P}_{\Sigma_A})
  = H^{*}_{0}(\mathbb{P}_{\Sigma_A})\oplus\bigoplus_{\nu\in\operatorname{Box}(\Sigma_A)\setminus\{0\}} H^0_\nu(\mathbb{P}_{\Sigma_A}) , \qquad H^0_\nu(\mathbb{P}_{\Sigma_A}) = \CC ,
\end{equation}
where the $\sum_{i=0}^{\ell}(d_{\alpha_i\alpha_{i+1}}-1)$ copies of~$\CC$ are all in cohomological degree~$0$.
The untwisted sector of the cohomology ring is
\begin{equation}
  H^{*}_{0}(\mathbb{P}_{\Sigma_A})=
  \frac{\mathbb{C}[D_{\alpha_i}\mid 0\leq i\leq\ell+1]}{\bigl<
    \sum_{i=0}^{\ell+1}q_{\alpha_i}D_{\alpha_i}, \ \sum_{i=0}^{\ell+1}p_{\alpha_i}D_{\alpha_i}, \
    \bigl\{D_{\alpha_i}D_{\alpha_j}\bigm| 2\leq\abs{j-i}\bigr\} \bigr>} .
\end{equation}
Since $d_{\alpha\beta}=(p_\alpha q_\beta-p_\beta q_\alpha)/n$ The linear relations imply $\sum_{i=0}^{\ell+1} d_{\beta\alpha_i}D_{\alpha_i}=0$ for all~$\beta$.
In addition, we can show by induction on~$i$ that all $D_{\alpha_i}^2=D_{\alpha_i}D_{\alpha_{i+1}}=0$.
The induction step is done by multiplying the linear relations by $D_{\alpha_i}$ and using the quadratic relations, which gives (after using the induction hypothesis)
\begin{equation}
  \begin{aligned}
    p_{\alpha_i} D_{\alpha_i}^2 + p_{\alpha_{i+1}} D_{\alpha_i} D_{\alpha_{i+1}} & = 0 \\
    q_{\alpha_i} D_{\alpha_i}^2 + q_{\alpha_{i+1}} D_{\alpha_i} D_{\alpha_{i+1}} & = 0
  \end{aligned}
\end{equation}
and since $p_{\alpha_i} q_{\alpha_{i+1}} - p_{\alpha_{i+1}} q_{\alpha_i} = n d_{\alpha_i\alpha_{i+1}} \neq 0$ we learn $D_{\alpha_i}^2=D_{\alpha_i}D_{\alpha_{i+1}}=0$.
Altogether, any product $D_\alpha D_\beta = 0$ so
\begin{equation}\label{cohomology0}
  H^{*}_{0}(\mathbb{P}_{\Sigma_A})=
  \frac{\CC\oplus\bigoplus_{i=0}^{\ell+1}\CC D_{\alpha_i}}{\bigl<
    \sum_{i=0}^{\ell+1}d_{\beta\alpha_i}D_{\alpha_i}, 0\leq\beta\leq r+1\bigr>} .
\end{equation}
This is expected since the top cohomology vanishes for noncompact surfaces: $H^{4}(\mathbb{P}_{\Sigma_A})=0$.

The complexified K\"ahler class is $\tau=\sum_{\alpha=1}^r \tau_\alpha\eta^\alpha$ (in all phases) where $\eta^\alpha$, $1\leq\alpha\leq r$, are defined by $\eta^\alpha= - \frac{1}{n} \sum_{\beta=1}^r d_{0\min(\alpha,\beta)}d_{\max(\alpha,\beta),r+1} D_\beta$ so that $\sum_{\alpha=1}^r Q^i_\alpha \eta^\alpha = D_i$ for all~$i$.

\subsection{Zero-instanton central charges in Hirzebruch-Jung models}

If $\mathcal{F}$ is a compactly-supported sheaf (or a compact complex) in a noncompact variety~$X$,
denote by $\mathcal{F}$ as well its K-theory class in~$K_{0}^{\text{c}}$.
Then, we define its central charge by
\begin{equation}\label{cc-ours}
  Z(\mathcal{F})= \int e^{\tau}\hbar^{c_{1}(X)/(2\pi i)}\widehat{\Gamma}(TX)\ch^{\text{c}}(\mathcal{F}) .
\end{equation}
The gamma class $\widehat{\Gamma}(TX)$, first Chern class $c_{1}(X)$ and complexified K\"ahler class~$\tau$ are elements of $H^{*}(X,\mathbb{C})$, hence they act on $\ch^{\text{c}}(\mathcal{F})\in H^{*}_{\text{c}}(X)$.
This yields another element of~$H^{*}_{\text{c}}(X)$, for which the integral map makes sense.

\paragraph{Brane-independent factors: Chern character and Gamma class.}

The Chern character of the tangent bundle $T\mathbb{P}_{\Sigma_A}$ can be determined from the Euler sequence
\begin{equation}
  0 \rightarrow \cO^{r} \rightarrow \bigoplus_{j=0}^{r+1}\cO(D_{j}) \rightarrow T\mathbb{P}_{\Sigma_A}\rightarrow 0 .
\end{equation}
In K-theory this sequence means that the K-theory class of $T\mathbb{P}_{\Sigma_A}$ is $-r+\sum_{j=0}^{r+1}R_j$.
The Chern character in different twisted sectors is found from~\eqref{chgamma} to be
\begin{equation}
  \ch_0(T\mathbb{P}_{\Sigma_A}) = 2 + \sum_{i=0}^{\ell+1} D_{\alpha_i} , \qquad
  \ch_\nu(T\mathbb{P}_{\Sigma_A}) = e^{2\pi i\nu_1} + e^{2\pi i\nu_2} \text{ for } \nu\neq 0 ,
\end{equation}
where $0\leq\nu_1,\nu_2<1$ are coordinates: $\nu = \nu_1 v_{\alpha_k}+\nu_2 v_{\alpha_{k+1}}$ for $\nu \in \operatorname{Box}(v_{\alpha_k},v_{\alpha_{k+1}})\setminus\{0\}$.

Since there is no (non-compact) cohomology beyond degree~$2$ in Hirzebruch-Jung models, the gamma class is exactly $\widehat{\Gamma} = 1 + \frac{\gamma}{2\pi i} c_1$.
We find
\begin{equation}
  \widehat{\Gamma}_0(T\mathbb{P}_{\Sigma_A}) = 1 + \frac{\gamma}{2\pi i} \sum_{i=0}^{\ell+1} D_{\alpha_i} , \qquad
  \widehat{\Gamma}_\nu(T\mathbb{P}_{\Sigma_A}) = 1 \text{ for } \nu\neq 0 .
\end{equation}

\paragraph{Central charge of (fractional) D0-brane.}

We consider here branes with K-theory class of the form
\begin{equation}\label{fractionalD0Ktheorycomplicated}
  \cF = \biggl( \prod_i R_i^{k_i} \biggr) G_{\alpha_j\alpha_{j+1}}
\end{equation}
for some $0\leq j\leq\ell$.
Note that $G_{\alpha_j\alpha_{j+1}}=(1-R_{\alpha_j}^{-1})G_{\alpha_{j+1}}$, where $1-R_{\alpha_j}^{-1}$ is the K-theory class of the structure sheaf of the divisor~$E_{\alpha_j}$, and likewise $G_{\alpha_j\alpha_{j+1}}=(1-R_{\alpha_{j+1}}^{-1})G_{\alpha_j}$, so these branes are supported on the intersection $E_{\alpha_j}\cap E_{\alpha_{j+1}}$.
Near this intersection point the geometry is a $\ZZ_{d_{\alpha_i\alpha_{i+1}}}$ orbifold of $\CC^2$,
which is just a smooth point if $d_{\alpha_i\alpha_{i+1}}=1$ or equivalently $\alpha_{i+1}=\alpha_i+1$.
These branes are thus (fractional) D0-branes.

The relations defining compact K-theory state that $R_i G_{\alpha_j\alpha_{j+1}} = G_{\alpha_j\alpha_{j+1}}$ for any $i\neq\alpha_j,\alpha_{j+1}$.
In addition, \eqref{K0toric}~with $m=(q_\beta,-p_\beta/n)$ and $\beta=1+\alpha_{j+1}$ and $\beta=\alpha_{j+1}$ gives
\begin{equation}
  R_{\alpha_{j+1}} = R_{\alpha_j}^{-d_{\alpha_j,1+\alpha_{j+1}}} \times (\text{other }R_i) , \qquad
  R_{\alpha_j}^{d_{\alpha_j,\alpha_{j+1}}} = (\text{other }R_i)
\end{equation}
where ``other $R_i$'' means a monomial in the $R_i$ with $i\neq\alpha_j,\alpha_{j+1}$.
The K-theory class~\eqref{fractionalD0Ktheorycomplicated} thus reduces to
\begin{equation}
  \cF_{j,s} \coloneqq (R_{\alpha_j})^s G_{\alpha_j\alpha_{j+1}} , \qquad s\in\ZZ_{d_{\alpha_j\alpha_{j+1}}} .
\end{equation}

To confirm the identification of these branes as fractional D0-branes, let us check that the sum of all of them gives a K-theory class independent of~$j$, namely check that a full D0 brane can move from one intersection $E_{\alpha_{j-1}}\cap E_{\alpha_j}$ to the next.
We compute
\begin{equation}
  \sum_{s=1}^{d_{\alpha_{j-1}\alpha_j}} \cF_{j-1,s}
  = \sum_{s=1}^{d_{\alpha_{j-1}\alpha_j}} (R_{\alpha_{j-1}})^s (1-R_{\alpha_{j-1}}^{-1}) G_{\alpha_j}
  = (R_{\alpha_{j-1}})^{d_{\alpha_{j-1}\alpha_j}} G_{\alpha_j} - G_{\alpha_j} .
\end{equation}
On the other hand, from $R_{\alpha_{j+1}} G_{\alpha_j\alpha_{j+1}} = d_{\alpha_j}^{-d_{\alpha_j,1+\alpha_{j+1}}} G_{\alpha_j\alpha_{j+1}}$ and
$\gcd(-d_{\alpha_j,1+\alpha_{j+1}},d_{\alpha_j\alpha_{j+1}})=\gcd(d_{\alpha_{j+1},1+\alpha_{j+1}},d_{\alpha_j\alpha_{j+1}})=\gcd(1,d_{\alpha_j\alpha_{j+1}})=1$, we learn that the set of K-theory classes $\cF_{j,s}$ is the same as the set of $(R_{\alpha_{j+1}})^s G_{\alpha_j\alpha_{j+1}}$, so
\begin{equation}
  \sum_{s=1}^{d_{\alpha_j\alpha_{j+1}}} \cF_{j,s}
  = \sum_{s=1}^{d_{\alpha_j\alpha_{j+1}}} (R_{\alpha_{j+1}})^s (1-R_{\alpha_{j+1}}^{-1}) G_{\alpha_j}
  = (R_{\alpha_{j+1}})^{d_{\alpha_j\alpha_{j+1}}} G_{\alpha_j} - G_{\alpha_j} .
\end{equation}
The two sums are equal because $\prod_{i=0}^{r+1} R_i^{d_{i\alpha_j}} = 1$ and $R_i G_{\alpha_j} = G_{\alpha_j}$ for $i\neq\alpha_{j-1},\alpha_j,\alpha_{j+1}$.

The compact Chern character of $\cF_{j,s}$ reads
\begin{equation}
  \begin{aligned}
    \ch^{\text{c}}_0(\cF_{j,s}) & = F_{\alpha_j\alpha_{j+1}} ,
    \\
    \ch^{\text{c}}_\nu(\cF_{j,s}) & = e^{2\pi i\nu_1 s} (1 - e^{-2\pi i\nu_1}) (1 - e^{-2\pi i\nu_2}) \overline{F}_{\emptyset} ,
    \qquad \nu\in\operatorname{Box}(v_{\alpha_j},v_{\alpha_{j+1}}) \setminus\{0\} ,
    \\
    \ch^{\text{c}}_\nu(\cF_{j,s}) & = 0, \qquad \nu\not\in\operatorname{Box}(v_{\alpha_j},v_{\alpha_{j+1}})
  \end{aligned}
\end{equation}
where $\nu=\nu_1v_{\alpha_j}+\nu_2v_{\alpha_{j+1}}$.

The untwisted sector contribution to the central charge is thus
\begin{equation}
  Z_0(\cF_{j,s}) = \int F_{\alpha_j\alpha_{j+1}} = \frac{1}{\abs{\Vol(\alpha_j,\alpha_{j+1})}} = \frac{1}{d_{\alpha_j\alpha_{j+1}}}
\end{equation}
which is independent of the charge~$s$ and matches the 0-instanton localization result.
On the other hand, each twisted sector gives a non-zero contribution
\begin{equation}
  Z_\nu(\cF_{j,s}) = \frac{1}{d_{\alpha_j\alpha_{j+1}}} e^{2\pi i\nu_1 s} (1 - e^{-2\pi i\nu_1}) (1 - e^{-2\pi i\nu_2}) .
\end{equation}
Preliminary computations (work in progress) suggest to identify these, up to a factor involving $\Gamma(\nu_1)\Gamma(\nu_2)$ and fractional powers of $e^{2\pi it_\alpha}$, with the residues at certain poles in the localization calculation.
These poles correspond physically to fractional instantons in the 2d gauge theory.
A similar identification is made for Landau-Ginzburg orbifolds in~\cite{fjrwhpf}.

\paragraph{Central charge of D2-brane.}

Now, we end the subsection by considering branes with K-theory class of the form
\begin{equation}\label{fractionalD2Ktheorycomplicated}
  \cF = \biggl( \prod_i R_i^{k_i} \biggr) G_{\alpha_j}
\end{equation}
for some $1\leq j\leq\ell$, namely branes supported on the $\alpha_j$-th exceptional divisor.

Relations defining compact K-theory state that $R_i G_{\alpha_j} = G_{\alpha_j}$ for any $i\neq\alpha_{j-1},\alpha_j,\alpha_{j+1}$, while $R_{\alpha_{j\pm 1}}^{-1} G_{\alpha_j} = G_{\alpha_j} - G_{\alpha_j\alpha_{j\pm 1}}$.
In addition, \eqref{K0toric} with $m=(q_\beta,-p_\beta/n)$ and $\beta=1+\alpha_j$ gives
\begin{equation}
  R_{\alpha_j} = \prod_{i\neq\alpha_j} R_i^{-d_{i,1+\alpha_j}} .
\end{equation}
The K-theory class~\eqref{fractionalD2Ktheorycomplicated} thus reduces to $G_{\alpha_j}$ plus fractional D0 branes studied previously.
Therefore, we focus on~$G_{\alpha_j}$.

The compact Chern character of $G_{\alpha_j}$ reads
\begin{equation}
  \begin{aligned}
    \ch^{\text{c}}_0(G_{\alpha_j}) & = (1-D_{\alpha_j}/2) F_{\alpha_j} ,
    \\
    \ch^{\text{c}}_\nu(G_{\alpha_j}) & = (1 - e^{-2\pi i\nu_j}) \overline{F}_{\emptyset} ,
    \qquad \nu\in\operatorname{Box}(v_{\alpha_{j-1}},v_{\alpha_j})\cup \operatorname{Box}(v_{\alpha_{j-1}},v_{\alpha_j}) \setminus\{0\} ,
    \\
    \ch^{\text{c}}_\nu(G_{\alpha_j}) & = 0, \qquad \nu\in\operatorname{Box}(v_{\alpha_i},v_{\alpha_{i+1}}), \ i\neq j,j-1
  \end{aligned}
\end{equation}
where $\nu=\nu_jv_{\alpha_j}+\nu_{j\pm 1}v_{\alpha_{j\pm 1}}$.

We know $D_i F_{\alpha_j} = 0$ for $i\neq\alpha_{j-1},\alpha_j,\alpha_{j+1}$.
Acting with $\sum_i d_{ki}D_i = 0$ on $F_{\alpha_i}$ we find
\begin{equation}
  0 = d_{k\alpha_{i-1}} F_{\alpha_{i-1}\alpha_i} + d_{k\alpha_i} D_{\alpha_i} F_{\alpha_i} + d_{k\alpha_{i+1}} F_{\alpha_i\alpha_{i+1}} .
\end{equation}
Applying this with $k=\alpha_i$, we find that all $d_{\alpha_i\alpha_{i+1}}F_{\alpha_i\alpha_{i+1}}$ are equal.
Denote by $\widehat{F}$ their common value.
It obeys $\int\widehat{F}=1$.
The same equation with $k=\alpha_{i-1}$ implies that $D_{\alpha_i} F_{\alpha_i} = - (d_{\alpha_{i-1}\alpha_{i+1}}/(d_{\alpha_{i-1}\alpha_i}d_{\alpha_i\alpha_{i+1}}))\widehat{F}$.

We find
\begin{equation}
  (1-D_{\alpha_j}/2) F_{\alpha_j}
  = F_{\alpha_j} + \frac{d_{\alpha_{j-1}\alpha_{j+1}}}{2d_{\alpha_{j-1}\alpha_j}d_{\alpha_j\alpha_{j+1}}} \widehat{F}
\end{equation}
and the expressions of $\sum_{\alpha=1}^r Q^i_\alpha \eta^\alpha F_{\alpha_j} = D_i F_{\alpha_j}$ in terms of $\widehat{F}$ let us work out that
\begin{equation}
  \eta^\alpha F_{\alpha_j} = \begin{cases}
    (d_{\alpha_{j-1}\alpha}/d_{\alpha_{j-1}\alpha_j})\widehat{F} & \text{for } \alpha_{j-1}\leq\alpha\leq\alpha_j \\
    (d_{\alpha\alpha_{j+1}}/d_{\alpha_j\alpha_{j+1}})\widehat{F} & \text{for } \alpha_j\leq\alpha\leq\alpha_{j+1} \\
    0 & \text{otherwise.}
  \end{cases}
\end{equation}
The untwisted sector contribution to the central charge is thus
\begin{equation}
  \begin{aligned}
    Z_0(G_{\alpha_j})
    = {} & \sum_{\alpha=\alpha_{j-1}}^{\alpha_j} \frac{d_{\alpha_{j-1}\alpha}}{d_{\alpha_{j-1}\alpha_j}} \tau_\alpha
    + \sum_{\alpha=\alpha_j+1}^{\alpha_{j+1}} \frac{d_{\alpha\alpha_{j+1}}}{d_{\alpha_j\alpha_{j+1}}} \tau_\alpha \\
    & + \frac{(\gamma+\log\hbar)(d_{\alpha_{j-1}\alpha_j}-d_{\alpha_{j-1}\alpha_{j+1}}+d_{\alpha_j\alpha_{j+1}})+i\pi d_{\alpha_{j-1}\alpha_{j+1}}}{2\pi i d_{\alpha_{j-1}\alpha_j}d_{\alpha_j\alpha_{j+1}}} .
  \end{aligned}
\end{equation}

We check that this matches perfectly with the localization result~\eqref{D2chargebigresult} (with $i,j,k\to\alpha_{j-1},\alpha_j,\alpha_{j+1}$), up to an overall factor $(\mathfrak{r}\Lambda)^{\hat{c}/2}$ that can be absorbed in a rescaling of the classes~$\eta_\beta$.  The identification is\footnote{Here we only write the leading term in the mirror map; there are instanton corrections of order $e^{2\pi i\tau_\alpha}$ for each~$\alpha$.  For this reason we only consider the pole at zero in the contour integral representation of the hemisphere partition function.  See \autoref{ssec:Bbraneabelian} for a discussion of some subtleties that distinguish the pole at zero from the Higgs branch contribution.}
\begin{equation}
  t^{\text{ren}}_{\beta}(\mathfrak{r}^{-1}) = -2\pi i\tau_{\beta}+(a_{\beta}-2)\log\hbar+O(e^{2\pi i\tau})
  \quad\text{and}\quad \hbar=\mathfrak{r}\Lambda .
\end{equation}
Checking the coefficient of $\log\hbar/(2\pi i)$ is non-trivial.  We use
$\sum_{\alpha=\alpha_{j-1}+1}^{\alpha_j} d_{\alpha_{j-1}\alpha}(2-a_\alpha)
= \sum_{\alpha=\alpha_{j-1}+1}^{\alpha_j} \bigl( -d_{\alpha_{j-1},\alpha-1}+2d_{\alpha_{j-1}\alpha}-d_{\alpha_{j-1},\alpha+1} \bigr)
= 1 + d_{\alpha_{j-1},\alpha_j} - d_{\alpha_{j-1},\alpha_j+1}$ and the similar relation
$\sum_{\alpha=\alpha_j+1}^{\alpha_{j+1}-1} d_{\alpha\alpha_{j+1}}(2-a_\alpha)
= - d_{\alpha_j,\alpha_{j+1}} + d_{\alpha_j+1,\alpha_{j+1}} + 1$ to work out the coefficient
\begin{equation}
  \begin{aligned}
    & \sum_{\alpha=\alpha_{j-1}}^{\alpha_j} \frac{d_{\alpha_{j-1}\alpha}}{d_{\alpha_{j-1}\alpha_j}} (2-a_\alpha)
    + \sum_{\alpha=\alpha_j+1}^{\alpha_{j+1}} \frac{d_{\alpha\alpha_{j+1}}}{d_{\alpha_j\alpha_{j+1}}} (2-a_\alpha)
    \\
    & = \frac{d_{\alpha_j\alpha_{j+1}} (1 + d_{\alpha_{j-1},\alpha_j} - d_{\alpha_{j-1},\alpha_j+1})
      + d_{\alpha_{j-1}\alpha_j} (- d_{\alpha_j,\alpha_{j+1}} + d_{\alpha_j+1,\alpha_{j+1}} + 1)}{d_{\alpha_{j-1}\alpha_j}d_{\alpha_j\alpha_{j+1}}}
    \\
    & = \frac{d_{\alpha_j\alpha_{j+1}} - d_{\alpha_{j-1}\alpha_{j+1}} + d_{\alpha_{j-1}\alpha_j}}{d_{\alpha_{j-1}\alpha_j}d_{\alpha_j\alpha_{j+1}}} .
  \end{aligned}
\end{equation}

\section{\label{sec:Band restriction rule}Band restriction rule in non-supersymmetric models}

Our discussion of B-branes and their central charges in \autoref{sec:hpf} (localization approach) and \autoref{sec:Ktheoreticaspects} (geometric approach) only determined the Higgs branch contributions for a restricted class of branes.
In this section we explain the band restriction rule that characterises that restricted class, and explain how the Higgs and mixed/Coulomb branch parts of a GLSM brane are distinguished in general.

We first review the gauge-decoupling limit of B-branes in $U(1)$ GLSMs through the lens of the hemisphere partition function (\autoref{ssec:U1reviewcareful}), stressing how different the Calabi-Yau and non-Calabi-Yau cases are.
The story for more general gauge groups partly relies on transporting a B-brane from one phase of a GLSM to another by continuously varying the FI-theta parameter $t_{\text{ren}}$ at some fixed energy scale~$\mu$.
To this aim, we explain transport of B-branes (\autoref{ssec:BbraneTransport}) through the example of $U(1)$ theories.

We then determine how transporting branes through a wall in general abelian models reduces to that of a gauge group $U(1)$ times a discrete factor whose effect we also explain, culminating in a statement of the band restriction rule (\autoref{ssec:wallcrossingHiggsing}).
We make our approach explicit for B-brane gauge-decoupling and B-brane transport in Hirzebruch-Jung models (\autoref{ssec:HJbranesHiggs}), reproducing the set of special representations of the orbifold group $\ZZ_n\subset GL(2,\CC)$ in the geometric McKay correspondence.
In a rank~$2$ example we find non-trivial monodromy in the FI-theta parameter space, which should prompt further study of the singular locus of theories with anomalous axial R-symmetry.

\subsection{\label{ssec:U1reviewcareful}B-branes in \texorpdfstring{$U(1)$}{U(1)} models [review with extra details]}

\subsubsection{Outline: the image of B-branes}

\paragraph{Model and its phases.}

As a practice and a building block for higher rank models, we discuss the $U(1)$ case in detail, following~\cite{Herbst:2008jq,Hori:2013ika}.
Consider a $U(1)$ GLSM with chiral multiplets of charges~$Q^j$ for $1\leq j\leq\dim V$ and no superpotential.
Let $N_{\pm}\geq 0$ be the total positive/negative charges, and $t_{\text{sh}} = \zeta_{\text{sh}}-i\theta_{\text{sh}}$ denote conveniently shifted FI-theta parameters
\begin{equation}
  \sum_j Q^j=N_+-N_-, \qquad \sum_j\abs{Q^j}=N_++N_-, \qquad
  t_{\text{sh}} = t_{\text{ren}}+\sum_j Q^j\log Q^j\bmod{2\pi i} .
\end{equation}
From this we read off for instance that $\theta_{\text{sh}}=\theta-N_-\pi$.

For $N_+=N_-$ this model is Calabi-Yau.
In FI-theta parameter space there is a singularity at $t_{\text{sh}}=0\bmod{2\pi i}$.
There are two phases $\zeta\gg 0$ and $\zeta\ll 0$, well-described by non-linear sigma models (NLSMs) on two different classical Higgs branches, which are line bundles on different weighted projective spaces.
The FI-theta parameter is shifted upon scale and axial R-symmetry transformations.  We fix these by considering the theory at some fixed complexified energy scale~$\mu$.
The physics at that scale is well-described for $(N_+-N_-)\zeta_{\text{ren}}\gg 0$ by an NLSM on the classical Higgs branch, and for $(N_+-N_-)\zeta_{\text{ren}}\ll 0$ by the direct sum of such an NLSM and of $|N_+-N_-|$ massive vacua.
These vacua are located on the classical Coulomb branch at\footnote{Coulomb branch vacua are such that $\partial\tW_{\text{eff}}/\partial\sigma=0$ modulo $2\pi i$, where the modulo is due to the periodicity of the theta angle.  On a disk, the boundary breaks quantization of $\int F$ hence breaks periodicity of theta, and we will see that different Coulomb branch vacua are selected for different UV branes.}
\begin{equation}
  \sigma \sim \mu \exp\biggl(\frac{-t_{\text{sh}}+2\pi ik}{|N_+-N_-|}\biggr), \qquad k\in\ZZ_{|N_+-N_-|} .
\end{equation}
Altogether there are two types of phases:
\begin{itemize}
\item \textbf{pure-Higgs phases} $\zeta_{\text{ren}}\gg 0$ with $N_+\geq N_-$, as well as $\zeta_{\text{ren}}\ll 0$ with $N_+\leq N_-$;
\item \textbf{phases with massive vacua} $(N_+-N_-)\zeta_{\text{ren}}\ll 0$, only in non-CY models.
\end{itemize}
This subsection is devoted to the image of (complexes of) Wilson line branes onto the Higgs branch and the massive vacua.

\paragraph{Pure-Higgs phase: the geometric image is correct.}

Consider the phase $\zeta_{\text{ren}}\gg 0$ in the case $N_+\geq N_-$ (the phase $\zeta_{\text{ren}}\ll 0$ for $N_+\leq N_-$ is related by charge conjugation).
This phase is pure-Higgs in the sense that it has no mixed/Coulomb branch.
Its Higgs branch is a quotient $X=(V\setminus\Delta)/\CC^*$ where $V$ is spanned by all chiral multiplets while the deleted set~$\Delta$ is the subspace where all positively charged chiral multiplets vanish.

A brane of the GLSM is described by a complex of Wilson lines (and by an admissible contour~$L$).
The Higgs branch NLSM is obtained by sending $g\to\infty$ at some fixed complexified energy scale~$\mu$.
This turns the auxiliary field~$D$ into a Lagrange multiplier and decouples the gauge degrees of freedom, thus quotienting $V$ to~$X$.
The complex is sent to a complex of line bundles on~$X$ obtained by restriction and pushforward from $V$ to~$X$.
In the language of \autoref{ssec:Bbraneabelian},
\begin{equation}
  F_{g\to\infty,\text{Higgs}}=F_{\text{geom}} \qquad \text{in a pure-Higgs phase.}
\end{equation}
We show the same statement for higher-rank gauge groups in \autoref{ssec:wallcrossingHiggsing}.

\paragraph{Phases with massive vacua: windows needed.}
Phases that are not pure-Higgs are more subtle: the gauge degrees of freedom do not fully decouple since they are responsible for the mixed/Coulomb branches.
As we explain in the rest of this subsection, the gauge-decoupling image of (complexes of) Wilson line branes~$\cW(q)$, depends on whether their charge lies in one of two windows.
\begin{equation}\label{U1windows}
  \begin{aligned}
    \textbf{Small window:} & & \abs{\theta+2\pi q} & < \min(N_-,N_+)\pi .
    \\
    \textbf{Big window:} & & \abs{\theta+2\pi q} & < \max(N_-,N_+)\pi .
  \end{aligned}
\end{equation}
A careful analysis of the contour integral expression of the central charge leads to the following conclusion~\cite{Hori:2013ika} for $\theta\neq N_{\pm}\pi\bmod{2\pi}$.
\begin{itemize}
\item Branes with charges~$q$ in the small window $\abs{\theta+2\pi q}<\min(N_-,N_+)\pi$ reduce purely to the Higgs branch in the gauge-decoupling limit, and $F_{g\to\infty,\text{Higgs}}$ and $F_{\text{geom}}$ coincide for these branes.  Correspondingly, the contour integral can be computed as a sum of residues, in the same way as for pure-Higgs phases.

\item The image of branes whose charges~$q$ lie in a (larger) \textbf{big window} consists of a Higgs branch contribution given by~$F_{\text{geom}}$, and some part supported by the massive vacua.  This latter (Coulomb branch) part is found by a saddle-point analysis.
  For a single Wilson line with $\min(N_-,N_+)\pi<\abs{\theta+2\pi q}<\max(N_-,N_+)\pi$, the contour integral has one saddle-point $\hat{\sigma}=\hat{\sigma}_q$, and is equal to the steepest descent contour integral starting from that saddle point, plus the residue sum that captures the Higgs branch contribution.

\item To find the image of an arbitrary complex of Wilson lines~$\cB$, one brings all charges to the big window by binding~$\cB$ to a collection of empty branes (complexes of Wilson lines that reduce to an empty brane in the gauge-decoupling limit).
  In fact, the naive Higgs branch image~$F_{\text{geom}}$ is generally incorrect for branes outside the big window.
  Correspondingly, the contour integral for $\max(N_-,N_+)\pi<\abs{\theta+2\pi q}$ cannot be evaluated by saddle-point analysis: instead, it is determined by first adding to the brane factor~$f_{\cB}$ the brane factor of an empty brane.  This latter brane factor annihilates poles of the contour integral, which makes its contribution vanish.
\end{itemize}

\subsubsection{Contours}
\label{sssec:contours}

\paragraph{Almost-universally admissible contours.}
Recall from \autoref{ssec:BbraneGLSM} that B-branes must come with the data of a Lagrangian contour for $\sigma\in\lie{t}_{\CC}$ which is a deformation of~$\lie{t}_{\RR}$ and that is admissible in the sense that the contour integral~\eqref{ZD2loc} converges.
It is typically enough to work with contours that can be written as graphs $\sigma=\tau+i\upsilon(\tau)$ of some function $\upsilon\colon\lie{t}_{\RR}\to\lie{t}_{\RR}$.
For our rank-one case the contour is simply a deformation of $\RR\subset\CC$.
We work out the integrand's behaviour for $\hat{\sigma}$ large away from the imaginary axis, for a Wilson line brane~$\cW(q)$:
\begin{equation}\label{Aqpotential}
  \begin{aligned}
    \log\abs{\text{integrand}} & = - A_q(\hat{\sigma}) + O(\log|\hat{\sigma}|) \\
    A_q(\hat{\sigma}=\hat{\tau}+i\hat{\upsilon}) & =
    \begin{aligned}[t]
      &
      \bigl( \zeta_{\text{sh}}+ (N_+-N_-) \bigl(\log\abs{\hat{\tau}+i\hat{\upsilon}}-1\bigr) \bigr) \hat{\upsilon} \\
      & + \biggl(
      \frac{\pi}{2}(N_++N_-)
      + (N_+-N_-) \arctan\frac{\hat{\upsilon}}{\abs{\hat{\tau}}}
      -(\sign\hat{\tau})(\theta+2\pi q)
      \biggr)
      \abs{\hat{\tau}}
    \end{aligned}
  \end{aligned}
\end{equation}
where we have split $\hat{\sigma}=\hat{\tau}+i\hat{\upsilon}$.
The integral converges provided $A_q(\hat{\sigma})\to+\infty$ fast enough at infinity in the contour.
For all cases except $(\zeta_{\text{sh}},N_+-N_-)=(0,0)$, we can take the contour\footnote{The choice of superlinear function $\hat{\tau}\mapsto \hat{\tau}^2$ is mostly arbitrary.}
\begin{equation}\label{contourU1}
  \begin{aligned}
    \hat{\sigma} & = \hat{\tau}+i\hat{\tau}^2 , \qquad \hat{\tau}\in\RR , \quad
    & & \text{in the CY case with } \zeta_{\text{sh}}>0 \text{, or any phase for } N_+>N_- , \\
    \hat{\sigma} & = \hat{\tau}-i\hat{\tau}^2 , \qquad \hat{\tau}\in\RR , \quad
    & & \text{in the CY case with } \zeta_{\text{sh}}<0 \text{, or any phase for } N_+<N_- . \\
  \end{aligned}
\end{equation}
Since $A_q\gtrsim\abs{\hat{\upsilon}}$ along this contour, it is admissible regardless of~$q$, hence it is also admissible for arbitrary B-branes of the GLSM\@.

The Calabi-Yau case $N_+=N_-$, with in addition $\zeta_{\text{sh}}=0$, deserves attention.
Then $A_q(\hat{\tau}+i\hat{\upsilon}) = \bigl(\pi N_+ -(\sign\hat{\tau})(\theta+2\pi q)\bigr) \abs{\hat{\tau}}$.  For Wilson lines $\cW(q)$ in the window
\begin{equation}\label{U1CYwindow}
  - \frac{N_+}{2} < \frac{\theta}{2\pi} + q < \frac{N_+}{2}
\end{equation}
even the contour $\RR$ is admissible (it is then also admissible away from $\zeta_{\text{sh}}=0$).  Outside this window, $A_q$ becomes arbitrarily negative for $\hat{\tau}\to+\infty$ or $-\infty$ hence no deformation of~$\RR$ is admissible.
This \textbf{window restriction rule}~\eqref{U1CYwindow} allows $N_+=N_-$ Wilson line branes for $\zeta_{\text{sh}}=0$ and $\theta\neq N_+\pi\bmod{2\pi}$.
As we vary $\theta$, the window changes discontinuously for $\theta=N_+\pi\bmod{2\pi}$, where the theory is singular.

\paragraph{Naive Higgs branch contribution.}

It is instructive to rewrite the hemisphere partition function from a contour integral to an infinite convergent series.
Focus for definiteness on the cases where \eqref{contourU1} is the contour $\hat{\sigma}=\hat{\tau}+i\hat{\tau}^2$ depicted in \autoref{fig:contour}, namely $N_+=N_-$ and $\zeta_{\text{sh}}>0$, or $N_+>N_-$.
Given the large~$\hat{\upsilon}$ asymptotics of~\eqref{Aqpotential}, we can close the integration contour to enclose poles on the positive imaginary axis, at $\hat{\sigma}=i(R_j/2+k)/Q^j$ for $Q^j>0$ and $k\in\ZZ_{\geq 0}$.  These poles are simple for generic~$R_j$ and we get\footnote{For non-generic $R_j$ there can be higher-order poles, in which case the explicit expression is more difficult to obtain.  However, the asymptotics are the same.}
\begin{equation}\label{ZD2HiggsphaseU1}
  \begin{aligned}
    & Z_{D^2}(\cB)
    = C(\mathfrak{r}\Lambda)^{\hat{c}/2} \int_{L} \mathrm{d}\hat{\sigma}  \prod_j \Gamma\biggl( iQ^j\hat{\sigma} + \frac{R_j}{2} \biggr) e^{it_{\text{ren}}\hat{\sigma}} f_{\cB}(\hat{\sigma}) \\
    & \! = C(\mathfrak{r}\Lambda)^{\hat{c}/2} \!\!\sum_{j\mid Q^j>0} \sum_{k=0}^{\infty}
    \frac{2\pi(-1)^k}{k!\,Q^j} e^{-t_{\text{ren}}(R_j/2+k)/Q^j} f_{\cB}\biggl(\frac{i(R_j/2+k)}{Q^j}\biggr)
    \prod_{i\neq j} \Gamma\biggl( \frac{R_i}{2}-\frac{Q^i}{Q^j}\biggl(\frac{R_j}{2}+k\biggr) \biggr) .
  \end{aligned}
\end{equation}
Using Stirling's formula and $\Gamma(x)\Gamma(1-x)=\pi/\sin\pi x$ we show that for a fixed $j$, and fixed $(k\bmod{Q^j})$, the summand behaves at large~$k$ according to
\begin{equation}\label{U1seriesasymptotic}
  \begin{aligned}
    & \log\abs{\text{summand}}
    = - \frac{\zeta_{\text{ren}}}{Q^j} k - \sum_i \frac{Q^i k}{Q^j} \log\abs*{\frac{Q^i k}{e\,Q^j}} + O(\log k) \\
    & \quad = -\frac{N_+-N_-}{Q^j}\bigl(k \log k - k - k \log\abs{Q^j}\bigr)
    - \frac{\zeta_{\text{ren}}+\sum_i Q^i\log\abs{Q^i}}{Q^j} k + O(\log k) .
  \end{aligned}
\end{equation}
For $N_+>N_-$ the sum converges regardless of~$t_{\text{ren}}$, as is consistent with the fact that there is no special value of~$t_{\text{ren}}$ in non-Calabi-Yau models.
For $N_+=N_-$ the sum converges in the phase $\zeta_{\text{sh}}=\zeta_{\text{ren}}+\sum_i Q^i\log\abs{Q^i}>0$.
The similar sum for $N_+<N_-$ defines an asymptotic series in (fractional) powers of $e^{-t_{\text{ren}}}$, which never converges.

By conjugating charges (hence working with the contour $\hat{\tau}-i\hat{\tau}^2$), we find that the same asymptotics hold when summing residues with $Q^j<0$ instead: in that case the series converges for $N_+<N_-$ and any~$\zeta_{\text{sh}}$, or $N_+=N_-$ and $\zeta_{\text{sh}}<0$, while it is an asymptotic series if $N_+>N_-$.

\begin{figure}[t]
  \centering
  \begin{tikzpicture}
    \draw[-{stealth}] (-2,0) -- (2,0) node [right] {$\hat{\tau}$};
    \draw[-{stealth}] (0,-2) -- (0,2) node [above] {$\hat{\upsilon}$};
    \draw plot[domain=-1.4:1.4] (\x,\x*\x) node [above] {$L$};
    \node at (-1.3,1) {\scriptsize (a)};
    \node at (-.7,-.3) {\scriptsize (b)};
    \node at (-.2,-1.1) {\scriptsize (c)};
    \node at (.7,-.3) {\scriptsize (d)};
    \node at (1.3,1) {\scriptsize (e)};
    \draw plot[domain=-1.4:-.48] (\x-.02,\x*\x-.02) -- (-.5,-0.9) -- (.5,-0.9)
    -- plot [domain=.48:1.4] (\x+.02,\x*\x-.02) node [below right=-3pt] {$L_\lambda$};
    \draw[densely dashed] (-.94,-0.9) -- (.94,-0.9) node [right] {$\hat{\upsilon}=-\lambda$};
    \foreach\y in {.1,.43,...,1.9}
      \filldraw (0,\y) circle (.05);
    \foreach\y in {-.155,-.7,...,-1.9}
      \filldraw (0,\y) circle (.05);
  \end{tikzpicture}
  \caption{\label{fig:contour}Universal contour~$L$ for $N_+>N_-$.  In the pure-Higgs phase $\zeta_{\text{ren}}\gg 0$, the integral picks up poles on the upper imaginary axis.  In the other phase it is useful to deform the contour to $L_\lambda$ while passing poles with $\Im\hat{\sigma}>-\lambda$, as described near~\eqref{poleswepass}.  We label five pieces of~$L_\lambda$ defined in~\eqref{abcde}.}
\end{figure}

\paragraph{How to identify Coulomb branch contributions.}

The Higgs branch series \eqref{ZD2HiggsphaseU1} is convergent in pure-Higgs phases, and it captures completely the brane's central charge.
In other phases there is both a Higgs branch and $\abs{N_+-N_-}$ massive vacua, so we need to determine the image of the brane on both branches.

We consider for definiteness the phase $\zeta_{\text{ren}}\ll 0$ in models with $N_+>N_-$.
Let us try to expand at large $(-\zeta_{\text{ren}})$ the hemisphere partition function with a Wilson line~$\cW(q)$.
Our experience with the pure-Higgs phase suggests one part of the hemisphere partition function should be a sum over poles along $-i\RR_{>0}$, but this sum is an asymptotic series.
For any finite~$\zeta_{\text{ren}}$ we should expect that a good approximation is given by a finite number of terms in the series: terms shrink until the term with
\begin{equation}\label{poleswepass}
  \frac{k}{\abs{Q^j}} \simeq \lambda, \qquad
  \lambda \coloneqq \exp\biggl( - \frac{\zeta_{\text{sh}}}{N_+-N_-}\biggr) ,
\end{equation}
then terms grow again.
This leads us to rewrite the contour integral over~$L$ into a sum of residues at $\hat{\sigma}=i(R_j/2+k)/Q^j$ for $Q^j<0$ and $k\in\ZZ_{\geq 0}$ bounded above according to $\Im(-\hat{\sigma})<\lambda$, plus a contour integral over some contour~$L_\lambda$ that crosses the imaginary axis at $\hat{\sigma}=-i\lambda$, as depicted in \autoref{fig:contour}.

Our key task is then to identify the contour integral over~$L_\lambda$ with a contribution from the Coulomb branch, or, on the contrary, simply with an error term.
To disentangle these cases, we estimate the $L_\lambda$ integral using a saddle-point analysis, and especially focus on the large-$\lambda$ asymptotics.
An important step is to ensure that the effective potential $A_q$ is sufficiently large throughout the contour.  This depends drastically on~$q$.

\subsubsection{Saddles or no saddles}
\label{sssec:saddles}

\paragraph{Small window: no saddle-point.}
The simplest case is a brane $\cW(q)$ in the \textbf{small window}
$\abs{\theta + 2\pi q} < \pi \min(N_-,N_+)$ defined in~\eqref{U1windows}.
Again we take $N_+>N_-$ and $\zeta_{\text{ren}}\ll 0$.
Based on the small window condition we can bound the coefficient of~$\hat{\tau}$ in~\eqref{Aqpotential} to get
\begin{equation}
  \frac{A_q(\hat{\sigma})}{(N_+-N_-)\lambda}
  \geq \biggl(\log\frac{\abs{\hat{\sigma}}}{\lambda}-1\biggr) \frac{\hat{\upsilon}}{\lambda}
  + \Bigl( \frac{\pi}{2} + \arctan\frac{\hat{\upsilon}}{\abs{\hat{\tau}}} \Bigr)
  \frac{\abs{\hat{\tau}}}{\lambda} ,
\end{equation}
which only depends on $\hat{\sigma}/\lambda$.  This motivates us to consider a contour~$L_\lambda$ that mostly only involves~$\lambda$ as a scale, except that it must coincide with~$L$ at infinity.  Specifically, we consider the contour~$L_\lambda$ that coincides with the parabolic contour $L$ for $\abs{\hat{\tau}}\geq\lambda$ and has two vertical lines along $\hat{\tau}=\pm\lambda$, connected by a horizontal line along $\hat{\upsilon}=-\lambda$:
\begin{equation}\label{abcde}
  \begin{aligned}
    \text{(a)} & & \hat{\sigma} & = \hat{\tau}+i\hat{\tau}^2 , \qquad & \hat{\tau}& \leq-\lambda , \\
    \text{(b)} & & \hat{\sigma} & = -\lambda+i\hat{\upsilon} , \qquad & \lambda^2\geq\hat{\upsilon} & \geq-\lambda , \\
    \text{(c)} & & \hat{\sigma} & = \hat{\tau}-i\lambda , \qquad & -\lambda\leq\hat{\tau} & \leq\lambda , \\
    \text{(d)} & & \hat{\sigma} & = \lambda+i\hat{\upsilon} , \qquad & -\lambda\leq\hat{\upsilon} & \leq\lambda^2 , \\
    \text{(e)} & & \hat{\sigma} & = \hat{\tau}+i\hat{\tau}^2 , \qquad & \lambda\leq\hat{\tau} & .
  \end{aligned}
\end{equation}

To bound the integral over~$L_\lambda$ we consider each segment of the contour, starting from the horizontal line~(c), which we parametrize as $\hat{\sigma}=\lambda(-i+\tan\varphi)$ for $\varphi\in[-\pi/4,\pi/4]$:
\begin{equation}\label{segment-c}
  \frac{A_q(\hat{\sigma})}{(N_+-N_-)\lambda} \geq 1 - \frac{1}{2}\log\bigl(1+\tan^2\varphi\bigr) + \varphi\tan\varphi \geq 1 ,
\end{equation}
where the last inequality comes from monotonicity since the derivative
\begin{equation}
  \partial_\varphi\biggl(1 - \frac{1}{2}\log\bigl(1+\tan^2\varphi\bigr) + \varphi\tan\varphi\biggr)
  = \frac{\varphi}{\cos^2\varphi}
\end{equation}
has the same sign as~$\varphi$.
The segments (b) and (d) have the same values of $A_q$ so let us focus on~(d), parametrized as $\hat{\sigma}=\lambda(1+i\nu)$ with $\nu\in[-1,\lambda]$.  There,
\begin{equation}
  \frac{A_q(\hat{\sigma})}{(N_+-N_-)\lambda}
  \geq \bigl(\log\abs{1+i\nu}-1\bigr) \nu
  + \frac{\pi}{2} + \arctan\nu ,
\end{equation}
whose $\nu$ derivative simplifies to $\log\abs{1+i\nu}\geq 0$.
This lower bound is thus bounded below by its value at $\nu=-1$, known from~\eqref{segment-c} to be at least~$1$.

Finally, segments (a) and (e) are symmetrical and we parametrize segment (e) using $\hat{\tau}\geq\lambda$:
\begin{equation}
  \frac{A_q(\hat{\sigma})}{(N_+-N_-)\lambda}
  \geq \biggl(\log\frac{\hat{\tau}}{\lambda} + \log\abs{1+i\hat{\tau}}-1\biggr) \frac{\hat{\tau}^2}{\lambda}
  + \Bigl( \frac{\pi}{2} + \arctan\hat{\tau} \Bigr)
  \frac{\hat{\tau}}{\lambda} ,
\end{equation}
whose $\hat{\tau}$ derivative simplifies to
$
\frac{1}{\lambda} \bigl(2\hat{\tau} \log\frac{\hat{\tau}}{\lambda} + 2\hat{\tau} \log\abs{1+i\hat{\tau}} + \frac{\pi}{2} + \arctan\hat{\tau} \bigr)\geq \log(1+\lambda^2)$.
This bound on the derivative implies a linear lower bound for~$A_q$.

Altogether, throughout the contour, we have found the inequality
\begin{equation}
  A_q(\hat{\sigma}) \geq (N_+-N_-)\lambda + C_\lambda \bigl(\abs{\hat{\tau}}-\lambda\bigr)^+ , \qquad
  C_\lambda = (N_+-N_-)\lambda \log(1+\lambda^2) > 0 ,
\end{equation}
where $(x)^+=\max(0,x)$ denotes the positive part.  The integral over $L_\lambda$ is then bounded by splitting it into contributions from (b), (c), (d), which involve a finite power of~$\lambda$, and those from (a), (e), which involve an exponentially decaying (hence bounded) integral\footnote{The term $|\hat{\tau}|^{O(1)}$ does not depend on~$\lambda$.}:
\begin{equation}
  \begin{aligned}
    & \int_{L_\lambda} e^{-A_q(\hat{\sigma})+O(\log|\hat{\sigma}|)} d\hat{\sigma}
    \leq e^{-(N_+-N_-)\lambda} \int_{L_\lambda} e^{-C_\lambda(\abs{\hat{\tau}}-\lambda)^+ + O(\log|\hat{\sigma}|)} d\hat{\sigma}
    \\
    & \quad \leq \lambda^{O(1)} e^{-(N_+-N_-)\lambda} + 2 e^{-(N_+-N_-)\lambda} \int_\lambda^{+\infty} |\hat{\tau}|^{O(1)} e^{-C_\lambda(\abs{\hat{\tau}}-\lambda)^+} d\hat{\tau}
    \leq \lambda^{O(1)} e^{-(N_+-N_-)\lambda}
  \end{aligned}
\end{equation}
as $\lambda\to+\infty$.
We deduce an asymptotic expansion of the form
\begin{equation}\label{smallwindowWilson}
  Z_{D^2}\bigl(\cW(q)\bigr)
  = \Biggl(\sum_{j\mid Q^j<0} \sum_{k=0}^{-R_j/2-Q^j\lambda} \bigl({\cdots}\bigr) \lambda^{-\frac{N_+-N_-}{\abs{Q^j}}k} \Biggr) + e^{-(N_+-N_-)\lambda+O(\log\lambda)} .
\end{equation}
We will soon check that the residual term is more exponentially suppressed than any of the $N_+-N_-$ possible Coulomb branch contributions.
Thus, a Wilson line brane in the small window~\eqref{U1windows} cannot have a Coulomb branch part and it must reduce purely to a Higgs branch brane.

\subsubsection{Big window and saddle-point}

\paragraph{Big window and saddle-point.}

We now find the image in the Higgs and Coulomb branches of any complex of Wilson lines $\cW(q)$ whose charges fit in the \textbf{big window}~\eqref{U1windows}, namely $\abs{\theta + 2\pi q} < \pi \max(N_-,N_+)$.
We call such branes \textbf{band-restricted}.

We continue with $N_+>N_-$ and $\zeta_{\text{ren}}\ll 0$ and assume $\theta\neq N_{\pm}\pi\bmod{2\pi}$ to avoid singularities.
The contour integral is now computed using a saddle-point analysis.
Away from the imaginary axis, the integrand obeys\label{U1logintegrand}
\begin{equation}\label{logintegrand}
  \log\bigl(\text{integrand}(\hat{\sigma})\bigr) =
  \begin{cases}
    (N_+-N_-) i\hat{\sigma}\bigl(\log(\hat{\sigma})-\ell_+-1\bigr) + O(\log\abs{\hat{\sigma}}) , & \Re\hat{\sigma}>0 \\
    (N_+-N_-) i\hat{\sigma}\bigl(\log(-\hat{\sigma})-\ell_--1\bigr) + O(\log\abs{\hat{\sigma}}), & \Re\hat{\sigma}<0 \\
  \end{cases}
\end{equation}
where
\begin{equation}
  \ell_{\pm} = - \frac{\zeta_{\text{sh}}}{N_+-N_-} + i \frac{\theta+2\pi q \mp \frac{\pi}{2}(N_++N_-)}{N_+-N_-}
\end{equation}
and we also note that $\Re\ell_+=\Re\ell_-=\log\lambda$.
These formulae generalize~\eqref{Aqpotential} by including the phase of the integrand.
Note that the two asymptotic expansions are only valid away from $\Re\hat{\sigma}=0$.
In fact they have different $\Re\hat{\sigma}\to 0$ limits:
at $\hat{\sigma}=i\lambda$ with $\lambda>0$ there is a jump by $-2\pi iN_+\lambda$;
at $\hat{\sigma}=-i\lambda$ with $\lambda>0$ there is a jump by $2\pi i N_- \lambda$.

In the region $\Re\hat{\sigma}>0$, the asymptotic expansion has a saddle-point at $\log(\hat{\sigma})=\ell_+$ if $\abs{\Im\ell_+}<\pi/2$ (to ensure $\Re\hat{\sigma}>0$).
In the other region there is a saddle point at $\log(-\hat{\sigma})=\ell_-$ if $\abs{\Im\ell_-}<\pi/2$ (to ensure $\Re\hat{\sigma}<0$).
If neither of these conditions hold there is no saddle point.
At a fixed~$\theta$, the collection of these saddle-points for $q\in\ZZ$ coincides with the set of $N_+-N_-$ Coulomb branch vacua, which are solutions of
\begin{equation}
  (i\hat{\sigma})^{N_+-N_-} = \exp(-t_{\text{sh}}) , \qquad t_{\text{sh}} = t_{\text{ren}} + \sum_i Q_i \log Q_i\bmod{2\pi i} .
\end{equation}
When the hemisphere partition function of a brane has a contribution from one of these saddle-points,
the brane has a non-trivial part along the corresponding massive vacuum.

\paragraph{Estimating the integral.}

Let us focus for definiteness on the first case $\abs{\Im\ell_+}<\pi/2$, namely $N_-\pi<\theta+2\pi q<N_+\pi$.
The steepest descent contour $L_{\text{steep}}$ passing through the saddle $\hat{\sigma}=\exp(\ell_+)$ is (one branch of) the locus where the integrand has the same phase and smaller norm than its value at the saddle.
Namely, expressing this condition in terms of ($i$~times) the logarithm~\eqref{logintegrand} of the integrand, the steepest descent contour is
\begin{equation}
  \bigl\{\hat{\sigma} \bigm| \hat{\sigma}\bigl(\log(\hat{\sigma})-\ell_+-1\bigr) + \exp(\ell_+) \in i\RR_{>0} \bigr\} .
\end{equation}
Simple numerics indicate\footnote{It would be pleasant to obtain an analytic proof.} that this contour, together with its continuation to the half-space $\Re\hat{\sigma}<0$, is homotopic to a parabola intersecting the imaginary axis at some point $-i\lambda\,c(\Im\ell_+)$ where $c(\varphi)$ is some function of the argument $\varphi\in(-\pi/2,\pi/2)$ that interpolates between $c(-\pi/2)=1$ and $c(\pi/2)=0$.

The original contour integral over the parabola~$L$ thus decomposes into a sum of $O(\lambda)$ residues at poles along $-i\RR_{>0}$, plus an integral over~$L_{\text{steep}}$.
The latter scales like
\begin{equation}
  \abs[\bigg]{\int_{L_{\text{steep}}}\!\!\!\!\!\mathrm{d}\hat{\sigma}\,(\text{integrand})}
  \sim \exp\biggl( (N_+-N_-) \cos\biggl(\frac{N_+\pi - (\theta + 2\pi q)}{N_+-N_-}\biggr) \lambda \biggr) ,
\end{equation}
where the cosine takes distinct values in $(-1,1)$ that increase with $\theta+2\pi q\in(N_-\pi,N_+\pi)$.
Together with analogous expressions for negative $\theta+2\pi q$, these asymptotic behaviours are pairwise distinct and differ from the Higgs branch contributions, hence they must correspond to all $|N_+-N_-|$ Coulomb branch vacua.
It was checked in~\cite{Hori:2013ika} that the exponentials match with the energy expected for these vacua.
Importantly, the cosine cannot be equal to $-1$, which means that all of these exponential behaviours are larger than the residual term we found for Wilson lines in the small window in~\eqref{smallwindowWilson}.
This confirms our previous conclusion that branes in the small window cannot have a Coulomb branch part.

\subsection{\label{ssec:BbraneTransport}B-brane transport in \texorpdfstring{$U(1)$}{U(1)} models [partly review]}

B-brane transport is defined in \cite[section 3.5]{Herbst:2008jq} and consists of following the fate of the boundary conditions as one deforms the bulk and boundary actions by  D-terms of the algebra~$\mathbf{2}_{B}$, namely bulk D-terms and twisted F-terms, and boundary D-terms.\footnote{This is akin to the more recently defined parallel transport of brane categories defined on the mirror side (A-branes in massive Landau--Ginzburg models) in~\cite{Gaiotto:2015zna,Gaiotto:2015aoa}.  To make the similarity manifest, one should consider in our context the interfaces obtained by varying FI-theta parameters abruptly to take different values on the two sides of the interface.  Then one should prove that B-brane transport functors defined in~\cite{Herbst:2008jq} coincide with the fusion of B-branes with that interface.}%
\textsuperscript{,}%
\footnote{Since B-brane transport allows adding D-terms in the boundary action, the resulting transport actually applies to equivalence classes of boundary conditions modulo (boundary) D-term deformations, rather than directly applying to boundary conditions.  We do not stress this subtlety in our presentation.}
In \autoref{sssec:transp-CY} we describe brane transport in the simplest case of non-anomalous models (including with rank $r>1$).
For anomalous models, we explain in \autoref{sssec:transp-gen} the construction of branes whose gauge-decoupling image is empty.  These are used in \autoref{sssec:transp-RG} to find how all $U(1)$ GLSM B-branes behave in the gauge-decoupling limit, before ending with the full story of B-brane transport in $U(1)$ GLSMs in \autoref{sssec:transp-full}.

\subsubsection{\label{sssec:transp-CY}Transport in the Calabi--Yau case}

\paragraph{Gauge group~$U(1)$.}

Let us explain first brane transport in non-anomalous $U(1)$ models (with $N_+=N_-$).
We recall that for a complex~$\cB$ of Wilson lines $\cW(q)$ in the window~\eqref{U1CYwindow}, namely $\mathbf{w}\coloneqq\{q\mid-N_+\pi<\theta+2\pi q<N_+\pi\}$, the contour~$\RR$ is admissible for all~$\zeta_{\text{ren}}$.
Such a band-restricted B-brane can thus be transported trivially from $\zeta_{\text{ren}}\gg 0$ to $\zeta_{\text{ren}}\ll 0$ by varying~$\zeta_{\text{ren}}$ without changing charges or morphisms in the complex.

Before discussing transport of more general branes, recall from \autoref{ssec:Bbraneabelian} that the Koszul complex
\begin{equation}
  \cK^+\coloneqq \bigotimes_{j\mid Q^j>0} \Bigl(\cW(-Q^j) \xrightarrow{X_j} \cW(0)\Bigr)
\end{equation}
is a resolution of the structure sheaf of the deleted set~$\Delta$, hence that complex
has a trivial geometric image on the Higgs branch, namely is an empty brane.
Its transport in any phase is still an empty brane.
The same holds for this complex with all charges shifted by tensoring with some~$\cW(Q')$.

Consider next a B-brane $\cB_+\in D^b(X_+)$ on the Higgs branch in the phase $\zeta_{\text{ren}}\gg 0$ (whence the subscript~$+$).
It can be written as the image of some B-brane $\cB_1\in D^b(V,U(1))$ of the GLSM\@.
The Koszul brane~$\cK^+$ that is empty in the phase $\zeta_{\text{ren}}\gg 0$ has charges from $-N_+$ to~$0$, so $\cB_1$~is equivalent to another complex~$\cB_2$ with grade-restricted charges $q\in\mathbf{w}$.
That brane can be transported to the phase $\zeta_{\text{ren}}\ll 0$ and projected to the Higgs branch category $D^b(X_-)$ there.
Brane transport from $\zeta_{\text{ren}}\gg 0$ to $\zeta_{\text{ren}}\ll 0$ only depends on~$\theta$ via the window~$\mathbf{w}$ of charges that are allowed at $\zeta_{\text{sh}}=0$.
This is consistent with the fact that brane transport only depends on the path in FI-theta parameter space up to homotopy.

Formally, for each interval $\mathbf{w}$ of $N_+$ consecutive integers one defines the window category $\cT_{\mathbf{w}}\subset D^b(V,U(1))$ consisting of complexes of Wilson lines with charges $q\in\mathbf{w}$.
Then one considers the functors $F_{\pm}\colon D^b(V,U(1))\to D^b(X_{\pm})$ that project to the Higgs branch in each phase.
Their restriction to any window category~$\cT_{\mathbf{w}}$ can be shown to be an equivalence of categories.
Brane transport $D^b(X_+)\to D^b(X_-)$ is then defined by composing these equivalences.
Pictorially,
\begin{equation}
  \begin{tikzpicture}
    \node(GLSM) at (0,0) {$D^b(V,U(1))$};
    \node at (0,-.4) {$\cup$};
    \node(window) at (0,-.8) {$\cT_{\mathbf{w}}$};
    \node(phase-) at (-2,-2) {$D^b(X_-)$};
    \node(phase+) at (2,-2) {$D^b(X_+)$};
    \draw[->](GLSM)--(phase-) node [pos=.6,above] {$F_-$};
    \draw[->](GLSM)--(phase+) node [pos=.6,above] {$F_+$};
    \draw[->](window)--(phase-) node [pos=.4,below] {$\cong$};
    \draw[->](window)--(phase+) node [pos=.4,below] {$\cong$};
  \end{tikzpicture}
\end{equation}

\paragraph{General rank Calabi--Yau case.}

For non-anomalous $U(1)^r$ GLSM the situation is very similar.
In this case, analyzed in~\cite{Herbst:2008jq}, the D-term deformations of the bulk action include marginal couplings of the IR SCFT, specifically twisted chiral couplings $e^{t}\in\mathcal{M}_{K}$, whose moduli space takes the form $\mathcal{M}_{K}=(\mathbb{C}^{*})^{r}\setminus \Delta_{\text{sing}}$ where $\Delta_{\text{sing}}$ is the locus where the theory is singular. Deep in the chambers of $\mathcal{M}_{K}$, where we have some better control of the IR SCFT it is possible to define boundary states that preserve B-type superconformal symmetry.  This provides a category of B-branes associated to each chamber in~$\mathcal{M}_{K}$.  Given a path $\gamma\colon[0,1]\to\mathcal{M}_{K}$ starting in a chamber (phase~I) and ending in another one (phase~II), brane transport provides a functor
\begin{equation}
  F_{[\gamma]}\colon\mathcal{D}_{\text{I}}\rightarrow \mathcal{D}_{\text{II}}
\end{equation}
between the categories of B-branes $\mathcal{D}_{\text{I}}$ and~$\mathcal{D}_{\text{II}}$.
One expects $F_{[\gamma]}$ to only depend on the homotopy class $[\gamma]$ of the path, so concatenating $\gamma$ with the reverse path gives the identity,
which in particular means that the functor $F_{[\gamma]}$ must be an equivalence of categories.
Even though explicit descriptions are only available for the initial and final boundary states deep in the phases, and not at intermediate points, $F_{[\gamma]}$ is interpreted as a categorical transport along the path~$\gamma$.
Branes in $\mathcal{D}_{\text{I}}$ and~$\mathcal{D}_{\text{II}}$ are conveniently described as IR limits of the UV boundary conditions $(\mathcal{B},L)$ explained in \autoref{ssec:BbraneGLSM}, so that determining~$F_{[\gamma]}$ reduces to finding how to transport IR images of Wilson line branes from a phase to another.
It is enough to treat the case of neighboring phases.

This is where the band restriction rule enters: B-brane transport turns out to depend on theta angles.
Consider the universal covering of $(\mathbb{C}^{*})^{r}$ with theta angles valued in $\mathbb{R}^{r}$ rather than a torus.
Since the wall has complex codimension~$1$, the set of angles~$\theta$ to which it asymptotes for large FI parameters (other than the transverse FI parameter) has real codimension~$1$ and
its complement in $\mathbb{R}^{r}$ has infinitely many connected components (``windows'').
Any constant-theta path from phase~I to phase~II through the wall passes through one of these windows.
As we will explain using the hemisphere partition function,
some Wilson lines (a different set for each window) are such that their IR image in phase~I maps under brane transport to their IR image in phase~II\@.
More precisely, the subcategory generated by these Wilson lines in the category of UV branes is equivalent to both categories $\mathcal{D}_{\text{I}}$ and $\mathcal{D}_{\text{II}}$ through the functor $F_{\text{geom},\text{I}}$ and $F_{\text{geom},\text{II}}$ coming from the GIT quotient construction.
In this setup, $F_{[\gamma]}\colon\mathcal{D}_{\text{I}}\to\mathcal{D}_{\text{II}}$ is the composition $F_{\text{geom},\text{II}}^{-1}\circ F_{\text{geom},\text{I}}$.
The aforementioned set of Wilson lines consists of those whose charges obey an inequality called band restriction, which depends on the window.

\subsubsection{\label{sssec:transp-gen}Generalities on transport, and empty branes}

\paragraph{Non-Calabi--Yau case.}

For the case of anomalous models, the situation becomes more complicated because in addition to the bare FI-theta parameters $t_{\text{bare}}$, we have the complexified energy scale~$\mu$. In this case we must consider paths in $(t_{\text{bare}},\mu)$ since Coulomb/mixed/Higgs branches can arise at fixed $t_{\text{bare}}$ but at different $\mu$. The IR theory can have not only B-branes corresponding to boundary states of the Higgs branch SCFT but also massive (Coulomb/mixed branch) vacua, and one must include all of these states in the category of IR boundary states in order for the categories at different points $(t_{\text{bare}},\mu)$ to be equivalent.
While B-brane transport still defines an equivalence $F_{[\gamma]}$ for a homotopy class of paths in the $(t_{\text{bare}},\mu)$ space, the band restriction rule works differently as we will see.
In addition we have to be careful when considering the projection of UV states to the IR\@.
We give more details on B-brane transport for both anomalous and non-anomalous models in the subsequent subsections.

Our main tool to study B-brane transport is the hemisphere partition function: it is independent of bulk D-terms and is holomorphic in twisted F-terms, hence B-brane transport amounts to analytically continuing the hemisphere partition function.

\paragraph{Empty brane.}
We return to the case of a $U(1)$ GLSM (without superpotential).
Consider the case $N_+\geq N_-$ and the phase $\zeta_{\text{ren}}\gg 0$.
As explained in \autoref{ssec:Bbraneabelian}, the Koszul complex
\begin{equation}\label{Koszulagain}
  \cK^+\coloneqq \bigotimes_{j\mid Q^j>0} \Bigl(\cW(-Q^j) \xrightarrow{X_j} \cW(0)\Bigr)
\end{equation}
is a resolution of the structure sheaf of the deleted set~$\Delta$, hence that complex
has a trivial geometric image on the Higgs branch.
Since the phase has no massive vacuum, the GLSM brane~$\cK^+$ is simply empty in that phase.

If $N_+>N_-$ (rather than $N_+=N_-$), then this (empty) B-brane can be transported to the other phase, where it is also empty because adding twisted F-terms does not spoil emptyness.
We can check this picture using the hemisphere partition function.
The integrand in~\eqref{ZD2locab} has poles along the positive and negative imaginary axes.
The integral picks up poles on the positive imaginary axis as in~\eqref{ZD2HiggsphaseU1}.
But the brane factor
\begin{equation}\label{KoszulbranefactorKplus}
  f_{\cK^+}(\hat{\sigma}) = \prod_{j\mid Q^j>0} \bigl(1 - e^{-2\pi Q^j\hat{\sigma}}\bigr)
\end{equation}
is such as to precisely cancel these poles, so $Z_{D^2}(\cK^+)=0$ exactly.
Brane transport analytically continues this vanishing partition function, which thus vanishes
in all phases, consistent with our statement in~\eqref{geom-empty-claim} that this GLSM brane is \textbf{empty in all phases}.

The situation is different for Calabi-Yau models ($N_+=N_-$): again $Z_{D^2}(\cK^+)=0$, but we cannot conclude that the same complex of Wilson lines~\eqref{Koszulagain} is empty in all phases.
Indeed, transporting the brane or its twists to another phase is not directly possible because none fit in the window~\eqref{U1CYwindow} that is required to transport the brane from one phase to the other.
The brane $\cK^+$ that is empty in one phase and its analogue $\cK^-$ with $Q^j<0$ that is empty in the other phase can be quite different in general.

\subsubsection{\label{sssec:transp-RG}Gauge-decoupling image of all branes.}

So far, we found the image of any GLSM brane~$\cB$ in a pure-Higgs phase.
It is
\begin{equation}
  F_{g\to\infty}(\cB) = F_{\text{geom}}(\cB) \in D^b(X)
\end{equation}
where the ``geometric'' functor $F_{\text{geom}}\colon D^b(V,U(1))\to D^b(X)$ projects complexes of $U(1)$-equivariant vector bundles from~$V$ to the Higgs branch $X=(V\setminus\Delta)/\CC^*$.
In phases with both a Higgs branch and massive vacua, we found the image of band-restricted branes, namely those that fit in the big window $\abs{\theta+2\pi q}<\max(N_+,N_-)\pi$.
We now generalize to all branes.

For definiteness take $N_+>N_-$.

The twist $\cK^+\otimes\cW(q)$ of the Koszul brane~\eqref{Koszulagain} is empty in the pure-Higgs phase $\zeta_{\text{ren}}\gg 0$ hence in all phases.
It is a complex with minimum and maximum charges $q_{\min}=q-N_+$ and $q_{\max}=q$.
We deduce that the brane $\cW(q)$ is equivalent in all phases to a complex with charges among $q-N_+,\dots,q-1$, and likewise to a complex with charges among $q+1,\dots,q+N_+$.

Consider now an arbitrary complex $\cB_1$ of Wilson lines.
Repeatedly replace every Wilson line~$\cW(q)$ with a charge outside the big window $\abs{\theta+2\pi q}<N_+$ by an equivalent complex of Wilson lines with charges closer to the big window.
This eventually ends and yields a complex $\cB_2$ of Wilson lines with charges in the big window.

\paragraph{Coulomb branch.}
The brane factors of $\cB_1$ and $\cB_2$ differ by a multiple of the brane factor~\eqref{KoszulbranefactorKplus} of~$\cK^+$.  The decomposition
\begin{equation}
  f_{\cB_1}(\hat{\sigma}) = f_{\cB_2}(\hat{\sigma}) + P(e^{2\pi\hat{\sigma}}) f_{\cK^+}(\hat{\sigma}),
\end{equation}
where~$P$ is some Laurent polynomial in~$e^{2\pi\hat{\sigma}}$, is essentially given by polynomial Euclidean division.  The remainder $f_{\cB_2}$~is unique and does not depend on details of the binding branes whose Higgs branch image is empty.

From its coefficients
\begin{equation}
  f_{\cB_2}(\hat{\sigma}) = \sum_{q=q_{\min}}^{q_{\max}} a_q e^{2\pi q\hat{\sigma}}
\end{equation}
where $q_{\min}$ and $q_{\max}$ are the bounds of the big window,
we can deduce the Coulomb branch part of the image of $\cB_1$ or equivalently $\cB_2$:
for each $q$ such that $N_-\pi<\abs{\theta+2\pi q}<N_+\pi$, there are $a_q$ branes (or $-a_q$ anti-branes if $a_q<0$) on the massive Coulomb vacuum at
\begin{equation}\label{CoulombpartofU1branes}
  i\hat{\sigma} = \exp\biggl(\frac{-\zeta_{\text{sh}}+ i(\theta+2\pi q -\sign(\theta+2\pi q) \pi N_-)}{N_+-N_-}\biggr) .
\end{equation}
Indeed, modulo a Higgs branch part, each Wilson line $\cW(q)$ in the big window goes to an empty brane for $\abs{\theta+2\pi q}<N_-\pi$ and otherwise goes to a single brane on one Coulomb branch vacuum.

\paragraph{Higgs branch.}
The Higgs branch image $F_{g\to\infty,\text{Higgs}}(\cB_1)$ cannot be read off from the brane factor only.
Since $F_{\text{geom}}$ and $F_{g\to\infty,\text{Higgs}}$ agree on band-restricted branes we have
\begin{equation}\label{HiggsbranchimageU1}
  F_{g\to\infty,\text{Higgs}}(\cB_1) = F_{g\to\infty,\text{Higgs}}(\cB_2) = F_{\text{geom}}(\cB_2),
\end{equation}
which differs in general from $F_{\text{geom}}(\cB_1)$.
The two functors only agree in general for band-restricted branes.

Let us say some words about the brane $\cK^-(q)\coloneqq\cW(q)\otimes\bigotimes_{j\mid Q^j<0} \bigl(\cW(-Q^j) \xrightarrow{X_j} \cW(0)\bigr)$, in which we selected the other sign of charges compared to the brane $\cK^+$~\eqref{Koszulagain} that is empty in all phases.
By construction the brane $\cK^-(q)$ is mapped to an empty brane by~$F_{\text{geom}}$.
In addition, $\cK^-(q)$ is band-restricted if $q+\theta/(2\pi)\in(N_--N_+/2,N_+/2)$,
in which case it has no Higgs part under RG flow.
This pure-Coulomb brane typically goes to several Coulomb branch vacua according to the discussion near~\eqref{CoulombpartofU1branes}, but one can take linear combinations of the brane factors of $\cK^-(q)$ such that a single coefficient outside the small window vanishes.
The corresponding direct sums of $\cK^-(q)$ branes are complexes that flow purely to a \textbf{Coulomb brane on one massive vacuum}.
Finally, despite having an empty image under~$F_{\text{geom}}$, branes $\cK^-(q)$ that are not band-restricted typically acquire a non-empty Higgs part, due to the round-about construction~\eqref{HiggsbranchimageU1}.

In conclusion, it should be stressed again that the effect of massive vacua on the gauge-decoupling limit of GLSM branes goes beyond simply adding a Coulomb branch part to the resulting brane: even the Higgs branch part is different from the natural geometric one.
Furthermore, the Higgs branch image depends on the theta angle because $\theta$~affects windows.

\subsubsection{\label{sssec:transp-full}Brane transport}

We now turn to brane transport between phases.

\paragraph{Non-Calabi-Yau case: starting from the pure-Higgs phase.}

For definiteness, $N_+>N_-$.  We use subscripts $+$ and $-$ for the phases $\zeta_{\text{ren}}\gg 0$ and $\zeta_{\text{ren}}\ll 0$, respectively.

Start from a brane $\cB_+\in D^b(X_+)$ on the Higgs branch~$X_+$ of the pure-Higgs phase.
Lift to a GLSM brane $\cB_1\in D^b(V,U(1))$ in the sense that $F_{\text{geom},+}(\cB_1)=\cB_+$.
Since there is no constraint on charges for an admissible contours to exist, the brane can be transported to the other phase.
However, finding the image in the other phase is delicate.

Using that the Koszul brane $\cK^+$~\eqref{Koszulagain} is empty in both phases and has charges from $-N_+$ to~$0$, the brane $\cB_1$ can be replaced by a band-restricted brane $\cB_2$ whose images in both phases are the same as those of~$\cB_1$.
Thanks to band restriction, the Higgs branch part of the image in the phase $\zeta_{\text{ren}}\ll 0$ is then $F_{\text{geom},-}(\cB_2)$.
The Coulomb branch (massive vacuum) part $F_{g\to\infty,\text{Coulomb}}(\cB_2)$ is deduced from the brane factor of~$\cB_2$.

The procedure simplifies if one starts from images $\cO(q)=F_{\text{geom},+}(\cW(q))$ of Wilson lines in the pure-Higgs phase, with $\abs{\theta+2\pi q}<N_+\pi$.  These generate $D^b(X_+)$.
The Wilson lines in the small window $\abs{\theta+2\pi q}<N_-\pi$ map to generators of $D^b(X_-)$.
The $N_+-N_-$ Wilson lines with $N_-\pi<\abs{\theta+2\pi q}<N_+\pi$ map to a combination of one massive vacuum and some component along $D^b(X_-)$.
Informally, one can say that these $N_+-N_-$ Wilson lines have ``gone away'' to the Coulomb branch, but to be more precise what goes to the Coulomb branch is a complex of these Wilson lines with some in the small window.

\paragraph{Non-Calabi-Yau case: going towards the pure-Higgs phase.}

For definiteness, $N_+>N_-$.
Start now from a brane in the phase $\zeta_{\text{ren}}\ll 0$.
This requires giving both a Higgs branch part $\cB_-\in D^b(X_-)$ and a Coulomb branch part~$\cC_-$.
The non-trivial step now is to find a complex of Wilson lines that reduces to $\cB_-$ and~$\cC_-$.

What is often easily available is a complex $\cB_1\in D^b(V,G)$ such that $F_{\text{geom},-}(\cB_1)=\cB_-$.
Then one uses the Koszul brane $\cK^-$ that has an empty image under $F_{\text{geom},-}$
to construct a complex $\cB_2\in D^b(V,G)$ built from Wilson lines with $\abs{\theta+2\pi q}<N_-\pi$, and such that $F_{\text{geom},-}(\cB_2)=\cB_-$.
Separately, one combines band-restricted branes that reduce to branes on single Coulomb branch massive vacua into a brane $\cB_3\in D^b(V,G)$ that reduces to~$\cC_-$.
The sought-after lift is then $\cB_2\oplus\cB_3$, which is then projected down to $F_{\text{geom},+}(\cB_2\oplus\cB_3)\in D^b(X_+)$ in the pure-Higgs phase.

The relevant diagram is as follows, where $\cT_{\mathbf{w}_{\pm}}$ denote window categories for the big and small windows.  The B-brane category in the $\zeta_{\text{ren}}\ll 0$ phase in fact has a semi-orthogonal decomposition $\langle C,D^b(X_-)\rangle$ into the category~$C$ of Coulomb branch branes (itself further decomposed into individual massive vacua) and the derived category of the Higgs branch~$X_-$.
It would be interesting to clarify the physical meaning of this semi-orthogonal decomposition: it only allows strings stretching in one direction between the Coulomb branch vacua and the Higgs branch.
\begin{equation}
  \begin{tikzpicture}
    \node[inner sep=1pt](GLSM) at (0,0) {$D^b(V,U(1))$};
    \node at (0,-.4) {$\cup$};
    \node(window+) at (0,-.8) {$\cT_{\mathbf{w}\mathrlap{_+}}$};
    \node at (0,-1.2) {$\cup$};
    \node(window-) at (0,-1.6) {$\cT_{\mathbf{w}\mathrlap{_-}}$};
    \node(CH) at (-3,-1.2) {$\langle C,D^b(X_-)\rangle$};
    \node at (-3,-1.6) {$\cup$};
    \node(phase-) at (-3,-2) {$D^b(X_-)$};
    \node(phase+) at (3,-2) {$D^b(X_+)$};
    \draw[->](GLSM)--(CH) node [pos=.6,above,xshift=-4pt] {$F_{g\to\infty}$};
    \draw[->](GLSM)--(phase+) node [pos=.6,above] {$F_+$};
    \draw[->](window-)--(phase-) node [pos=.4,below] {$\cong$};
    \draw[->](window+)--(CH) node [pos=.4,below] {$\cong$};
    \draw[->](window+)--(phase+) node [pos=.6,below] {$\cong$};
  \end{tikzpicture}
\end{equation}

\subsection{\label{ssec:wallcrossingHiggsing}Band restriction rule}

Given a B-brane in a $U(1)^r$ GLSM with no superpotential, we can ask for its image in some phase.
More precisely, as we explained, we take the gauge-decoupling limit $g\to\infty$ while keeping the complexified energy scale~$\mu$ and FI-theta parameters fixed, so that the phase gives a good description of the physics at scale~$\mu$.

Deep in a pure-Higgs phase, at every point on the Higgs branch all continuous gauge symmetries are Higgsed and the vector multiplets have mass $\gtrsim e$.
The massive vector multiplets can thus be integrated out.
Their effect is thus to impose D-term equations and quotient out by gauge symmetry, so that the theory is well-described by an NLSM on the Higgs branch $X_\zeta=V/\!/\!_\zeta\, G$.
A B-brane $\cB\in D^b(V,G)$ of the GLSM, namely a complex of equivariant vector bundles on~$V$, is mapped by these steps to the brane obtained by restricting the vector bundles then pushing forward to the quotient $X_\zeta\simeq(V\setminus\Delta)/G_{\CC}$.  In other words,
\begin{equation}\label{gdeqgeom}
  F_{g\to\infty} = F_{\text{geom}} \colon D^b(V,G) \to D^b(X_\zeta)
\end{equation}
in pure-Higgs phases, for instance in all phases of a Calabi-Yau model.

To integrate out the gauge degrees of freedom it was essential to have no mixed or Coulomb branches.
As we saw in $U(1)$ models, the existence of such branches affects even the Higgs branch image of GLSM branes.
For other phases, the general strategy that we apply to Hirzebruch-Jung models in \autoref{ssec:HJbranesHiggs} is to transport B-branes starting from a pure-Higgs phase.

In this subsection we focus on transporting B-branes through a single wall, deep in that wall, far from other walls.
First we give a physical explanation for band restriction rules, then we justify it using the hemisphere partition function.

\subsubsection{Band restriction rule from Higgsing to a \texorpdfstring{$U(1)$}{U(1)} model}

\paragraph{Geometric aspects.}

The wall is a codimension~$1$ cone in~$\lie{g}^*$.
Let $\lie{h}\subseteq\lie{g}$ be the one-dimensional subspace orthogonal to the hyperplane containing the wall,
and let $I$ be the set of flavours~$i$ such that $Q^i\in\lie{h}^\perp$.
Since the wall is spanned by some charge vectors, $\lie{h}$~contains a non-zero vector with rational coordinates so it generates a compact subgroup isomorphic to $U(1)$ inside $G=U(1)^r$.
We let $u\in\lie{h}$ be the generator normalized such that $\exp(2\pi i\alpha u)=1\in G$ $\Leftrightarrow$ $\alpha\in\ZZ$ and defined up to a sign.

The D-term equation expresses $\zeta_{\text{ren}}$ as a positive linear combination of charge vectors.
Away from walls, this linear combination must involve charge vectors that span all of~$\lie{g}^*$.
As $\zeta_{\text{ren}}$ touches the wall, it belongs to the cone of some $Q^i$, $i\in I$, which means that the Higgs branch has a locus $P\subset X_\zeta$ at which only $X^i$, $i\in I$, are non-zero.
In other words, $P$~consists of common zeros of all chiral multiplets charged under~$\lie{h}$.
Configurations in~$P$ do not Higgs the gauge algebra completely, but rather Higgs it to~$\lie{h}$ (never a larger subalgebra because we take $\zeta_{\text{ren}}$~generic in the wall).
Besides the $U(1)$ factor generated by $\lie{h}$, the unbroken gauge group~$H_p$ at $p\in P$ can have a discrete factor and may depend on~$p$.

Let us assume for simplicity that $P=\{p\}$ is a single point, as this is the situation for Hirzebruch-Jung models.
D-term equations fix $(X_i)_{i\in I}$ completely at~$p$, hence they also fix $(X_i)_{i\in I}$ as functions of $(X_j)_{j\not\in I}$ near~$p$.
Their generic vev in $\lie{h}^\perp$ gives masses to all vector multiplets except those along~$\lie{h}$.
We integrate them out, as well as fluctuations for the $(X_i)_{i\in I}$, in a neighborhood of~$p$.
This gives a \textbf{local model}, with
\begin{equation}\label{localmodelgenericstory}
  \begin{aligned}
    & \text{gauge group } H_p\simeq U(1)\times\Gamma, \qquad \Gamma \text{ discrete,} \\
    & \text{FI-theta parameter } t_{\text{ren}}\cdot u , \\
    & \text{chiral multiplets } (X_j)_{j\not\in I} \text{ of charges } Q^j\cdot u .
  \end{aligned}
\end{equation}
Of course the local model is only a good description near~$p$, but this is precisely the neighborhood whose topology changes upon crossing the wall.
We thus expect wall-crossing to only affect branes in the region near~$p$, which is well-described by the local model.
The two phases $\zeta_{\text{ren}}\cdot u\gtrless 0$ correspond to the two sides of the wall in the full model.

\paragraph{Band restriction rule.}

Assume first that the local model is Calabi-Yau, namely that $\sum_{j\not\in I} Q^j\cdot u=0$, itself equivalent to $Q^{\text{tot}}\cdot u=0$, namely the wall is Calabi-Yau.
In each phase, the local model only has a Higgs branch.
A Wilson line with some charge under $U(1)\subset H_p$ can be transported between the phases $\zeta_{\text{ren}}\cdot u\gg 0$ and $\zeta_{\text{ren}}\cdot u\ll 0$ provided its $U(1)$ charge is band-restricted.
Translating back to the full GLSM, \textbf{band-restricted} branes are $\cW(q)$ such that~\cite{Herbst:2008jq}
\begin{equation}
  \abs[\big]{(\theta\cdot u) + 2\pi (q\cdot u)} < \pi N_{u,+} = \pi N_{u,-} , \qquad
  N_{u,\pm} = \sum_{j} (Q^j\cdot u)^{\pm}
\end{equation}
where $(x)^{\pm}\coloneqq(\abs{x}\pm x)/2$ and we extended the sum to all~$j$ since $Q^j\cdot u=0$ for $j\in I$.
Complexes of band-restricted Wilson lines are transported unchanged through the wall.

If the local model is not Calabi-Yau, we choose the sign of~$u$ such that $Q^{\text{tot}}\cdot u>0$.
The local model then has a pure-Higgs phase at $\zeta_{\text{ren}}\cdot u\gg 0$ and a phase with a Higgs branch and some massive vacua at $\zeta_{\text{ren}}\cdot u\ll 0$.
These massive vacua lie at $Q^{\text{tot}}\cdot u$ values on the classical Coulomb branch, on which the discrete abelian group~$\Gamma$ acts trivially, so twisted sectors increase the number of massive vacua to $\abs{\Gamma}\,Q^{\text{tot}}\cdot u$.
Brane transport from $\zeta_{\text{ren}}\cdot u\gg 0$ to $\zeta_{\text{ren}}\cdot u\ll 0$ now involves two nested windows, which translate to bands in the full GLSM:
\begin{equation}\label{bandsanomalous}
  \begin{alignedat}{4}
    & \textbf{small band} & \qquad &
    \abs[\big]{(\theta\cdot u) + 2\pi (q\cdot u)} < \pi \min(N_{u,+},N_{u,-}) , \\
    & \textbf{big band} & &
    \abs[\big]{(\theta\cdot u) + 2\pi (q\cdot u)} < \pi \max(N_{u,+},N_{u,-}) ,
  \end{alignedat}
\end{equation}
where $N_{u,\pm} = \sum_j (Q^j\cdot u)^{\pm}$ are defined as before.
We call \textbf{band-restricted} a brane whose charges are in the big band.
A Wilson line in the small band gets transported to the Higgs branch,
while a Wilson line in the big band but not the small one is transported to a combination of the Higgs branch and one massive vacuum.

\paragraph{Possible generalizations.}

The charge of the Wilson line under $\Gamma\subset H\subset G$ only plays a role in determining which massive vacuum appears.  We defer to later work a careful analysis of the contribution from massive vacua.

If the locus~$P$ with unbroken gauge symmetry has non-zero dimension, the analogue of these massive vacua is a collection of mixed branches on which~$\sigma\in\lie{h}$ and the chiral multiplets $(X_i)_{i\in I}$ both get a vev.
A natural conjecture is that the band restriction rule still applies: a Wilson line in the small band is transported to a brane supported purely on the Higgs branch, while a Wilson line in the big band maps to a combination of a Higgs branch brane and a brane supported on a specific mixed branch.
The fact that the discrete part of the unbroken subgroup~$H_p$ can depend on $p\in P$ manifests itself in the possible presence of orbifold singularities in the locus~$P$ and in the mixed branch to which the Wilson line maps.

We caution that the approximations we made are only valid for crossing a single wall in an asymptotic regime.  If we wish to cross multiple walls, each wall crossing will be in a different asymptotic regime and thus involve a different $U(1)$ subgroup and hence lead to a different band restriction rule.
Crossing several walls in a row gives successive band restriction rules and one should consider the intersection of these various bands.

\subsubsection{Band restriction rule from the hemisphere partition function}

In \autoref{ssec:BbraneGLSM} we discussed the hemisphere partition function $Z_{D^2}(\cB)$ with a B-brane.
In Calabi-Yau models it can be converted by closing contours to an infinite sum of residues~\eqref{ZD2locabHiggs}.

For $U(1)$ models that are not Calabi-Yau, say with $Q^{\text{tot}}>0$,
we reviewed how brane transport can be understood by comparing two expansions of~$Z_{D^2}(\cB)$.
The first, corresponding to the phase $\zeta_{\text{ren}}\gg 0$, is to write $Z_{D^2}(\cB)$ as a sum of residues~\eqref{ZD2locabHiggs} at poles of chiral multiplet one-loop determinants with $Q^j>0$.
The second expansion corresponds to the phase $\zeta_{\text{ren}}\ll 0$, and it can only be carried out straightforwardly for branes in the big window.
It consists of decomposing $Z_{D^2}(\cB)$ into a Higgs branch contribution from residues with $Q^j<0$, and a massive vacuum contribution from a steepest descent contour.
For branes outside the big window, the sum of residues~\eqref{ZD2locabHiggs} does not correctly give the Higgs branch contribution.

We generalize these ideas now to $U(1)^r$ non-Calabi-Yau models.
In generic phases we expect one Higgs branch and a collection of mixed and Coulomb branches.
A mixed branch is characterized by the space $\lie{q}^\perp\subset\lie{g}$ in which $\hat{\sigma}$ varies, and correspondingly by the set $I=\{i\mid Q^i\in\lie{q}\}$ of flavours that are not given a mass by the vev of~$\hat{\sigma}$.
The mixed branch is roughly a product of some Higgs branch for $(X_i)_{i\in I}$ and some Coulomb branch vacua with $\hat{\sigma}\in\lie{q}^\perp$.
The mixed branch is found by integrating out all chiral multiplets $(X_j)_{j\not\in I}$ to get the effective twisted superpotential for $\hat{\sigma}\in\lie{q}^\perp$, then by solving a D-term equation for the remaining~$(X_i)_{i\in I}$.
How is $\hat{\sigma}$ reduced to $\lie{q}^\perp$?
Physically, this reduction is due to the non-zero vevs of $(X_i)_{i\in I}$.
Computationally, the reduction is done by closing some of the integrals to pick up residues at common poles of $\dim\lie{q}$ chiral multiplet one-loop determinants of $(X_i)_{i\in I}$.
Indeed, the leading pole\footnote{Other poles are related to vortex configurations and are subleading in the series over poles.} of such a one-loop determinant $\Gamma(iQ^i\cdot\hat{\sigma}+R_i/2)$ is at $Q^i\cdot\hat{\sigma}=iR_i/2$, which is close to the constraint imposed by the non-zero vev of~$X_i$.
Intersections of $\dim\lie{q}$ hyperplanes impose $\hat{\sigma}$ approximately in~$\lie{q}^\perp$, as we want.
Our $U(1)$ experience then suggests to compute the integral by a saddle-point approximation at large generic $\abs{\hat{\sigma}}$ within $\hat{\sigma}\in\lie{q}^\perp$.
In summary,
\begin{itemize}
\item pick up residues at common poles of $\dim\lie{q}$ chiral multiplet one-loop determinants;
\item find saddles of the integral over $\hat{\sigma}\in\lie{q}^\perp$.
\end{itemize}

This process of decomposing $Z_{D^2}(\cB)$ into contributions of various branches is quite difficult to carry out from first principles.
Instead, we relate decompositions upon wall-crossing.
For simplicity we focus on the Higgs branch.

In pure-Higgs phases, the partition function is given by the convergent series of residues \eqref{ZD2locabHiggs}, which gives the Higgs branch (and only) contribution.
Starting from such a phase, we cross the walls from $\zeta_{\text{ren}}\cdot u\gg 0$ to $\zeta_{\text{ren}}\cdot u\ll 0$, where the sign of~$u$ is chosen as above such that $Q^{\text{tot}}\cdot u\geq 0$.
At each wall, provided the brane ``goes through the wall'', namely fits in the big band~\eqref{bandsanomalous}, we find that the sum of residues~\eqref{ZD2locabHiggs} splits into the analogous sum for the other phase, and a mixed branch contribution (with $\lie{q}^\perp=\lie{h}$ in the notation above).
If the brane does not fit in the big band, the sum of residues does not give the correct Higgs branch contribution.
The picture that emerges is that the sum of residues only correctly gives the Higgs branch contribution, in some given phase, for branes that go through all walls between a pure-Higgs phase and that phase.
It is not clear whether for every phase there should exist a collection of branes that go through all walls and generates the Higgs branch B-brane category.
For (the resolved phase of) Hirzebruch-Jung models, there is.

Let us thus start with the sum of residues~\eqref{ZD2locabHiggs}, ranging over collections $J$ of $r$ flavours such that $\zeta_{\text{ren}}\in\Cone_J$.
Some collections appear on both sides of the wall and are uninteresting for us: these are analogous to parts of the Higgs branch whose topology doesn't change under wall-crossing.
Collections that are only allowed on one side of the wall must take the form $J=\{j\}\cup J'$, where charge vectors $(Q^i)_{i\in J'}$ lie in the wall, hence $J'\subset I$.
The sum of residue then takes the form
\begin{equation}\label{ZD2locabHiggswall}
  Z_{D^2,\text{residue}}(\cB) \simeq \sum_{J'\subset I,} \sum_{k:J'\to\ZZ_{\geq 0}} \Biggr(
 \sum_{j\not\in I\mid\zeta_{\text{ren}}\in\Cone_{J'\cup\{j\}}} \sum_{k_j\geq 0}
  \pm \res_{i\hat{\sigma}=i\hat{\sigma}_{J'\cup\{j\},k}} ({\cdots}) \Biggr) + \dots ,
\end{equation}
where the trailing dots denote other collections~$J$, and we recall that $\hat{\sigma}_{J,k}$ is the common solution of $iQ^i\cdot\hat{\sigma}+R_i/2=-k_i$ for $i\in J$.

The sums over $j$ and $k_j$ recombine into a one-dimensional contour integral which can be found in several ways.
The simplest way is to start from the original contour integral and select the residue at $iQ^i\cdot\hat{\sigma}+R_i/2=-k_i$ for all $i\in J'$.
These conditions on $Q^i\cdot\hat{\sigma}$ force $\hat{\sigma}$ to belong to~$\lie{h}$ up to some imaginary offset $r_{J',k}>0$, so $\hat{\sigma}=ir_{J',k}+su$ with an integral over~$s$.
We get schematically
\begin{equation}
  Z_{D^2,\text{residue}}(\cB) \simeq \sum_{J'\subset I,} \sum_{k:J'\to\ZZ_{\geq 0}} \int_{\hat{\sigma}=ir_{J',k}+su} \mathrm{d}s
  \Bigl( \pm \res_{i\hat{\sigma}=ir_{J',k}+su} ({\cdots}) \Bigr) + \dots ,
\end{equation}
Each Gamma function $\Gamma\bigl(iQ^i\cdot\hat{\sigma}+R_i/2\bigr)$ in the integrand with $i\not\in J'$ becomes $\Gamma\bigl(i(Q^i\cdot u)s+\text{real}\bigr)$, namely one of the one-loop determinants of the local model.
Other parts simplify similarly and the resulting one-dimensional contour integral is itself a hemisphere partition function: that of the local model~\eqref{localmodelgenericstory} with various R-charge assignments.
Picking up residues on one side or the other of the contour gives~\eqref{ZD2locabHiggswall} on either side of the wall.
This was expected since performing all $r$ contour integrals should be the same as preforming $r-1$ and then the last one.

As we reviewed extensively in \autoref{ssec:U1reviewcareful}, if the local model is not Calabi-Yau, the sums of residues on both sides of the wall differ by a contour integral that should be evaluated by a saddle-point calculation.
This reproduces the band-restricted rule in a more rigorous way than we did previously.

\subsection{\label{ssec:HJbranesHiggs}Brane transport in phases of Hirzebruch-Jung models}

We determine here the Higgs branch image of Wilson lines in various phases of Hirzebruch-Jung models.
For Calabi-Yau cases, this image is given in each phase by a geometric functor~\eqref{gdeqgeom}, so we only consider non-Calabi-Yau models.
The strategy is then to transport branes from the orbifold phase (which is pure-Higgs) to the phase of interest, taking into account the band restriction rules~\eqref{bandsanomalous}.
For the general $\CC^2/\ZZ_{n(p)}$ model we focus on reaching the fully resolved phase, by crossing walls in a particular order: for each~$\theta$ there are $n$ band-restricted Wilson line branes and we determine their Higgs branch images in the fully resolved phase.  We reproduce the notion of ``special'' representations of~$\ZZ_n$~\cite{riemenschneider1987characterization} as corresponding to cases where the brane has no mixed or Coulomb component.  Then we explore all phases of a rank~$2$ example and find that brane transport depends on the order of wall crossings.

\subsubsection{Contour and empty branes}

In $U(1)$ non-Calabi-Yau models there is a universal contour for~$\hat{\sigma}$ that is admissible for all branes.
Let us show in our Hirzebruch-Jung model that the contour\footnote{We find it useful to switch back and forth between basis I and basis II of the GLSM throughout this discussion, hence we state explicitly which basis is used.}
\begin{equation}\label{universalHJcontour}
  \Bigl\{ \bigl(\hat{\tau}_1-i(\hat{\tau}_1)^2,\dots,\hat{\tau}_r-i(\hat{\tau}_r)^2\bigr) \text{ in basis II} \Bigm| \hat{\tau}\in\RR^r\Bigr\} \subset \lie{g}_\CC
\end{equation}
is a deformation of $\RR^r$ that ensures convergence of the hemisphere partition function for Wilson lines $\cW(q)$ with arbitrary $\zeta_{\text{ren}}$, $\theta$ and~$q$.

The condition to be a deformation of~$\RR^r$ is that no pole of one-loop determinants are encountered when deforming from $\RR^r$ to the contour.
To be precise, we must turn on small positive R-charges to avoid the contour pinching discussed in \autoref{ssec:regZU1} onwards.
The poles are at $Q^j\cdot\hat{\sigma}\in iR_j/2+i\ZZ_{\geq 0}\subset i\RR_{>0}$, so the condition to avoid crossing any pole is that none of the $Q^j\cdot\hat{\sigma}$ should touch the line $i\RR_{>0}$ anywhere on the contour.
For $1\leq j\leq r$ we easily check $Q^j\cdot\hat{\sigma}=-n\hat{\tau}_j+in\hat{\tau}_j^2\not\in i\RR_{>0}$.  For $j=0$ (and $j=r+1$ by replacing $p_\alpha\to q_\alpha$) we check
\begin{equation}
  \Im(Q^0\cdot\hat{\sigma}) = \Im\biggl(\sum_{\alpha=1}^r p_\alpha (\hat{\tau}_\alpha-i\hat{\tau}_\alpha^2)\biggr)
  = - \sum_{\alpha=1}^r p_\alpha\hat{\tau}_\alpha^2 \leq 0 .
\end{equation}

Next, we study the integrand of the hemisphere partition function far along the contour.
For this we write $\hat{\tau}_\alpha=\lambda\hat{n}_\alpha$ for real $\lambda\gg 0$ and the direction~$\hat{n}$ normalized in an arbitrary way, say, $\sum_\alpha\hat{n}_\alpha^2=1$.
The arguments of Gamma functions are quadratic in~$\lambda$ so we need the asymptotics
\begin{equation}
  \log\abs{\Gamma(\alpha\lambda^2+i\beta\lambda+\gamma)}
  = 2\alpha\lambda^2\log(\lambda) + O(\lambda^2)
\end{equation}
for $(\alpha,\beta)\in\RR^2$, $\gamma\in\CC$, except obviously for $\alpha=\beta=0$ and $\gamma\in\ZZ_{\leq 0}$.
We deduce
\begin{equation}
  \log\abs{\text{integrand}}
  = \sum_{j=0}^{r+1} 2(Q^j\cdot\hat{n}^2) \lambda^2\log(\lambda) + O(\lambda^2)
  \leq - 2 \lambda^2\log(\lambda) + O(\lambda^2)
\end{equation}
where we denoted abusively $\hat{n}^2$ the vector whose components in basis II are $\hat{n}_\alpha^2$.
To show the inequality, note that components of $Q^{\text{tot}}=\sum_j Q^j$ in basis~II are $p_\alpha+q_\alpha-n\leq -1$ (except in the Calabi-Yau case), and we normalized~$\hat{n}$.
We thus find that the integrand is exponentially suppressed at infinity along the contour.
The FI-theta and Wilson line contributions are subleading (of order~$\lambda^2$) hence the same contour works for all $t_{\text{ren}}$ and~$q$, hence also arbitrary complexes of Wilson lines.

As in $U(1)$ models, we now find empty branes as Koszul resolutions in a pure-Higgs phase.

Pure-Higgs phases are those whose closure contains~$Q^{\text{tot}}$ (see \autopageref{para:pureHiggs}).
For our Hirzebruch-Jung model, components $p_\alpha+q_\alpha-n$ of~$Q^{\text{tot}}$ in basis~II are negative, hence $Q^{\text{tot}}$ lies in the interior of the orbifold phase~$\RR_{<0}^r$ (see \autoref{ssec:HiggsGeom}).
The orbifold phase is thus the only pure-Higgs phase.
It has Higgs branch $\CC^2/\ZZ_{n(p)}$.
The gauge group is broken to the $\ZZ_n$ subgroup that leaves $X_1,\dots,X_r$ fixed and acts on $P=X_0$ and $Q=X_{r+1}$ with charges $p$ and~$1$, respectively.
The $n$-th root of unity~$\omega$ embeds in $U(1)^r$ with coordinates $\omega^{p_\alpha}$ in basis~I\@.
The Higgs branch image of a Wilson line with charges $(b_1,\dots,b_r)$ in basis~I is thus an equivariant line bundle on $\CC^2/\ZZ_{n(p)}$ with $\ZZ_n$ charge
\begin{equation}\label{ZnchargeWilsonline}
  \sum_{\alpha=1}^r p_\alpha b_\alpha .
\end{equation}

By construction of the $\ZZ_n$ subgroup, a Wilson line with the same charges as~$X_\alpha$ for some $1\leq\alpha\leq r$ becomes an equivariant bundle with vanishing $\ZZ_n$~charge.
This simply restates the fact that the Koszul branes ($1\leq\alpha\leq r$)
\begin{equation}\label{KalphaHJ}
  \cK_\alpha\coloneqq \Bigl( \cW(\dots,0,-1,a_\alpha,-1,0,\dots) \xrightarrow{X_\alpha} \cW(0) \Bigr)
\end{equation}
are empty in the orbifold phase.
The brane can then be transported to an arbitrary phase since the contour~\eqref{universalHJcontour} remains admissible for arbitrary~$\zeta_{\text{ren}}$.
The image of this same complex~$\cK_\alpha$ in each phase remains empty.\footnote{The situation is quite different in Calabi-Yau models, where the contour may stop being admissible (and has no admissible deformation) as we cross a wall because some of the Wilson lines constituting~$\cK_\alpha$ cannot be defined at the wall itself.}
At the level of hemisphere partition functions we are simply stating that the analytic continuation of a function that is identically zero is zero.

Given any complex of Wilson line, our first step to find its image in any given phase is to bind it with the empty branes~$\cK_\alpha$ to reduce all charges to a fundamental domain of the quotient
\begin{equation}\label{Zr/Q=Zn}
  \ZZ^r \Bigm/ \Bigl( \Span_{\ZZ} \bigl\{Q^j\bigm| 1\leq j\leq r\bigr\} \Bigr) \simeq \ZZ_n .
\end{equation}
Our choice of fundamental domain will be guided by the band restriction rule: we wish to find some Wilson lines that go through all walls in the sense of being in the big band upon crossing each wall.
This ensures that the Higgs branch images of these Wilson lines match their images under~$F_{\text{geom}}$ in all phases.
Since the bands depend on~$\theta$, our choice of fundamental domain should depend on~$\theta$.

\subsubsection{Going to the fully resolved phase}

We transport branes from the orbifold phase to the fully resolved phase through a particular sequence of walls, which we choose to be blowing up exceptional divisors in the order $E_1,\dots,E_r$.

Consider a Wilson line with charges $(b_1,\dots,b_r)$ in basis~I\@.
The local model describing how some $E_j$ is blown up was given in~\eqref{localdivisorchargesU1Zm}.
In our present case $i=j-1$ and $k=r+1$, the local model is a $U(1)$ GLSM with chiral multiplets $X_i$, $X_j$, $X_k$ of $U(1)$ charges $d_{jk}=p_j$, $-d_{ik}=-p_{j-1}$, and $d_{ij}=1$.
The embedding $U(1)\subset U(1)^r$ means that $\cW(b)$ has charge
\begin{equation}
  b_{U(1)} = \sum_{\alpha=j}^r p_\alpha \, b_\alpha .
\end{equation}
To write the bands~\eqref{bandsanomalous} we work out $N_{u,+}=p_j+1\leq N_{u,-}=p_{j-1}$.
The bands are:
\begin{equation}\label{bandsHJ}
  \begin{alignedat}{4}
    & \text{small band} & \qquad &
    \abs[\bigg]{\sum_{\alpha=j}^r p_\alpha(\theta_\alpha+2\pi b_\alpha)} < \pi (p_j+1) , \\
    & \text{big band} & &
    \abs[\bigg]{\sum_{\alpha=j}^r p_\alpha(\theta_\alpha+2\pi b_\alpha)} < \pi p_{j-1} .
  \end{alignedat}
\end{equation}
Let us show that the $r$ big band inequalities~\eqref{bandsHJ} define a fundamental domain (ignoring boundaries, which play no role for generic~$\theta$) of the quotient~\eqref{Zr/Q=Zn}.  Namely let us show that for every $b\in\ZZ^r$ there is a unique choice of integers~$c_j$ such that the big band inequalities~\eqref{bandsHJ} hold with $b$ shifted to $b-\sum_{j=1}^r c_jQ^j$.  For $1\leq k\leq r$, note that $\sum_{\alpha=j}^r p_\alpha Q^k_\alpha$ vanishes, by the recursion relation above~\eqref{pqqpmodn}, except for $j=k+1$ where it is $p_{k+1}$, and $j=k$, where it is $-p_{k-1}$.  This means that $c_k$ only affects the big band inequalities for $j=k$ and $j=k+1$.  Importantly, for $j=k$ it shifts the left-hand side from an interval $(-\pi p_{k-1},\pi p_{k-1})$ of width $2\pi p_{k-1}$ to the next.  This lets us solve recursively for~$c_j$ starting from~$c_1$:
\begin{equation}
  c_j = \operatorname{round} \biggl( \frac{p_j c_{j-1}}{p_{j-1}} - \sum_{\alpha=j}^r \frac{p_\alpha}{p_{j-1}}\biggl(\frac{\theta_\alpha}{2\pi}+b_\alpha\biggr) \biggr)
\end{equation}
where ``round'' stands for the closest integer (since we take generic~$\theta$, there is no need to specify how half-integers are rounded).

To summarize, one finds the Higgs branch image of a complex of Wilson lines as follows.
Bind the complex to empty branes~$\cK_\alpha$ (and their twists by $\cO(\dots)$) so as to reduce charges to the fundamental domain defined by the big bands~\eqref{bandsHJ}.  The Higgs branch image of the original brane is then equal to that of the new brane, which coincides with the image of the new brane under the GIT quotient.
For example in the fully resolved phase that quotient construction maps a Wilson line brane $\cW(b_1,\dots,b_r)$ in the fundamental domain to a line bundle whose pull-back to the $j$-th exceptional divisor~$\CP^1$ is $\cO(b_j)$ (see the discussion near~\eqref{DbHJpullback}).
In the orbifold phase, the image is instead an equivariant line bundle with charge $\sum_\alpha p_\alpha b_\alpha$ according to~\eqref{ZnchargeWilsonline}.

So far we have only discussed Higgs branch images.

Next, let us find Wilson line branes whose IR image is supported entirely on the Higgs branch in the fully resolved phase, so that our description above specifies the image completely.  We choose a convenient (albeit somewhat baroque) value of~$\theta$: other values would give different transport functors.  Namely we take $\theta$ to be the solution of
\begin{equation}\label{particulartheta}
  \sum_{\alpha=j}^r p_\alpha\theta_\alpha = -\pi (p_j+1) + 2\pi\varepsilon
  \text{ for }1\leq j\leq r,
\end{equation}
for some unimportant $\varepsilon\in(0,1)$.
This makes small bands quite nice: the $j$-th one is then
\begin{equation}\label{jthband}
  \sum_{\alpha=j}^r p_\alpha b_\alpha \in [0,p_j] .
\end{equation}
Since all $p_\alpha\leq p_j$ for $\alpha\geq j$, it is clear that the $r+1$ Wilson lines (in basis~I)
\begin{equation}\label{smallWilsonlines}
  \cW(0,\dots,0) \text{ and } \cW(\dots,0,1,0,\dots)
\end{equation}
with either zero or one non-zero entries~$1$ go through all small bands.
The converse holds: any Wilson line $\cW(b)$ that fits in all small bands~\eqref{jthband} must be one of these.
If $b=0$ we are done.
Otherwise, let $1\leq j\leq r$ be the position of the last non-zero entry of~$b$, namely $b_j\neq 0$ and $b_{j+1}=\dots=b_r=0$.
The $j$-th small band restriction rule simplifies to $p_j b_j\in\{0,\dots,p_j\}$ hence $b_j=1$ (we assumed $b_j\neq 0$).
If $b_j$ is the only non-zero entry we are done: the brane is one of~\eqref{smallWilsonlines}.
Otherwise, let $1\leq i<j$ be the position of the last non-zero entry before~$b_j$, namely $b_i\neq 0$ and $b_{i+1}=\dots=b_{j-1}=0$.
The $i$-th small band restriction rule simplifies to $p_ib_i+p_j\in\{0,\dots,p_i\}$.  The only solution is $b_i=0$ because $p_i>p_j>0$.

Since they fit in small bands, the image of these Wilson lines~\eqref{smallWilsonlines} in any of the phases that we visited is purely along the Higgs branch, with no massive vacuum part.

In the fully resolved phase, the image of $\cW(0)$ is the structure sheaf, while the image of $\cW(\dots,0,1,0,\dots)$ is a line bundle whose pull-back to each exceptional divisor~$\CP^1$ except one is $\cO(0)$, and the last one~$\cO(1)$ (see the discussion near~\eqref{DbHJpullback}).
In the orbifold phase, the image of $\cW(0)$ is the trivial line bundle on $\CC^2/\ZZ_{n(p)}$, while $\cW(\dots,0,1,0,\dots)$ maps to an equivariant line bundle with charge $p_\alpha$ according to~\eqref{ZnchargeWilsonline}.
Thus we see that the branes which can be passed through all the walls to the large volume phase are precisely the fractional branes corresponding to
\begin{enumerate}
\item the trivial representation and
\item the ``special'' representations of~$\ZZ_n$ in the language of~\cite{riemenschneider1987characterization}
\end{enumerate}
We refer to the notes~\cite{riemenschneider2003special} for further mathematical references.

\subsubsection{\label{ssec:C2Zn2 brane transport}Non-trivial monodromy in the \texorpdfstring{$\mathbb{C}^2/\mathbb{Z}_{n(2)}$}{C2/Zn(2)} model}

We now illustrate how crossing walls in different orders can give different transport functors.
Specifically we consider the two-parameter model $\CC^2/\ZZ_{n(2)}$ with $n=2k-1\geq 5$.  Its charge matrices in basis I and~II are respectively as follows.
\begin{equation}
  \begin{array}{cccccc}
    \toprule
           & X_0 & X_1 & X_2 & X_3 \\
    \midrule
      U(1)_1 & 1 & -k & 1 & 0 \\
      U(1)_2 & 0 & 1 & -2 & 1 \\
    \bottomrule
  \end{array}
  \qquad
  \begin{array}{cccccc}
    \toprule
           & X_0 & X_1 & X_2 & X_3 \\
    \midrule
      U(1)_1' & 2 & -n & 0 & 1 \\
      U(1)_2' & 1 & 0 & -n & k \\
    \bottomrule
  \end{array}
\end{equation}
We carefully analysed its phases starting in~\eqref{twoparameter}, finding all Coulomb branch and mixed branch vacua.
For convenience we reproduce the phase diagram of \autoref{fig:phasesC2Zn2}: the left diagram is in basis~I and the right one in basis~II and we have indicated different data on the two sides to avoid clutter, such as the number of massive vacua in each phase.
We transport a Wilson line $\cW(b_1,b_2)$ (in basis~I\@) from $A=\emptyset$ to $A=\{1,2\}$ through either $A=\{1\}$ or $A=\{2\}$.
We recall that $A$~is the set of blown-up divisors.
\[
  \phasesofGLSM
\]

As argued near~\eqref{KalphaHJ}, since the model has a single pure-Higgs phase, the branes that are empty in that phase are empty everywhere.  These are the two branes
\begin{equation}\label{K12example}
  \begin{aligned}
    \cK_1 & \coloneqq  \Bigl( \cW(k,-1) \xrightarrow{X_1} \cW(0,0) \Bigr) , \\
    \cK_2 & \coloneqq  \Bigl( \cW(-1,2) \xrightarrow{X_2} \cW(0,0) \Bigr) , \\
  \end{aligned}
\end{equation}
By binding a Wilson line brane $\cW(b_1,b_2)$ with these branes~$\cK_\alpha$ and more generally with $\cK_\alpha\otimes\cW(c_1,c_2)$, one can shift the charge $(b_1,b_2)$ to any fundamental domain of the quotient $\ZZ^2/((k,-1)\ZZ+(-1,2)\ZZ)$.  Such binding does not affect the image of the Wilson line brane in the gauge-decoupling limit.

\paragraph{First transport.}

The band restriction rules for passing from the orbifold phase $A=\emptyset$ to the phase $A=\{1\}$ where $E_1$ is blown up are~\eqref{bandsHJ}
\begin{equation}
  \abs[\bigg]{2b_1+b_2+\frac{2\theta_1+\theta_2}{2\pi}} <
  \begin{cases}
    3/2 \quad & \text{small band,} \\
    n/2 \quad & \text{big band.}
  \end{cases}
\end{equation}
The band restriction rules for passing then to the fully resolved phase $A=\{1,2\}$ are the same for small and big bands:
\begin{equation}
  \abs[\bigg]{b_2+\frac{\theta_2}{2\pi}} < 1 .
\end{equation}
Taking $\theta_1,\theta_2>0$ small for definiteness (this choice affects the brane transport functor)\footnote{Note that we use a different value of $\theta_1,\theta_2$ than the value~\eqref{particulartheta} taken in the previous subsubsection.}, the Wilson lines that go through both small bands are
\begin{equation}\label{C2Zn2small1}
  \cW(0,0), \quad
  \cW(0,-1), \quad
  \cW(1,-1),
\end{equation}
while the $n$ Wilson lines going through both big bands are
\begin{equation}\label{C2Zn2big1}
  \cW(b_1,0) \text{ for } 1-\lceil k/2\rceil \leq b_1\leq \lceil k/2\rceil-1
  \text{ and }
  \cW(b_1,-1) \text{ for } 1-\lfloor k/2\rfloor\leq b_1\leq\lfloor k/2\rfloor .
\end{equation}
As we have discussed for general models, the IR images of the Wilson lines~\eqref{C2Zn2small1} are supported on the Higgs branch (in fact they generate the derived category of the resolved geometry) while all other Wilson lines in~\eqref{C2Zn2big1} have a Coulomb branch and a Higgs branch components.
As explained in \autoref{ssec:Bbraneabelian}, the Higgs branch component of each Wilson line in~\eqref{C2Zn2big1} is given by the geometric projection~$F_{\text{geom}}(\cW(b_1,b_2))$.
For more general charges, one first binds with the Koszul branes~\eqref{K12example}, and in terms of $\ell=(b_1+kb_2\bmod{n})$ with $0\leq\ell\leq n-1=2k-2$ one finds
\begin{equation}\label{example-FH1}
  \cW(b_1,b_2)
  \xrightarrow[\text{restrict to Higgs branch}]{\text{transport via phase $A=\{1\}$}}
  \begin{cases}
    F_{\text{geom}}(\cW(\ell,0)) & \text{if $\ell\leq \lceil k/2\rceil-1$,} \\
    F_{\text{geom}}(\cW(\ell-k+1,-1)) & \text{if $\lceil k/2\rceil\leq\ell\leq \lfloor k/2\rfloor+k-1$,} \\
    F_{\text{geom}}(\cW(\ell-2k+1,0)) & \text{if $\lfloor k/2\rfloor+k\leq\ell$.}
  \end{cases}
\end{equation}

\paragraph{Second transport.}

If instead we go from $A=\emptyset$ to $A=\{2\}$ to $A=\{1,2\}$, the bands are
\begin{equation}
  \abs[\bigg]{b_1+kb_2+\frac{\theta_1+k\theta_2}{2\pi}} <
  \begin{cases}
    (k+1)/2 \quad & \text{small band,} \\
    n/2 \quad & \text{big band,}
  \end{cases}
\end{equation}
and
\begin{equation}
  \abs[\bigg]{b_1+\frac{\theta_1}{2\pi}} <
  \begin{cases}
    1 \quad & \text{small band,} \\
    k/2 \quad & \text{big band.}
  \end{cases}
\end{equation}
Taking small $\theta_1,\theta_2>0$ for definiteness, we find that only the $n$ Wilson lines
\begin{equation}
  \begin{aligned}
    \cW(b_1,-1) & \text{ for } 1\leq b_1\leq \lceil k/2\rceil - 1, \\
    \cW(b_1,0) & \text{ for } {-\lfloor k/2\rfloor}\leq b_1\leq \lceil k/2\rceil - 1, \\
    \cW(b_1,1) & \text{ for } {-\lfloor k/2\rfloor}\leq b_1\leq -1, \\
  \end{aligned}
\end{equation}
go through both big bands, while only the Wilson lines $\cW(0,0)$ and $\cW(-1,0)$ go through both small bands.  IR limits of these two are thus supported purely on the Higgs branch, while IR limits of the other Wilson lines are combinations of Coulomb branch and Higgs branch branes.
We can determine the Higgs branch component for general charges $(b_1,b_2)$: denoting $\ell=(b_1+kb_2\bmod{n})$ with $0\leq\ell\leq n-1=2k-2$ as in~\eqref{example-FH1}, one finds
\begin{equation}\label{example-FH2}
  \cW(b_1,b_2)
  \xrightarrow[\text{restrict to Higgs branch}]{\text{transport via phase $A=\{2\}$}}
  \begin{cases}
    F_{\text{geom}}(\cW(\ell,0)) & \text{if $\ell\leq \lceil k/2\rceil-1$,} \\
    F_{\text{geom}}(\cW(\ell-k,1)) & \text{if $\lceil k/2\rceil\leq\ell\leq k-1$,} \\
    F_{\text{geom}}(\cW(\ell-k+1,-1)) & \text{if $k\leq\ell\leq \lceil k/2\rceil+k-2$,} \\
    F_{\text{geom}}(\cW(\ell-2k+1,0)) & \text{if $\lceil k/2\rceil+k-1\leq\ell$.}
  \end{cases}
\end{equation}
The resulting objects in the derived category of the Hirzebruch-Jung resolution of~$\CC^2/\ZZ_{n(2)}$ differ from~\eqref{example-FH1} in general: for $0\leq\ell\leq\lceil k/2\rceil-1$ and $k\leq\ell\leq\lceil k/2\rceil+k-2$ and $\lfloor k/2\rfloor+k\leq\ell\leq 2k-2$ the results coincide, whereas for $\lceil k/2\rceil\leq\ell\leq k-1$, and for $\ell=3k/2-1$ (with $k$~even) the resulting Higgs branch images differ.

As we just saw, the transport functors corresponding to paths from $A=\emptyset$ to $A=\{1,2\}$ through either region $A=\{1\}$ or $A=\{2\}$ at fixed $\theta_1,\theta_2$ are different.
This means that the two paths are not homotopic, which indicates that the four walls depicted in the phase diagram join non-trivially in the $\theta$ directions.  It may be interesting to investigate carefully the topology of the complexified K\"ahler moduli space of this model, for instance by determining singularities in hemisphere partition functions.
In more general models, developping efficient tools to work out all Coulomb/mixed/Higgs branch components after transport would be worthwhile.

\section*{Acknowledgments}

We would like to thank L. Borisov, T. Okuda and D. Pomerleano for enlightening discussions and we would like to thank especially G. Moore for suggesting this project and providing references.
The work largely occurred while BLF was employed at the Princeton Center for Theoretical Science and then at the Philippe Meyer Institute of \'Ecole Normale Sup\'erieure.
JC and BLF thank the IH\'ES for hospitality during the school ``Supersymmetric localization and exact results''.
MR would like to thank Rutgers University, Steklov Mathematical Institute, 
Higher School of Economics and Heidelberg University for hospitality while part 
of this work has been performed. MR acknowledges support from the National Key 
Research and Development Program
of China, grant No.\ 2020YFA0713000, the Research Fund for International
Young Scientists, NSFC grant No.\ 1195041050 and the support of the 
Institute for Advanced Study, DOE grant DE-SC0009988 and the Adler Family Fund.

\bibliographystyle{fullsort}
\bibliography{bibliography}
\end{document}